\documentclass[]{aastex61}
\usepackage{epstopdf}

\newcommand\aastex{AAS\TeX}

\received{December 31, 2017}
\revised{xx}
\accepted{xx}
\submitjournal{ApJ}

\shorttitle{\aastex\ sample article}
\shortauthors{Cai}
\begin{document}

\title{Numerical Analysis of Nonlocal Convection --- Comparison with Three-dimensional Numerical Simulations of Efficient Turbulent Convection}

\correspondingauthor{TAO CAI}
\email{tcai@must.edu.mo,caitao7@mail.sysu.edu.cn}

\author[0000-0003-3431-8570]{TAO CAI}
\affil{Lunar and Planetary Science Laboratory, Macau University of Science and Technology, Macau, P.R.China}
\affil{School of Mathematics, Sun Yat-sen University, No. 135 Xingang Xi Road, Guangzhou, 510275, P.R.China}



\begin{abstract}
We compare 1D nonlocal turbulent convection models with 3D hydrodynamic numerical simulations. We study the validity of closure models and turbulent coefficients by varying the Prandtl number, the P$\acute{e}$clet number, and the depth of the convection zone. Four closure models of the fourth-order moments are evaluated with the 3D simulation data. The performance of the closure models varies among different cases and different fourth-order moments. We solve the dynamic equations of moments together with equations of the thermal structure. Unfortunately, we cannot obtain steady-state solutions when these closure models of fourth-order moments are adopted. The numerical solutions of the down-gradient approximations of the third-order moments, on the other hand, are robust. We calibrate the coefficients of the 1D down-gradient model from the 3D simulation data. The calibrated coefficients are more robust in the cases of deep convection zones. Finally we have compared the 1D steady-state solutions with the 3D simulation results. The 1D model has captured many features appearing in the 3D simulations : (1) $\nabla-\nabla_{a}$ has a U-shape with a minimum value at the lower part of the convection zone. (2) There exists a bump for $\nabla-\nabla_{a}$ near the top of the convection zone when the P$\acute{e}$clet number is large. (3) The temperature gradient can be sub-adiabatic due to the nonlocal effect. Apart from these similarities, the prediction on the kinetic energy flux, however, is unsatisfactory.
\end{abstract}




\section{Introduction} \label{sec:intro}
One-dimensional (1D) turbulent convection models are widely used in the calculations of stellar structure and evolution. The most popular one is the mixing length theory (MLT) \citep{1958ZfA..46...108}. It assumes that the flow blobs travel a distance proportional to the pressure scale height in the convectively unstable zone. The energy can be transported efficiently through the heat carried by the blobs. In this theory, the mixing length parameter $\alpha_{conv}$ needs to be calibrated. One important phenomena in stellar convection is overshooting. The stellar evolutionary track can be seriously affected when overshooting is taking into account \citep{2001ApJ..122...1602,2004ApJS..155...667}. In the MLT, the overshooting distance is usually tackled by introducing another overshooting parameter $\alpha_{ov}$ \citep{1965ApJ..142...1468,1973ApJ..184...191,1975AA..40...303}. This treatment, however, has been criticized for its lack of physical consistence \citep{1987AA..188...49}.

Overshooting is actually a nonlocal effect of turbulent convection. \citet{1978CAA..2...118} is among the first to use the nonlocal Reynolds stress model to treat the overshooting phenomenon. The Reynolds stress model derives dynamic equations of the second- or higher-order moments from the Navier-Stokes equations. The system of dynamic equations governing these moments, however, is incomplete unless closure models are provided. In the earlier version of Xiong's model \citep{1978CAA..2...118,1989AA..209...126}, the down-gradient forms of the third-order moments are adopted. Later, the model is extended to include dynamic equations of the third-order moments, and a quasi-normal approximation of the fourth-order moments is adopted as closure relations \citep{1997ApJS..108...529}. Similar Reynolds stress models have been derived by \citet{1992ApJ..392...218,1993ApJ..416...331,1998ApJ..493...834,2001JPO..31...1413,2007MNRAS..375...388,2012ApJ..756...37}. In their models, nonlocal models of the dissipation rate ($\varepsilon$ or $\omega$) are introduced, and the closure relations are different. In all the Reynolds stress models, coefficients associated with turbulent convection and diffusion need to be determined. These coefficients can be calibrated from the observational data of solar convection \citep{2006ApJ..643...426,2007MNRAS..375...388}, or from numerical simulations and experimental results \citep{1998ApJ..493...834}. In stars the turbulent convection is affected by many factors, such as the depth of the convection zone, the P$\acute{e}$clet number, the Prandtl number, and so on. The validity of  the coefficients and closure relations across a wide parameter space of these factors should be examined.

In recent years, growing computational resources has made three-dimensional simulation a powerful method in studying the stellar turbulent convection \citep{2009LRSP..6...2,2015SR..196...303,2017LRCA..3...1}. One convenient way to test the validity is to compare the three-dimensional simulation results with the one-dimensional model. However, most of the simulations were compared with the mixing length theory. Only a few were compared with the Reynolds stress model. \citet{1993ApJS..89...361} investigated the closure models with a generalized smoothed particle hydrodynamics simulations, but the background structure was almost homogeneous. \citet{1996ApJ..466...372} performed a 3D simulation of deep convection zone and tested the algebraic closures for high-order moments. Limited by the computing resource at that time, only one simulation case was studied. \citet{2002JAS..59...2729} tested their closure model based on a two-scale mass-flux approach against large eddy simulation data of free convective conditions. They found that model predictions showed good agreement with simulation data. \citet{2004GRL..31...23301} tested the same model with data from numerical simulations of free convection in the ocean. They found that the model was far more accurate than Gaussian parameterizations. \citet{2007MNRAS..374...305,2007IAUS..239...80,2007IAUS..239...83,2007IAUS..239...92} tested different closure relations for the fourth-order moments in the 3D simulations of turbulent convection in the Sun and a K dwarf with realistic settings, and in efficient and inefficient convection zones for idealized microphysics. They found that in the quasi-adiabatic convection zone, these new closure models provided a significant improvement over the quasi-normal approximation. \citet{2010MNRAS..407...2451} tested closure models with Rayleigh-B$\acute{e}$nald simulations, and later \citet{2015AN..336..32} investigated the rotation effects with the same closure relations. They have found that the closure relations work fairly well with slow rotation but the quality decreases for fast rotation. The increasing computational power allows us to perform numerical simulations with higher resolutions and thus a much larger range of achievable Rayleigh and Prandtl numbers.

When the 1D Reynolds stress model is applied to stellar convection, the equations of thermal structure should be solved with the dynamic equations of moments. Most of the work on the closure relations mentioned above did not account for the effect from the interaction with thermal structures. When the equations of thermal structure are included, numerical instability appears in some cases. With a down-gradient approximation of the third-order moments, we have investigated the numerical stability by considering a simplified version of Xiong's nonlocal model \citep{2014MNRAS..443...3703}. It seems that the numerical schemes are robust across different types of stars. When the model is extended to include the dynamic equations of the third-order moments, six additional partial differential equations regarding the third-order moments are included. Unfortunately, we are not able to obtain the numerical results with current closure models of fourth-order moments. \citet{1993ApJ..416...331} proposed a trimmed-down version of the stationary limit of the full third-order moment equations. Different from Xiong's nonlocal model, \citet{1993ApJ..416...331} expressed each third order moments with linear algebraic forms of all second order moments. With this down-gradient model, \citet{1999ApJ..526...L45} compared the 1D down-gradient approximation model with 3D numerical simulations. Two simulation cases with depths of 4.2 and 4.8 pressure scale were performed. He has shown that the agreement between 1D model and 3D result is promising. In a later paper, \citet{2001JPO..31...1413} discussed that eddy damping has to be added to improve both accuracy and stability for the previous model. With eddy damping included, this full form model was solved for the case of A-stars\citep{2002MNRAS..330...6} and the case of DA white dwarfs\citep{2004MNRAS..350...267}. In \citet{2007IAUS..239...83}, it was reported that this approach does not work for deep, efficient convection. They identified the cross-relations as culprits. Then in \citet{2007IAUS..239...92}, they replaced the cross-relations in the third order moment closure by \citet{2001JPO..31...1413} with those from \citet{2002JAS..59...2729}, but instability still occurred.

In this paper, we discuss the validity of the closure models and coefficients with varying parameters : the Prandtl number, the Peclet number, and depth of the convection zone. In section 2, we describe our 3D numerical methods and parameter settings. In section 3, we report the flow structure of our 3D numerical simulations. The statistical results regarding the closure relations of the fourth-order moments are studied. In section 4, we describe the 1D nonlocal model, calibrate the coefficients, and compare with 3D simulation results. In section 5, we summarize our findings.

\section{The 3D Simulations}
Based on the stratified approximations, \citet{2016JComP..310...342} has developed a semi-implicit spectral model for performing three-dimensional numerical simulations in Cartesian geometry. The method solves a stratified form of the following compressible hydrodynamic equations :
\begin{eqnarray}
\partial_{t}\rho&=&-\mathbf{\nabla}\cdot \mathbf{M} ~,\\
\partial_{t}\mathbf{M}&=&-\mathbf{\nabla}\cdot (\mathbf{MM}/\rho)+\mathbf{\nabla}\cdot \mathbf{\Sigma}-\mathbf{\nabla} p+\rho \mathbf{g} ~,\\
\partial_{t}E&=&-\nabla \cdot[(E+p)\mathbf{M}/\rho-\mathbf{M}\cdot\mathbf{\Sigma}/\rho+\mathbf{f}]+\mathbf{M}\cdot \mathbf{g} ~.
\end{eqnarray}
where $\rho$ is the density, $\mathbf{M}=\rho \mathbf{v}$ is the mass flux, $\mathbf{v}$ is the velocity, $E$ is the total energy, $p$ is the pressure, $\mathbf{g}$ is the gravitational acceleration, $\mathbf{f}$ is the diffusive heat flux, and $\Sigma$ is the viscous tensor
\begin{equation}
\mathbf{\Sigma}= 2\mu \mathbf{\sigma}+\lambda (\mathbf{\nabla} \cdot \mathbf{V}) \mathbf{I} ~,
\end{equation}
where $\mathbf{\sigma}$ is the strain rate tensor. Two assumptions are made in the stratified approximations. First, the variation of density is ignored in the viscous terms and nonlinear advection terms in the momentum equation. Second, the terms containing higher order horizontal variations of hydrodynamic variables are ignored in the energy equation. One advantage of the stratified approximation is that the acoustic and gravity waves are retained. The stratified approximation induces an error of $O(Ma^2)$ in both momentum and energy equations. For the numerical scheme, a Fourier spectral scheme is used in the horizontal directions, and a second-order finite-difference scheme is used in the vertical directions. For time advancing, a second-order Crank-Nicholson scheme is adopted for linear terms, and a third-order Adams-Bashforth scheme is adopted for nonlinear terms. The boundary conditions are : $v_{z}=\partial_{z}v_{x}=\partial_{z}v_{y}=0$, $T$=constant at the top; and $v_{z}=\partial_{z}v_{x}=\partial_{z}v_{y}=0$, $F_{tot}$=constant at the bottom.

In this paper, we consider the case of efficient convection for a compressible ideal gas in a box. The aspect ratio of the box is $L_{x}:L_{y}:L_{z}=6:6:1$. The initial state of the ideal gas is configured as
\begin{eqnarray}
&&T/T_{top}=1+Z(L_{z}-z)/L_{z} \\
&&\rho/\rho_{top}=(T/T_{top})^{n}\\
&&p/p_{top}=(T/T_{top})^{n+1}
\end{eqnarray}
where the subscript $top$ represents the value at the top of the simulation domain; $Z$ is the ratio measuring the depth of the fluid; $n$ is the polytropic index. When $n$ is smaller than the adiabatic polytropic index $n_{ad}$, the flow is convectively unstable. In our simulations, we set $n_{ad}=1.5$ and $n=1$, therefore the flow is convectively unstable in the whole simulation domain. We have run a total of 20 cases with the depths of convection zones spanning from 2.19 to 5.67 pressure scale heights. The parameters are given in Table~\ref{tab:table1}, and the non-dimensional variables have the following definition:
the Prandtl number is
\begin{equation}
{\rm{Pr}}=c_{p}\mu/\kappa~,
\end{equation}
where $c_{p}$ is the specific heat capacity at constant pressure, $\mu$ is the dynamic viscosity, $\kappa$ is the conductive coefficient.
The Rayleigh number is
\begin{equation}
{\rm Ra(z)}=\frac{[1-(\gamma-1)n]{g\rm Pr}  L_{z}^3 \rho^2 Z}{\gamma \mu^2}~,
\end{equation}
where $\gamma$ is the ratio of specific heats.
The Reynolds number is
\begin{equation}
{\rm Re(z)}=\frac{\rho v_{rms} L_{z}}{\mu}~,
\end{equation}
where $v_{rms}$ is averaged the root-mean-square velocity. And the P$\acute{e}$clet number is
\begin{equation}
{\rm Pe(z)}={\rm Re(z)}{\rm Pr}~.
\end{equation}
We run each of the simulations for a long period until the flow reaches a thermal equilibrium state.
In all the simulations, the deviations of the averaged total flux from the input flux are less than 0.5 percent and thermal equilibrium states are fairly achieved. The different types of energy fluxes are evaluated from the following formulas :
\begin{eqnarray}
&&F_{r}=- \kappa \frac{\partial \overline{T}}{\partial z}~,\\
&&F_{c}= c_{p}\overline{\rho w \theta}~,\\
&&F_{k}= \frac{1}{2}\overline{\rho w v^2}~,
\end{eqnarray}
where $F_{r}$ is the radiative (or conductive) flux, $F_{c}$ is the convective flux, and $F_{k}$ is the kinetic energy flux, $w$ is the vertical velocity, and $\theta$ is the temperature perturbation.

\startlongtable
\begin{deluxetable}{cccccccccccc}
\tablecaption{Parameters of numerical simulations \label{tab:table1}}
\tablehead{
 Case & $N_{x}\times N_{y}\times N_{z}$ & $Z$ & $\mu$ & $F_{tot}$ & Ra & Pr & $v_{rms}$ & Re & Pe & $\delta_{\kappa}$ & $\delta_{\nu}$
 }
\startdata
 A1 & $256\times 256\times 201$ & $2$    &   $5\times 10^{-3}$      & 0.25   &   $2.83\times 10^{4}$   & $0.1$    & 0.325  & 130  & 13.0  & 0.137 & 0.043\\
 A2 & $256\times 256\times 201$ & $2$    &   $2.5\times 10^{-3}$    & 0.125  &   $1.13\times 10^{5}$   & $0.1$    & 0.295  & 235  & 23.4  & 0.173 & 0.055\\
 A3 & $256\times 256\times 201$ & $2$    &   $1.25\times 10^{-3}$   & 0.0625 &   $4.60\times 10^{5}$   & $0.1$    & 0.250  & 395  & 39.5  & 0.068 & 0.022\\
 A4 & $256\times 256\times 201$ & $2$    &   $2.5\times 10^{-3}$    & 0.25   &   $5.67\times 10^{4}$   & $0.05$   & 0.351  & 281  & 14.1  & 0.137 & 0.031\\
 A5 & $256\times 256\times 201$ & $2$    &   $1.25\times 10^{-3}$   & 0.25   &   $1.13\times 10^{5}$   & $0.025$  & 0.369  & 591  & 14.8  & 0.137 & 0.022\\
 B1 & $256\times 256\times 201$ & $4$    &   $5\times 10^{-3}$      & 0.5    &   $2.78\times 10^{5}$   & $0.1$    & 0.458  & 271  & 27.1  & 0.077 & 0.025\\
 B2 & $256\times 256\times 201$ & $4$    &   $2.5\times 10^{-3}$    & 0.25   &   $1.11\times 10^{6}$   & $0.1$    & 0.392  & 458  & 45.8  & 0.055 & 0.017\\
 B3 & $256\times 256\times 201$ & $4$    &   $1.25\times 10^{-3}$   & 0.125  &   $4.44\times 10^{6}$   & $0.1$    & 0.326  & 751  & 75.1  & 0.039 & 0.012\\
 B4 & $256\times 256\times 201$ & $4$    &   $2.5\times 10^{-3}$    & 0.5    &   $5.56\times 10^{5}$   & $0.05$   & 0.478  & 561  & 28.1  & 0.078 & 0.017\\
 B5 & $256\times 256\times 201$ & $4$    &   $1.25\times 10^{-3}$   & 0.5    &   $1.11\times 10^{6}$   & $0.025$  & 0.487  & 1139 & 28.5  & 0.078 & 0.012\\
 C1 & $512\times 512\times 201$ & $8$    &   $5\times 10^{-3}$      & 1.0    &   $6.70\times 10^{6}$   & $0.1$    & 0.568  & 540  & 54.0  & 0.035 & 0.011\\
 C2 & $512\times 512\times 201$ & $8$    &   $2.5\times 10^{-3}$    & 0.5    &   $2.68\times 10^{7}$   & $0.1$    & 0.468  & 876  & 87.6  & 0.025 & 0.008\\
 C3 & $512\times 512\times 201$ & $8$    &   $1.25\times 10^{-3}$   & 0.25   &   $1.07\times 10^{8}$   & $0.1$    & 0.367  & 1365 & 136.5 & 0.018 & 0.006\\
 C4 & $512\times 512\times 201$ & $8$    &   $2.5\times 10^{-3}$    & 1.0    &   $1.34\times 10^{7}$   & $0.05$   & 0.577  & 1092 & 54.6  & 0.035 & 0.008\\
 C5 & $512\times 512\times 201$ & $8$    &   $1.25\times 10^{-3}$   & 1.0    &   $2.68\times 10^{7}$   & $0.025$  & 0.583  & 2214 & 55.3  & 0.035 & 0.006\\
 D1 & $512\times 512\times 201$ & $16$   &   $5\times 10^{-3}$      & 2.0    &   $4.63\times 10^{7}$   & $0.1$    & 0.619  & 1034 & 103.5 & 0.022 & 0.007\\
 D2 & $512\times 512\times 201$ & $16$   &   $2.5\times 10^{-3}$    & 1.0    &   $1.85\times 10^{8}$   & $0.1$    & 0.504  & 1615 & 161.5 & 0.015 & 0.005\\
 D3 & $512\times 512\times 201$ & $16$   &   $1.25\times 10^{-3}$   & 0.5    &   $7.40\times 10^{8}$   & $0.1$    & 0.402  & 2550 & 255.0 & 0.011 & 0.003\\
 D4 & $512\times 512\times 201$ & $16$   &   $2.5\times 10^{-3}$    & 2.0    &   $9.25\times 10^{7}$   & $0.05$   & 0.621  & 2063 & 103.1 & 0.022 & 0.005\\
 D5 & $512\times 512\times 201$ & $16$   &   $1.25\times 10^{-3}$   & 2.0    &   $1.85\times 10^{8}$   & $0.025$  & 0.635  & 4196 & 104.9 & 0.022 & 0.003\\
 \enddata
\tablecomments{$N_{x},N_{y},N_{z}$ are the grid numbers in x, y, and z directions, respectively. $Z$ is a ratio measuring the depth of the convection zone. $\mu$ is the dynamic viscosity. $F_{tot}$ is the total flux. Ra is the averaged Rayleigh number. Pr is the Prandtl number. $v_{rms}$ is the averaged root-mean-square velocity. Re is the averaged Reynolds number. Pe is the averaged P$\acute{e}$clet number. $\delta_{\kappa}$ is the estimated thickness of the thermal boundary layer, and $\delta_{\nu}$ is the estimated thickness of the viscous boundary layer. All the averaged values are taken both spatially and temporally.}
\end{deluxetable}

\section{Results}
\subsection{Flow structures of 3D models}
\begin{figure}[ht!]
\plotone{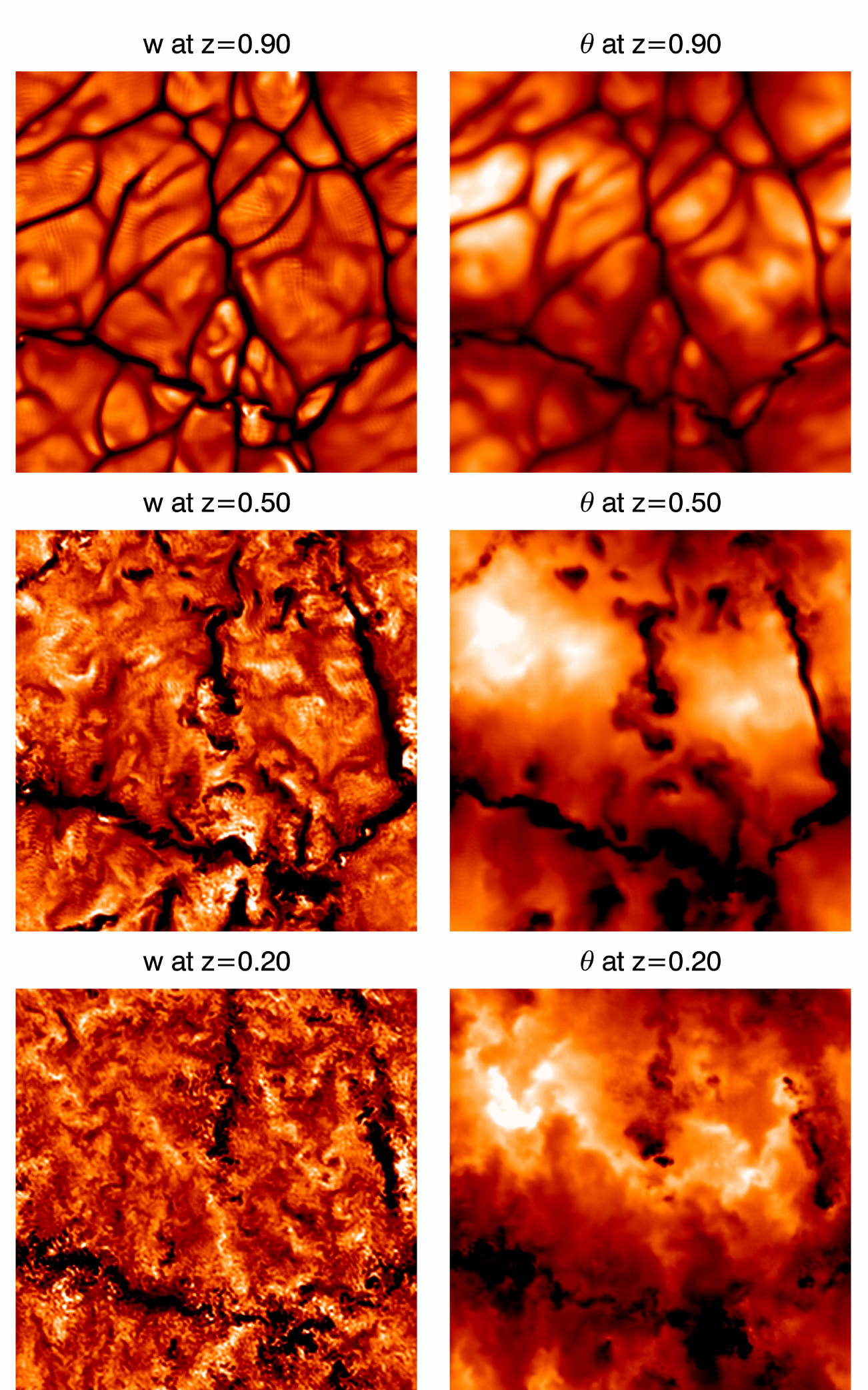}
\caption{Contours of vertical velocity (left panel) and temperature fluctuation (right panel) at different heights from case D4 ($Z=16$, $\mu=2.5\times 10^{-4}$, $Pr=0.05$, $F_{tot}=2.0$). Dark and bright colors represent negative and positive values, respectively. \label{fig:flowD4}}
\end{figure}

\begin{figure}[ht!]
\plotone{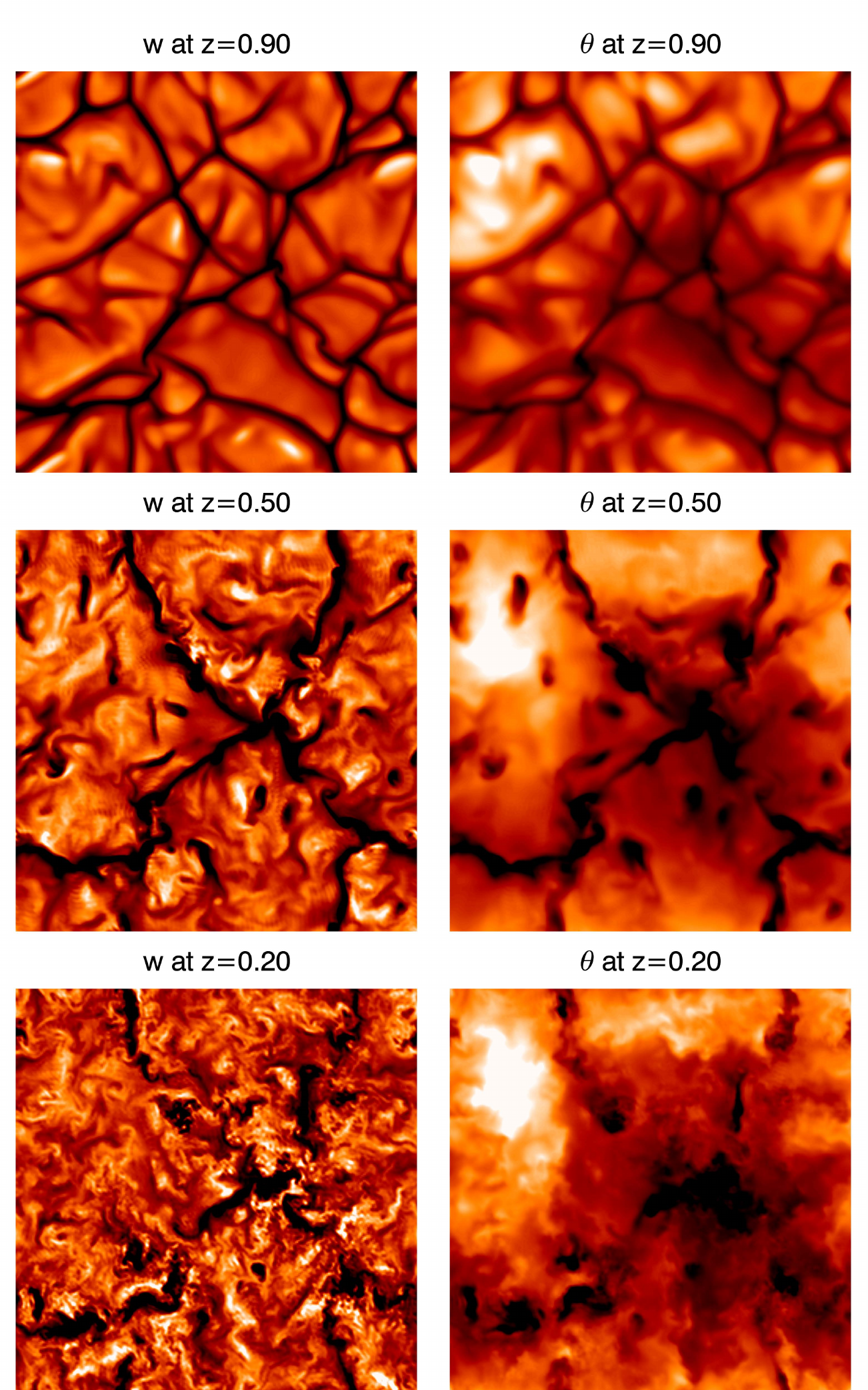}
\caption{Contours of vertical velocity (left panel) and temperature fluctuation (right panel) at different heights from case D1 ($Z=16$, $\mu=5\times 10^{-3}$, $Pr=0.1$, $F_{tot}=2.0$). Dark and bright colors represent negative and positive values, respectively.\label{fig:flowD1}}
\end{figure}

Fig.~\ref{fig:flowD4} shows the contours of vertical velocities and temperature fluctuations at the top, middle, and bottom of the convection zones for the case D4. The panels on the left show the structure of vertical velocities, and the panels on the right show the corresponding temperature fluctuations. The darker colors represent negative values, and the bright colors represent positive values. Note that cellular network of strong downflow lanes is formed at the top of the convection zone. Deep into the convection zone, the network of downflow lanes breaks up or merges into two major perpendicular downflow lanes in the middle of the convection zone. These two major downward lanes are strong enough to penetrate all the way to the bottom of the convection zone. \citet{2000ApJs..127...159} found that the perpendicular downward lanes tended to be aligned with the periodic directions in their simulations by an aspect ratio of $L_{x}:L_{y}:L_{z}=2:2:1$. However, it is not always the situation in our simulations. Fig.~\ref{fig:flowD1} shows case D1 where the downward perpendicular flows are aligned at about $45^{\circ}$ to the periodic directions in the middle of the convection zone. This difference can be caused by the different aspect ratios we have specified. The large scale flows are restricted by the width of the box. If the box is not wide enough to contain one large cell of downward lanes, then these lanes are forced to be aligned with the periodic directions, taking advantage of the periodic boundary conditions in the horizontal directions. But when the box is wide enough to contain several cells of downward lanes, the lanes can be aligned freely in other directions. Apart from the large scale motions, roll structures of small scale turbulence are present in the middle and bottom of the convection zone. D4 has the same input flux as D1, but the viscosity is two times smaller than that one of D1. The motions of small scale turbulence of case D4 are more vigorous than that of D1, which means that viscosity plays an important role in small scale motions. On the right panels of Fig.~\ref{fig:flowD4}-\ref{fig:flowD1}, structures of temperature fluctuations are presented. The negative temperature fluctuations are highly correlated with the downward flows, which means that convective fluxes are transported efficiently by downward flows. The correlations between positive temperature fluctuations and upward flows, however, are not as significant as those of downward flows.

\begin{figure}[ht!]
\plotone{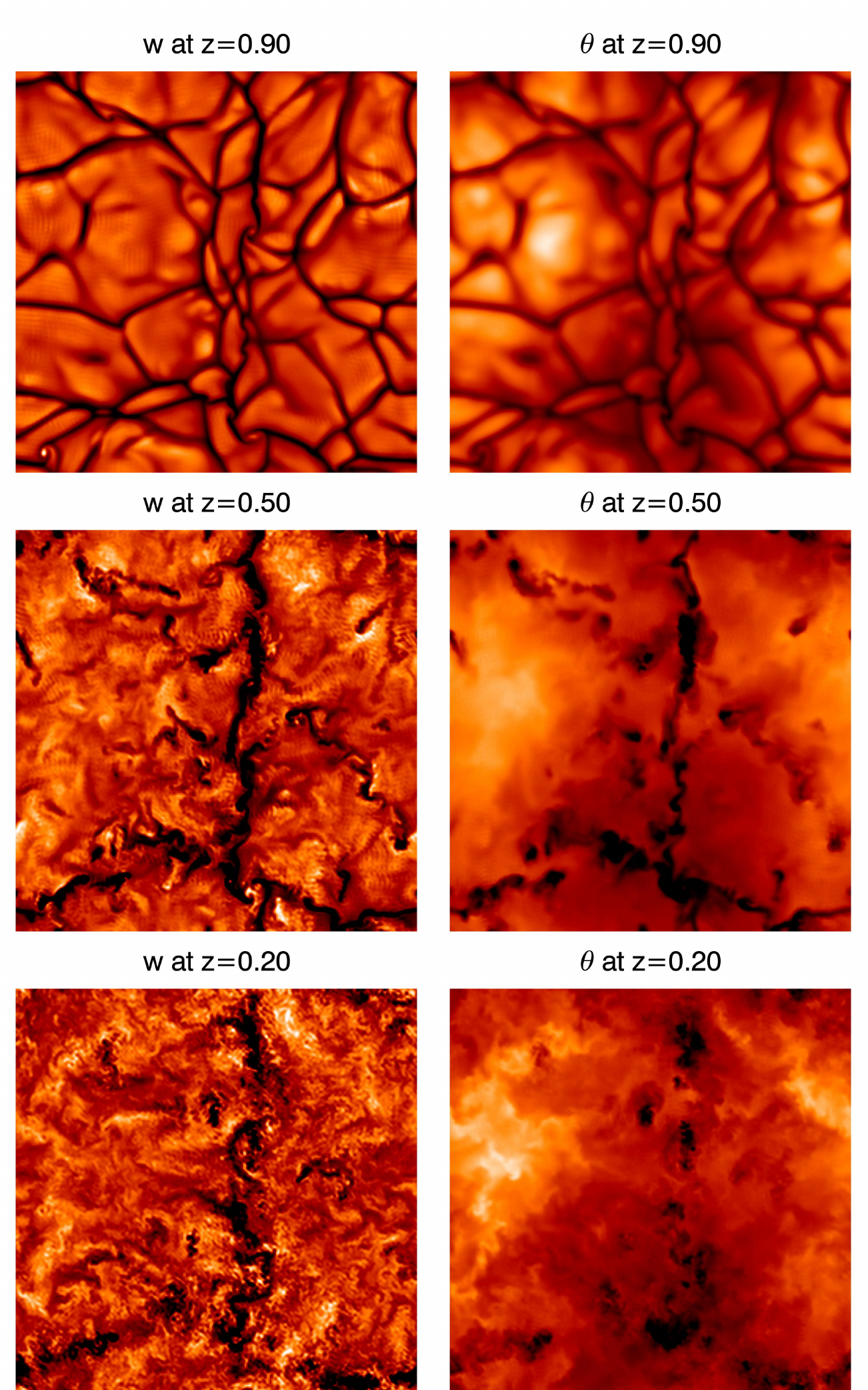}
\caption{Contours of vertical velocity (left panel) and temperature fluctuation (right panel) at different heights from case D2 ($Z=16$, $\mu=2.5\times 10^{-3}$, $Pr=0.1$, $F_{tot}=1.0$). Dark and bright colors represent negative and positive values, respectively.\label{fig:flowD2}}
\end{figure}

Fig.~\ref{fig:flowD2} shows the contours of vertical velocities and temperature fluctuations for case D2. Case D2 has the same viscosity as case D4, but the total flux is two times smaller. At the top, a similar cellular network of strong downward lanes is present. In the middle, however, the downward lanes tend to form block cells instead of being perpendicular flows. In this case, the box is wide enough to contain several cells, thus the effect from the side boundary is alleviated. Since the total flux is smaller than case D4, the downward lanes are weaker when penetrating into the bottom. It is more obvious if one looks at temperature fluctuations at the bottom, where the large cell of cold flows could hardly be identified in D2.

\begin{figure}[ht!]
\plotone{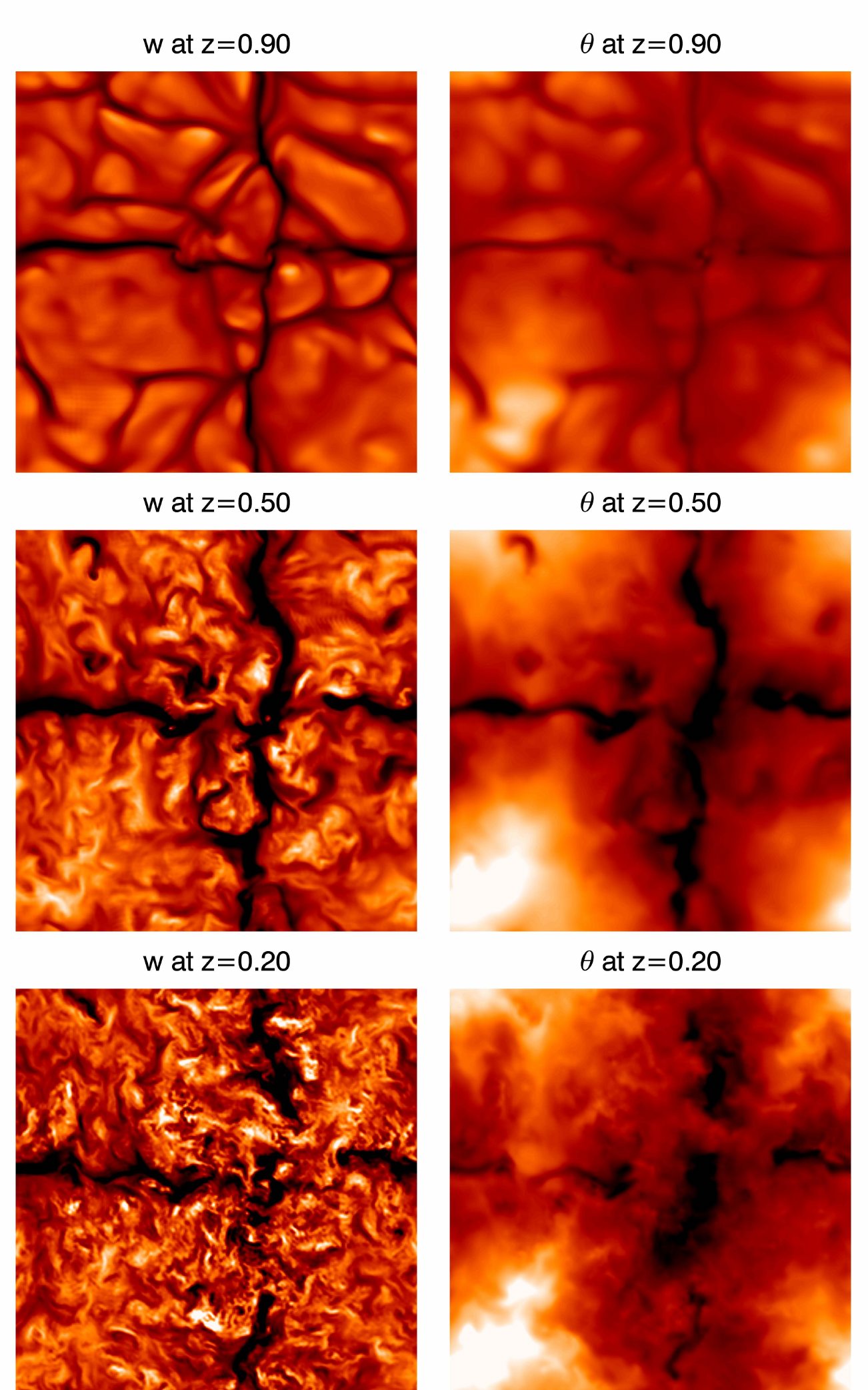}
\caption{Contours of vertical velocity (left panel) and temperature fluctuation (right panel) at different heights from case C4 ($Z=8$, $\mu=2.5\times 10^{-3}$, $Pr=0.05$, $F_{tot}=1.0$). Dark and bright colors represent negative and positive values, respectively. \label{fig:flowC4}}
\end{figure}

Fig.~\ref{fig:flowC4} shows the flow structures in a more shallow convection zone, case C4. The viscosity and total flux are the same as in case D2, but the depth in number of pressure scale height is much smaller. The depths are 4.39 and 5.66 pressure scale heights for case C4 and D2, respectively. In case C4, the cellular network is subtly present in the top. And two perpendicular downward lanes are much stronger than other lanes. These two downward lanes develop in the middle and penetrate into the bottom of the convection zone. Compared with Fig.~\ref{fig:flowD2}, the magnitudes of both vertical velocities and temperature fluctuations are greater in case C4. This can be explained using mixing length theory. When the turbulent convection is very efficient, the temperature gradient is very close to the adiabatic temperature gradient. Therefore the flux transported by conduction is almost identical to the one transported by adiabatic temperature gradient $F_{ad}=[(m+1)(\gamma-1)/\gamma] F_{tot}$, and the flux transported by convection can be estimated by $F_{c}=F_{tot}-F_{ad}$. According to the mixing length theory, the convective flux has a relationship with the root mean square velocity of $F_{c}\sim \rho (\overline{w^{2}})^{3/2}$. The fluxes carried by convection are almost the same in cases C4 and D2. Since the average density is higher in D2, the velocity should be smaller.

\begin{figure}[ht!]
\plotone{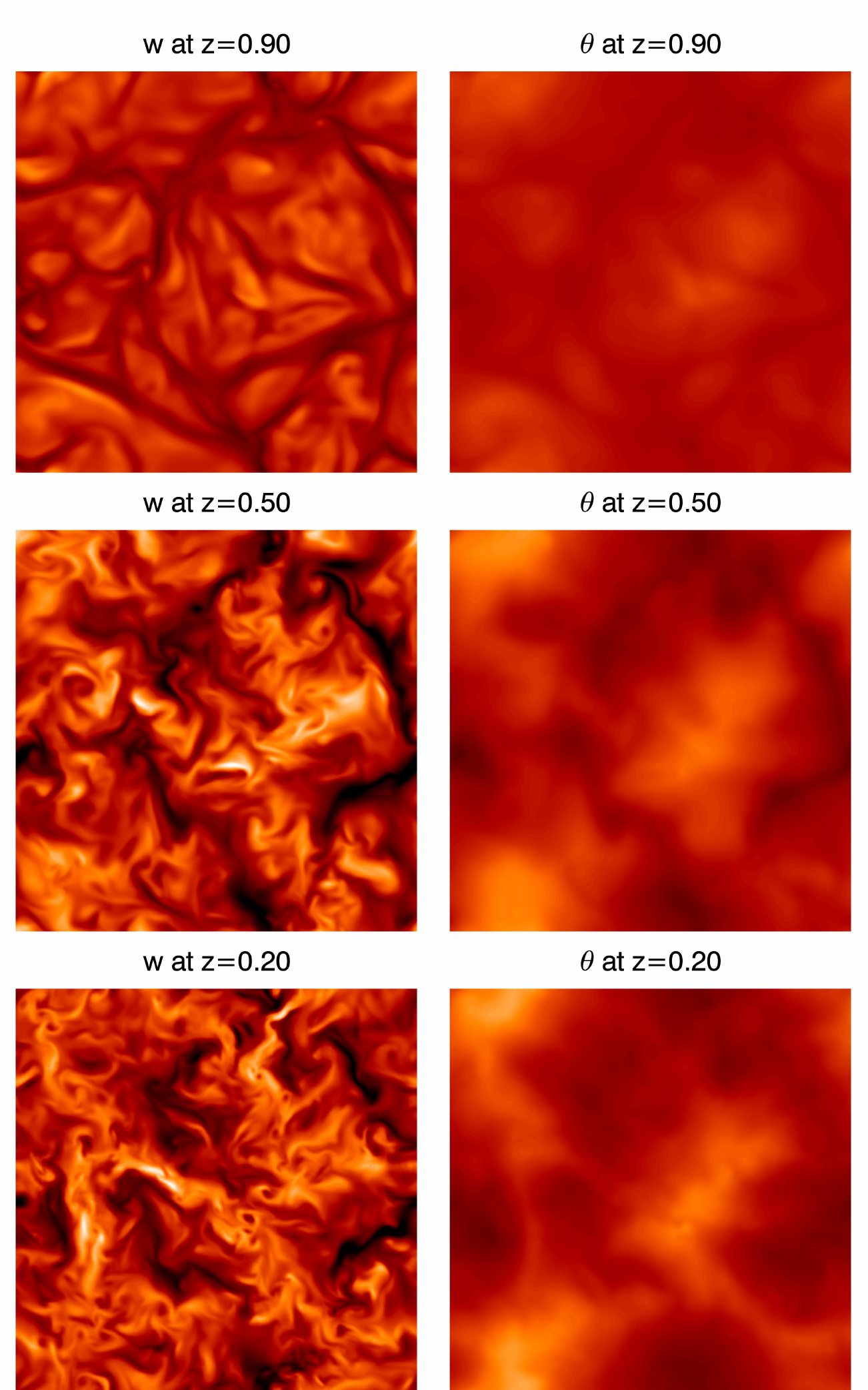}
\caption{Contours of vertical velocity (left panel) and temperature fluctuation (right panel) at different heights from case A5 ($Z=2$, $\mu=1.25\times 10^{-3}$, $Pr=0.025$, $F_{tot}=0.25$). Dark and bright colors represent negative and positive values, respectively. \label{fig:flowA5}}
\end{figure}

Fig.~\ref{fig:flowA5} shows the flow structures in the shallowest case, A5. The depth of the convection zone is about 2.19 pressure scale height. At the top, the upward flows are comparable to the downward flows in magnitudes, and the cellular structure is present but not significant. Similar to the deep cases, strong downward lanes are formed in the middle and penetrate to the bottom of the convection zone. The small scale fluid motions, however, are much weaker than those in the deep cases. It is more obvious if we compare the profile of temperature perturbations. There is significant difference of the flow structures between the shallow and deep convection zones, therefore it is worthwhile to take this difference into account when developing turbulent theories.

The flow structure can be affected if the mesh resolution is inadequate to resolve the small scale turbulence \citep{2011JComP..230...8698}. It is necessary to check whether the mesh resolution is able to cover the inertial and dissipative ranges in the power spectrum of kinetic energy. The turbulent motion is aperiodic in the vertical direction in the stratified model, so it is more appropriate to compute the kinetic energy spectrum on the horizonal directions only. As we use a Fourier spectral method on the horizontal directions, the kinetic energy spectrum can be easily evaluated in the spectral space.
\begin{equation}
\int P_{2}(k)dk = \int \int \frac{1}{2}\rho v^{2} (x,y) dxdy = \frac{1}{2\rho}\sum_{m=0}^{M} \sum_{n=0}^{N} [M_{z,mn}M_{z,mn}^{*} + \frac{1}{(m\eta_{x})^2+(n\eta_{y})^2}(\delta_{mn}\delta_{mn}^{*}+\xi_{mn}\xi_{mn}^{*})]~,
\end{equation}
where $P_{2}(k)$ is the two-dimensional power spectral of the kinetic energy, $M$ and $N$ are the maximum truncated spectral numbers on the horizontal spectral space, $M_{z}$ is the vertical momentum density, $\delta$ is the horizontal divergence of the horizonal momentum density, $\xi$ is the curl of the horizonal momentum density, $\eta_{x}=2\pi/L_{x}$ and $\eta_{y}=2\pi/L_{y}$, the subscripts $m,n$ represent the spectral numbers on the horizontal directions, and the superscript $*$ represents the conjugate of the corresponding variable.

\begin{figure*}
\gridline{\fig{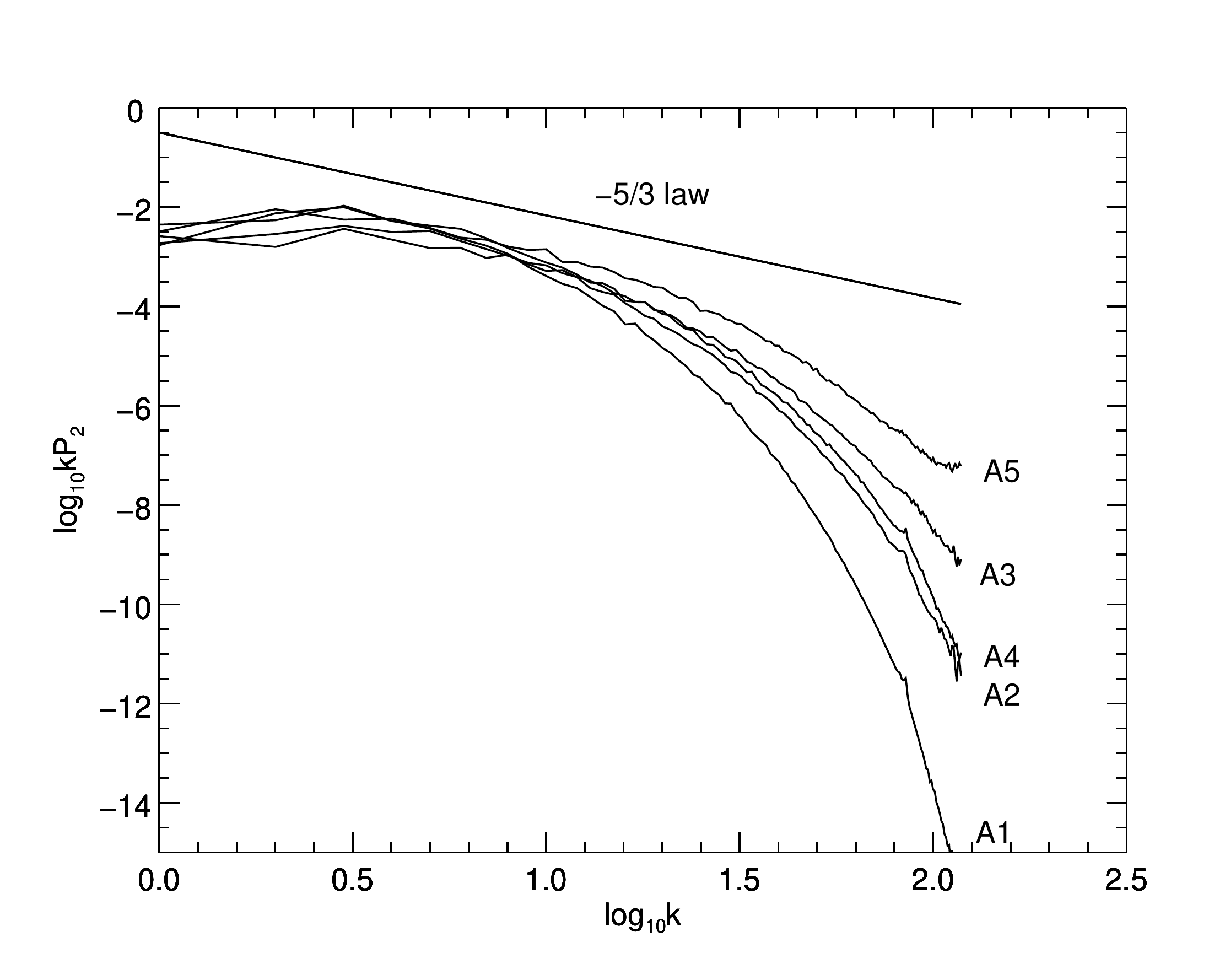}{0.5\textwidth}{(a)}
          \fig{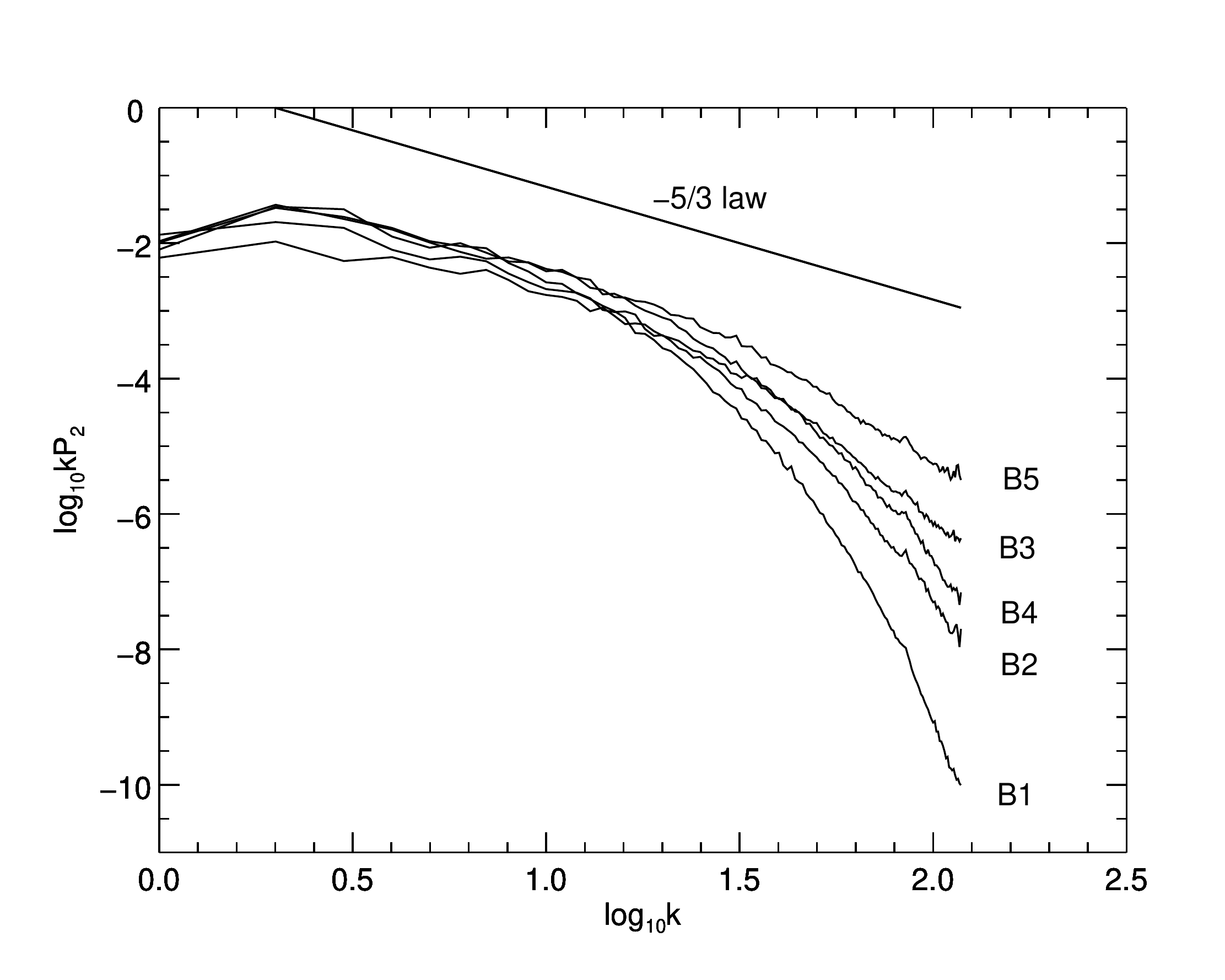}{0.5\textwidth}{(b)}
          }
\gridline{\fig{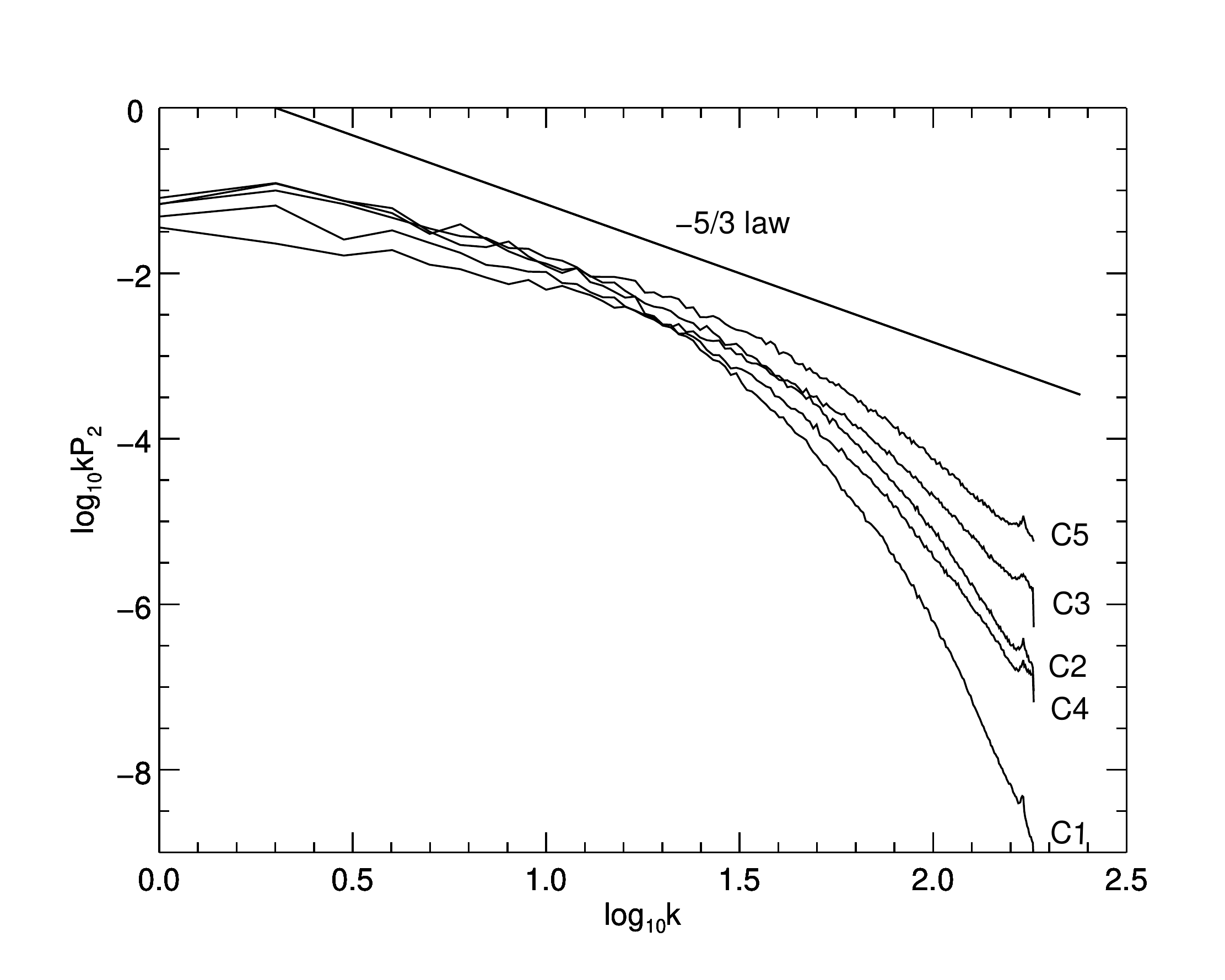}{0.5\textwidth}{(c)}
          \fig{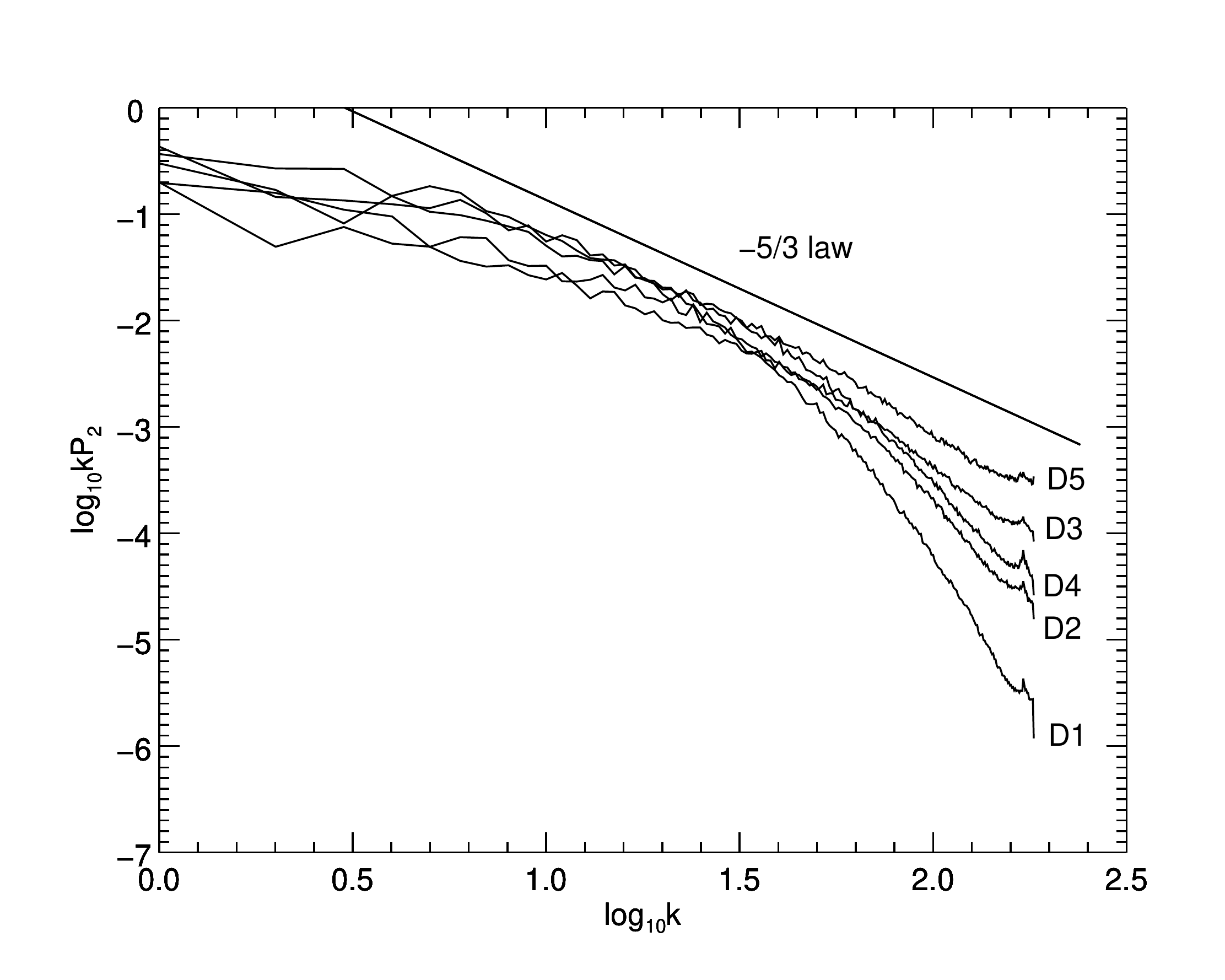}{0.5\textwidth}{(d)}
          }
\caption{Power spectra of the kinetic energy $P_{2}(k)k$ as a function of wavenumber $k$. The power spectra are integrated on the horizontal plane z=0.5. The Kolmogorov's $-5/3$ power law is depicted by a straight line.\label{fig:power_spectrum} }
\end{figure*}

Fig.~\ref{fig:power_spectrum} shows the power spectrum of the kinetic energy $P_{2}(k)k$ as a function of the wavenumber $k$ at $z=0.5$. The straight line shows the $-5/3$ slope, which is the Kolmogorov's power law in the inertial range. For all the simulation cases, the wave numbers cover the inertial range and extend to the dissipative range. The inertial range increases as the depth of the convection zone increases, therefore a higher resolution is required to resolve the small scale turbulence for the cases with deeper convection zones. The required numerical resolutions can be estimated from the well-known scaling relation in thermal convection $Nu\sim Ra^{1/4}$\citep{2000JFM..407...27}. Based on the mixing length theory, \citet{1993AA..253...131} derived a scaling relation of $Nu\sim (RaPr)^{1/4}$ for the efficient turbulent convection in astrophysical flows. From the scaling relation, one can estimate the thickness of the thermal boundary layer from $\delta_{\kappa}/L_{z}\sim (RaPr)^{-1/4}$, and the thickness of the viscous boundary layer from $\delta_{\nu}/L_{z}\sim Ra^{-1/4}Pr^{1/4}$. The last two columns in Table~\ref{tab:table1} list the estimated values of $\delta_{\kappa}$ and $\delta_{\nu}$. In most of our simulation cases (except D3 and D5), $\delta_{\nu}$ is greater than the numerical vertical grid size $\Delta_{z}=0.005$. The numerical vertical grid size is barely enough for D3 and D5. The numerical horizontal grid size is $\Delta_{x}=\Delta_{y}=0.023$ in groups A and B, and $\Delta_{x}=\Delta_{y}=0.012$ in groups C and D, respectively. Compared with $\delta_{\nu}$, the horizontal numerical resolution is adequate for group A, marginal for groups B and C, and inadequate for group D. It would be better to increase the mesh resolutions further for group D so that smaller scale turbulent diffusion can be resolved. Unfortunately, the computer is out of memory when we double the horizontal mesh resolutions. We admit that the simulation results in group D may suffer from the lack of numerical resolutions. Accordingly, the achieved Reynolds numbers in group D should be lower than the estimated values reported in Table~\ref{tab:table1}. As the numerical resolution in case D1 seems marginally sufficient to resolve the small scale turbulent diffusion (Fig.~\ref{fig:power_spectrum}), we focus on this case when discussing the statistical results of this group.

\subsection{Statistical results of 3D models}
\subsubsection{The fourth-order moments}
\citet{2007MNRAS..374...305} have made a comprehensive study of four different fourth-order closure models: the QN model \citep{1941DANS..32...615} , the GH model \citep{2002JAS..59...2729}, the GN3 and GN4 models \citep{1993ApJS..89...361}. Their discussion was based on large eddy simulations of turbulent convection in the Sun and a K dwarf star. \citep{2003MNRAS..340...923}. Here we have computed a total of 20 cases of direct numerical simulations (DNS) with different combination of parameters, hence it is possible for us to perform a parametric analysis, for moderate Ra numbers and Pr numbers, as achievable by DNS. For the convenience of comparison, we list the closure models mentioned in \citet{2007MNRAS..374...305}. The first one is the QN model :
\begin{eqnarray}
&&\overline{w^4} = 3\overline{w^2}^2~,\\
&&\overline{\theta^4} = 3\overline{\theta^2}^2~,\\
&&\overline{w^3 \theta} =3 \overline{w^2}\overline{w\theta}~,\\
&&\overline{w \theta^3} =3 \overline{\theta^2}\overline{w \theta}~,\\
&&\overline{w^2 \theta^2}=\overline{w^2}\overline{\theta^2}+2\overline{w\theta}^2~.
\end{eqnarray}
The second one is the GH model:
\begin{eqnarray}
&&\overline{w^4}=3(1+\frac{1}{3}S_{w}^2)\overline{w^2}^2~,\\
&&\overline{\theta^4}=3(1+\frac{1}{3}S_{\theta}^2)\overline{\theta^2}^2~,\\
&&\overline{w^3 \theta} =3(1+\frac{1}{3}S_{w}^2) \overline{w^2}\overline{w\theta}~,\\
&&\overline{w \theta^3} =3(1+\frac{1}{3}S_{\theta}^2) \overline{\theta^2}\overline{w \theta}~,\\
&&\overline{w^2 \theta^2}=\overline{w^2}\overline{\theta^2}+2\overline{w\theta}^2+S_{w}S_{\theta}\overline{w\theta}\sigma_{w}\sigma_{\theta}~.
\end{eqnarray}
The third one is the GN3 model:
\begin{eqnarray}
&&\overline{w^4}=(2.3+1.8S_{w}^2)\overline{w^2}^2~,\\
&&\overline{\theta^4}=(2.3+1.8S_{w}^2)\overline{\theta^2}^2~,\\
&&\overline{w^3 \theta} =(2.3+1.8S_{w}^2) \overline{w^2}\overline{w\theta}~,\\
&&\overline{w \theta^3} =(2.3+1.8S_{w}^2) \overline{\theta^2}\overline{w \theta}~,\\
&&\overline{w^2 \theta^2}=(2.3+1.8S_{w}^2)(\frac{1}{3}\overline{w^2}\overline{\theta^2}+\frac{2}{3}\overline{w\theta}^2)~.
\end{eqnarray}
And the last one is the GN4 model:
\begin{eqnarray}
&&\overline{w^4}=(2.3+2.1S_{w}^2)\overline{w^2}^2~,\label{eq:GN4s}\\
&&\overline{\theta^4}=(2.3+2.1S_{\theta}^2)\overline{\theta^2}^2~,\\
&&\overline{w^3 \theta} =(2.3+2.1\frac{\overline{w^2\theta}^2}{\overline{\sigma_{w}^4 \sigma_{\theta}^2}}) \overline{w^2}\overline{w\theta}~,\\
&&\overline{w \theta^3} =(2.3+2.1\frac{\overline{w\theta^2}^2}{\overline{\sigma_{w}^2 \sigma_{\theta}^4}}) \overline{\theta^2}\overline{w\theta}~,\\
&&\overline{w^2 \theta^2}=(2.3+2.1\frac{\overline{w^2\theta}^2}{\overline{\sigma_{w}^4 \sigma_{\theta}^2}})\frac{1}{3}\overline{w^2}\overline{\theta^2}+(2.3+2.1\frac{\overline{w\theta^2}^2}{\overline{\sigma_{w}^2 \sigma_{\theta}^4}})\frac{2}{3}\overline{w\theta}^2~\label{eq:GN4e}
\end{eqnarray}
In these formulas, $S_{w}$ and $S_{\theta}$ are the kurtoses of velocity and temperature, and $\sigma_{w}$ and $\sigma_{\theta}$ are the standard deviation of velocity and temperature, respectively.

\begin{figure*}
\gridline{\fig{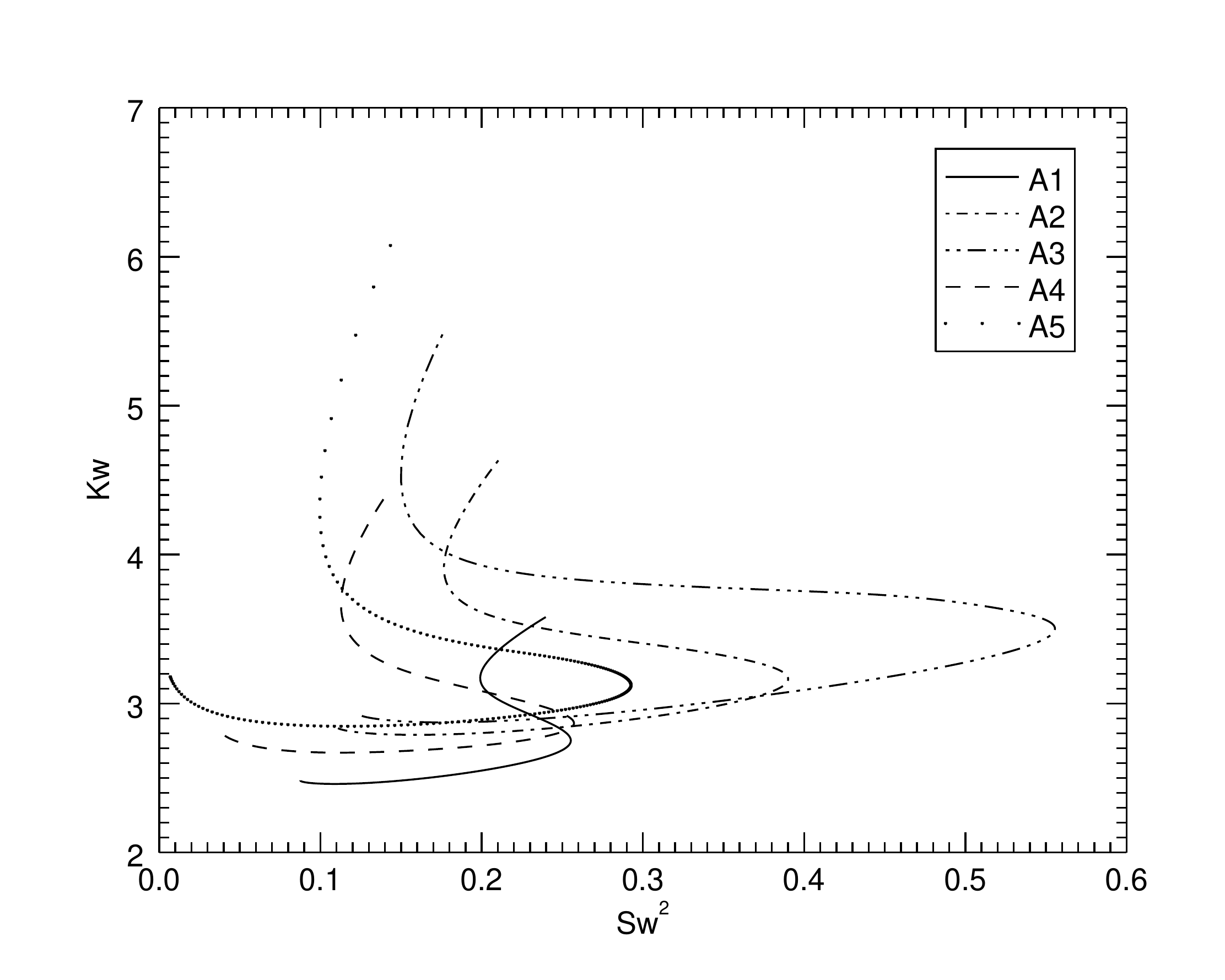}{0.5\textwidth}{(a)}
          \fig{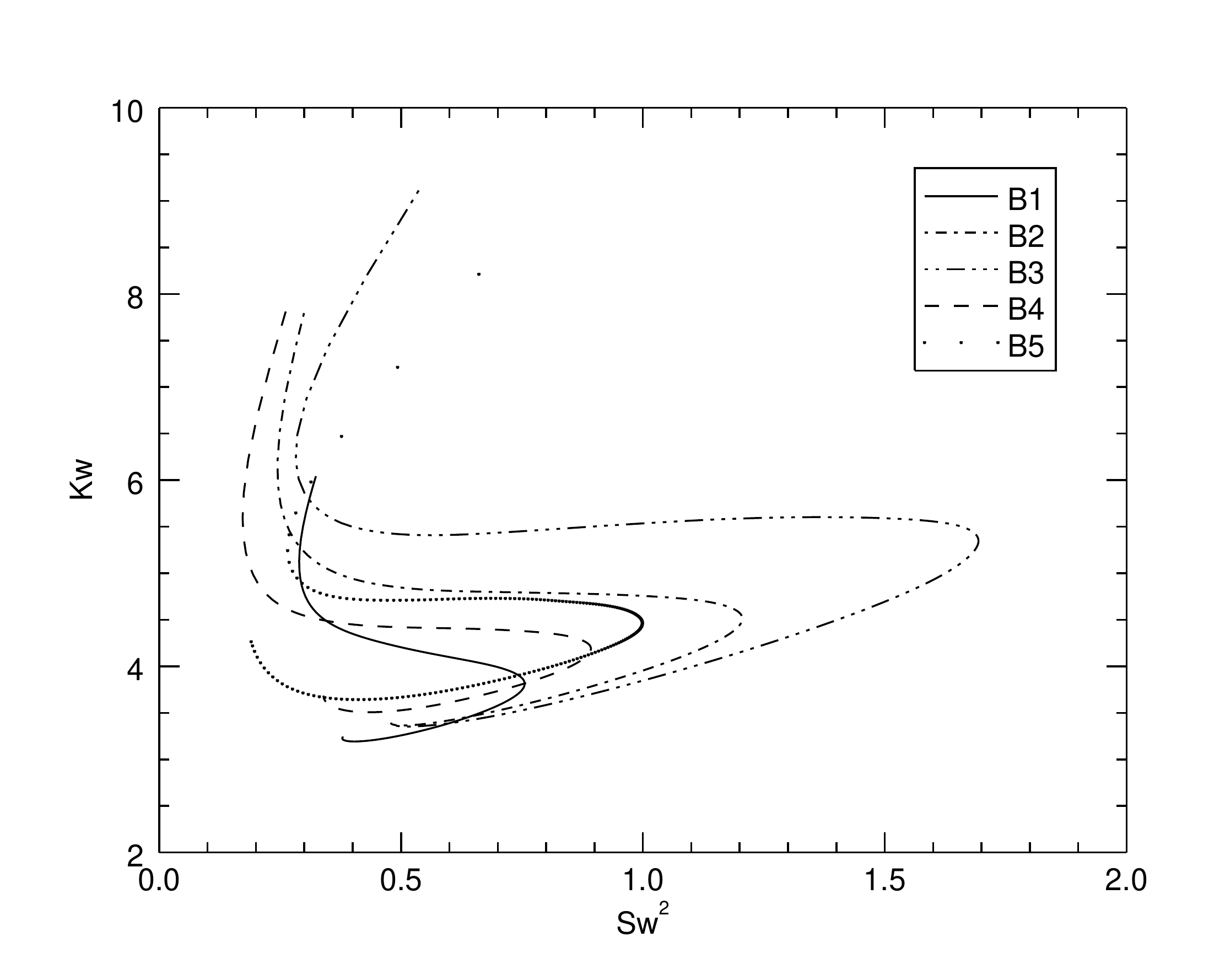}{0.5\textwidth}{(b)}
          }
\gridline{\fig{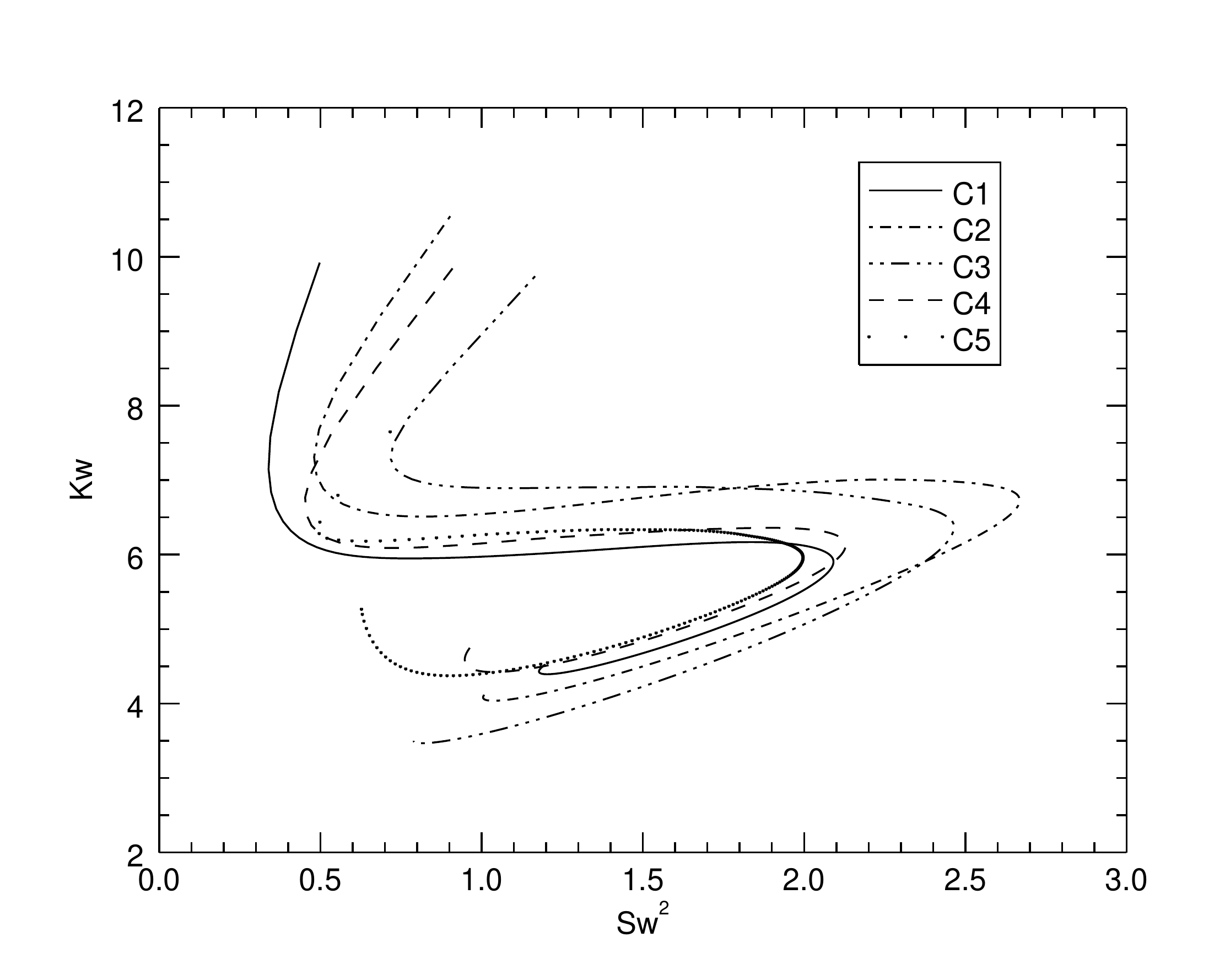}{0.5\textwidth}{(c)}
          \fig{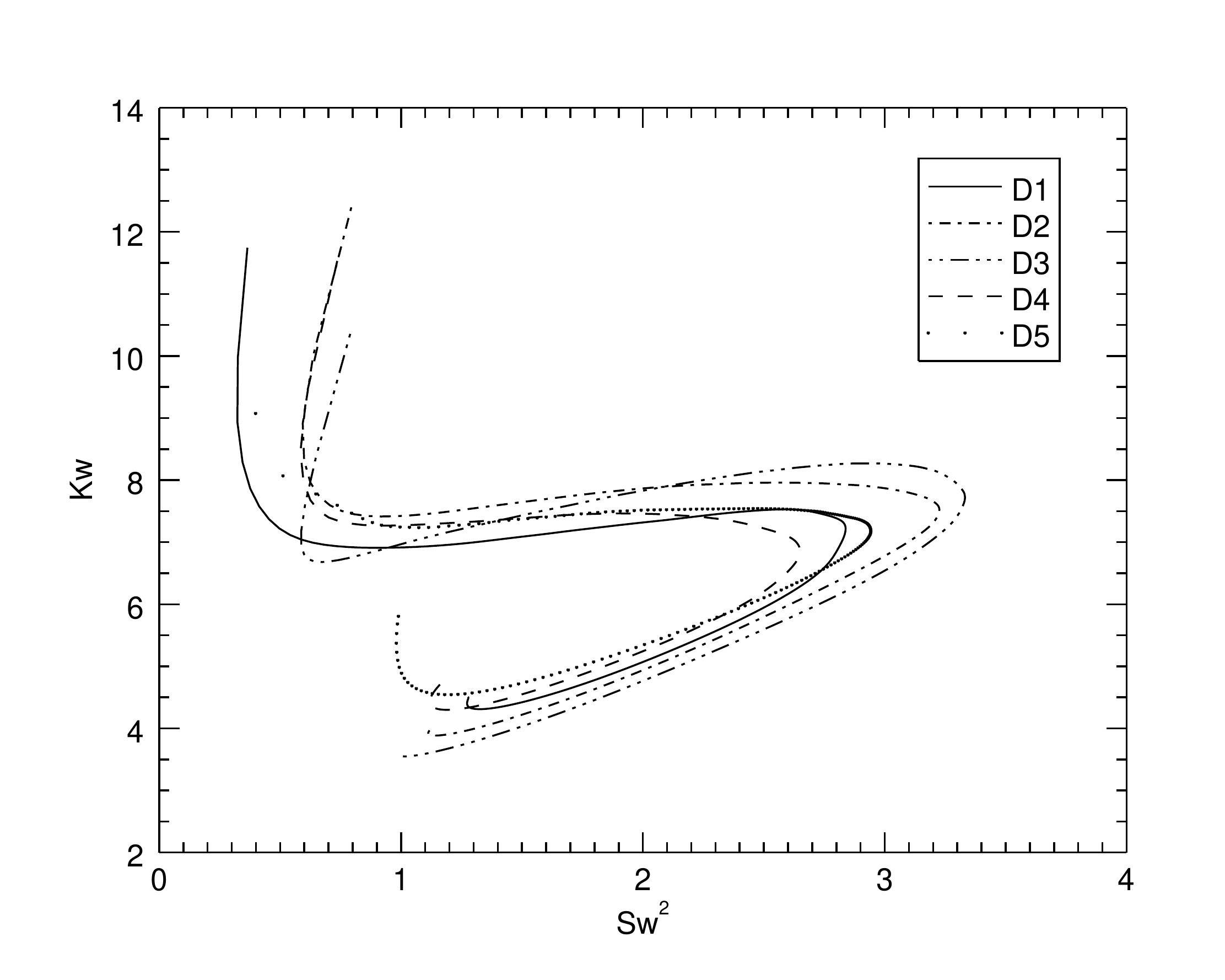}{0.5\textwidth}{(d)}
          }
\caption{The relationships between $S_{w}^2$ and $K_{w}$. Panel (a)-(d) are four different groups with increasing depth of convection zone.\label{fig:quasinormal_w}}
\end{figure*}

\begin{figure*}
\gridline{\fig{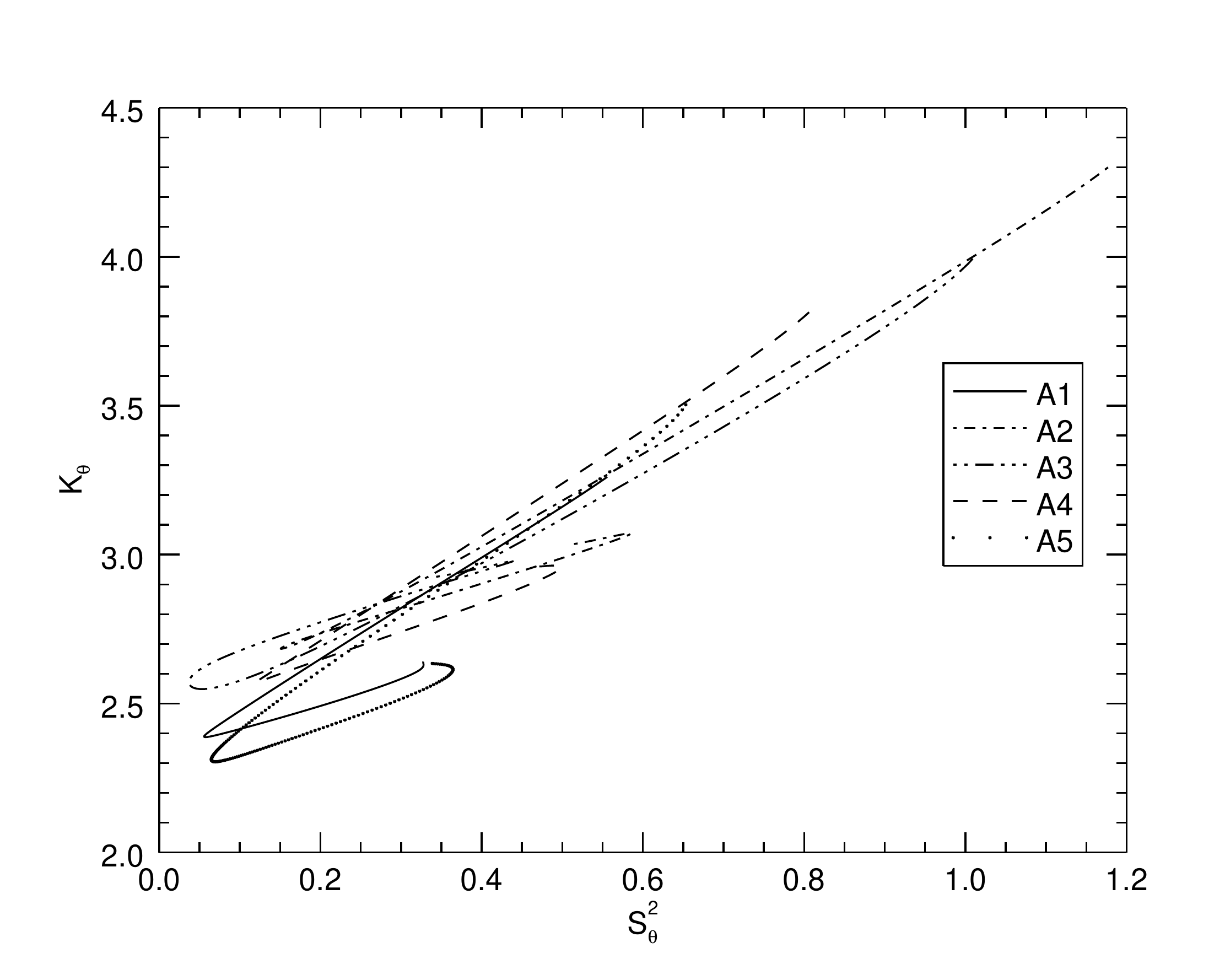}{0.5\textwidth}{(a)}
          \fig{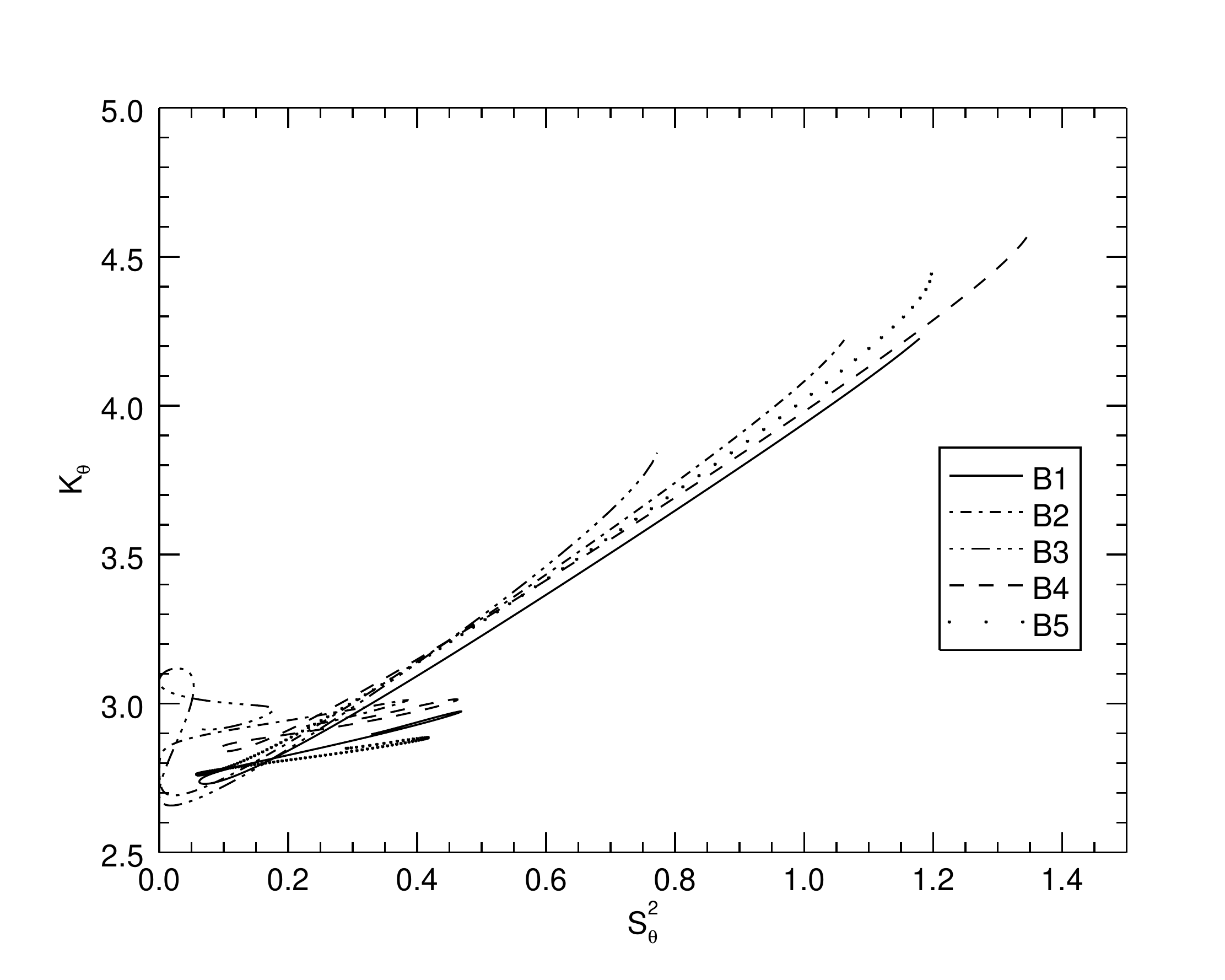}{0.5\textwidth}{(b)}
          }
\gridline{\fig{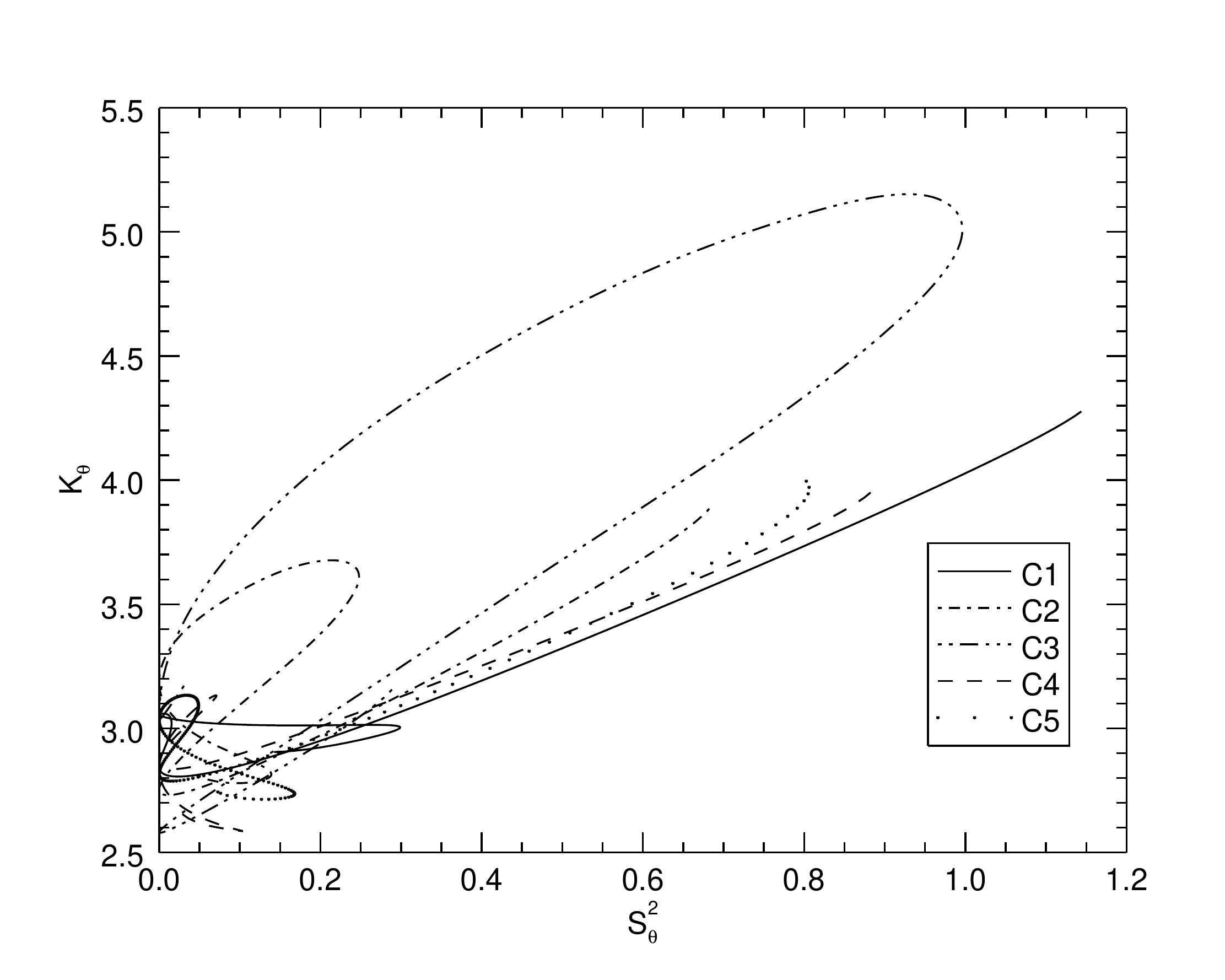}{0.5\textwidth}{(c)}
          \fig{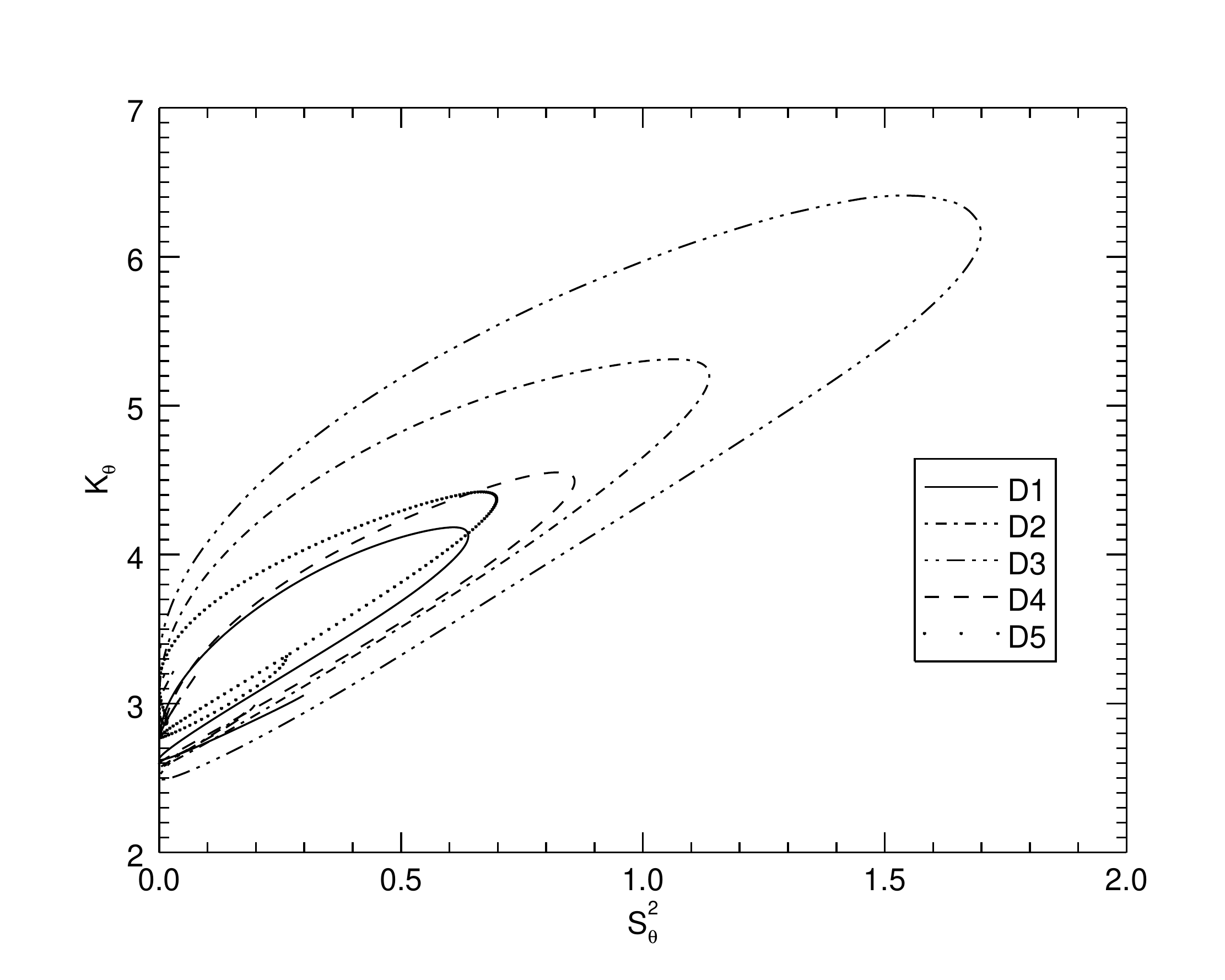}{0.5\textwidth}{(d)}
          }
\caption{The relationships between $S_{\theta}^2$ and $K_{\theta}$. Panel (a)-(d) are four different groups with increasing depth of convection zone.\label{fig:quasinormal_theta}}
\end{figure*}

In the QN model, the fluctuations of temperature and velocity around their mean values are assumed to be Gaussian distributions, so that the fourth-order moments can be expressed in terms of second-order moments \citep{1941DANS..32...615}. Under the quasi-normal approximation, the kurtoses of temperature and velocity fluctuations $K_{\theta}$ and $K_{w}$ are exactly 3. \citet{1976JAS..33...476,1976JAS..33...482} suggested that the QN model could be improved through the clipping method to avoid unphysical states which the QN otherwise leads to, and the following constraints between kurtoses and skewnesses must be satisfied \citep{2005JAS..62...2632}
\begin{eqnarray}
&&K_{w} \geq 1+S_{w}^2~,\\
&&K_{\theta} \geq 1+S_{\theta}^2~.
\end{eqnarray}
If $S_{w}^2 \geq 2$ or $S_{\theta}^2\geq 2$, then $K_{\theta}$ and $K_{w}$ would excess the limitation value of 3, and the assumption of quasi-normal approximation becomes untenable. Figure~\ref{fig:quasinormal_w} plots the relationships between $K_{w}$ and $S_{w}^2$ for all the simulation cases. First, we find that the maximum value of $S_{w}^2$ is achieved in the middle of the convection zone, and increases with the Rayleigh number. Since the downward flow network developed at the top merges into stronger downward lanes in the middle of the convection, it is expected that the maximum value of skewness achieves in the middle of convection zone, because of the asymmetry of upward and downward flows. The asymmetry of upward and downward flows is more significant as Rayleigh number increases. When the Rayleigh number is greater than $10^7$, the value of $S_{w}^2$ can be greater than 2 and the quasi-normal approximation is expected to have poor performance. Second, we note that in most cases the kurtoses are higher than 3 in a wide range of the simulation domain. It indicates that the probability density function of velocity fluctuations has a heavy tail compared to the standard normal distribution. For the shallowest cases in panel (a), the kurtoses are close to 3 except in the top of the convection zone where strong downward flows are generated. For the case A1 with smallest Rayleigh number, the kurtosis could be smaller than 3 and the fluid motion tends to be in a quasi-laminar state. For the cases with Rayleigh number higher than $10^7$, the kurtoses in the middle of the convection zone are basically higher than 5, thus the quasi-normal approximation of velocity fluctuations should be rejected. Companion with Figure~\ref{fig:quasinormal_w}, Figure~\ref{fig:quasinormal_theta} shows the relationship between $K_{\theta}$ and $S_{\theta}^2$. Firstly, unlike the velocity fluctuations, the inequality $S_{\theta}^2 \leq 2$ holds well for all the simulation cases. In our simulations, we have observed that $S_{\theta}^2\leq S_{w}^2$  almost holds for all the cases with Rayleigh numbers higher than $10^{7}$. Secondly, the kurtoses $K_{\theta}$ are smaller than 5, and the inequality $K_{\theta}\leq K_{w}$ almost holds in all the cases.

\begin{figure*}
\gridline{\fig{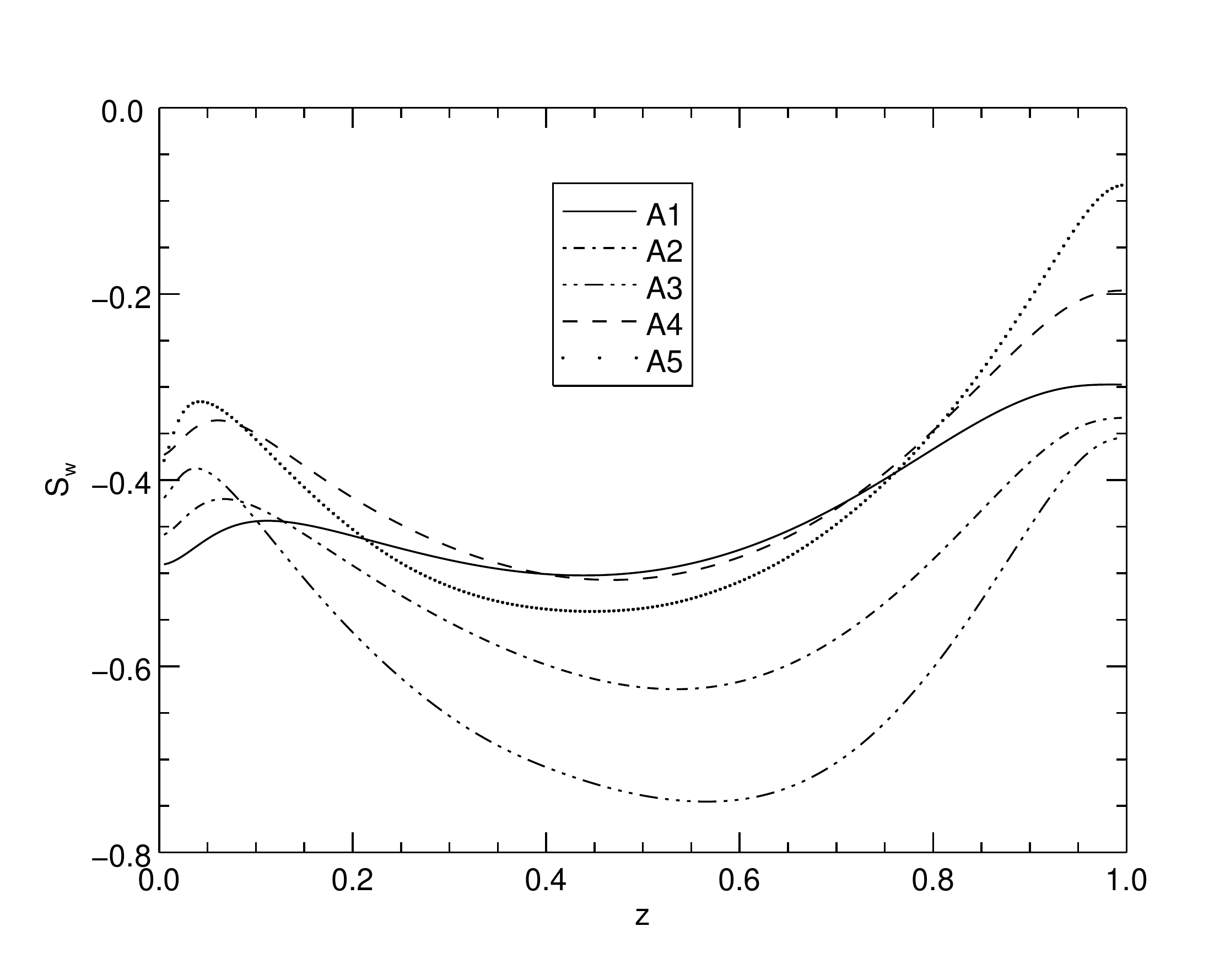}{0.5\textwidth}{(a)}
          \fig{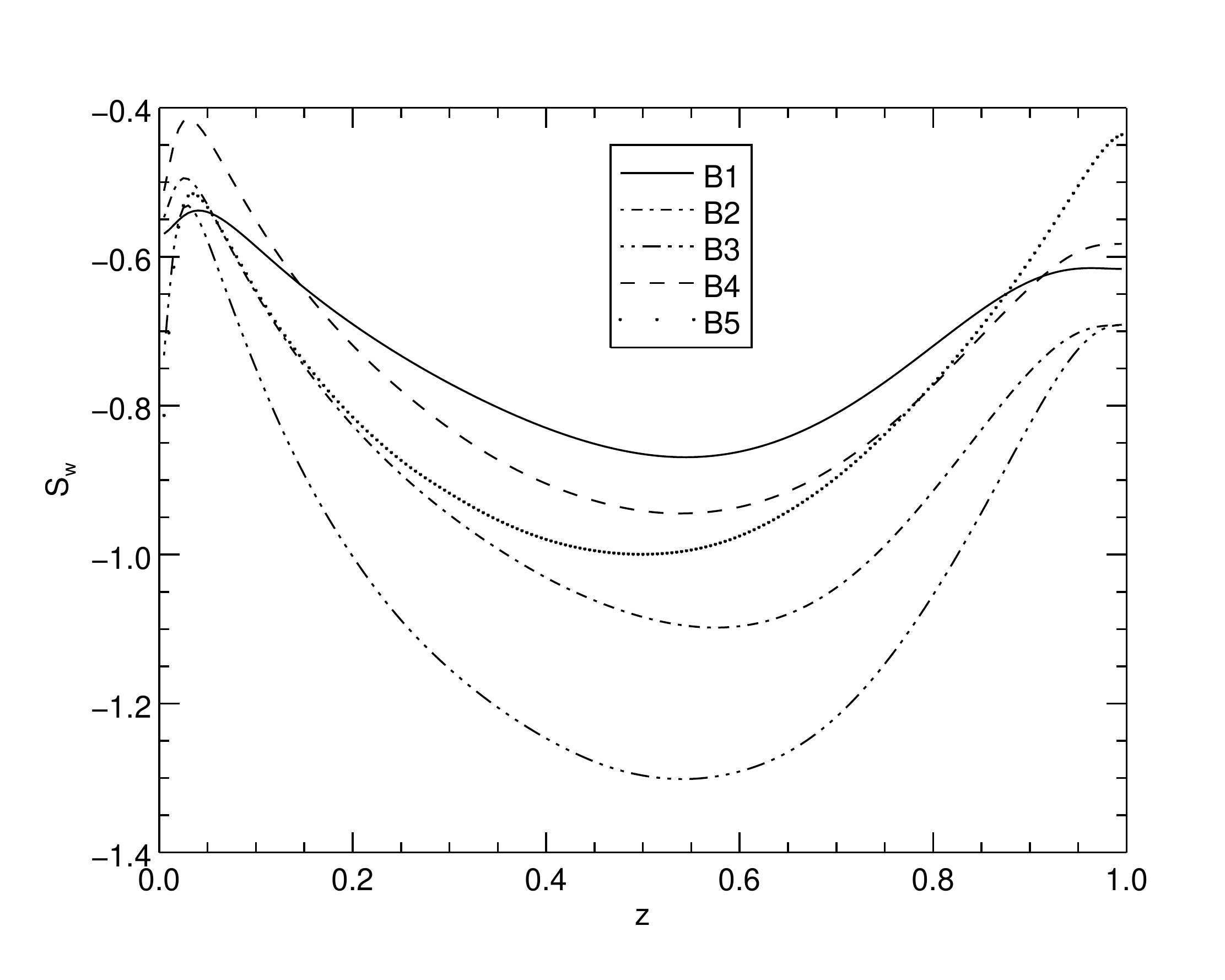}{0.5\textwidth}{(b)}
          }
\gridline{\fig{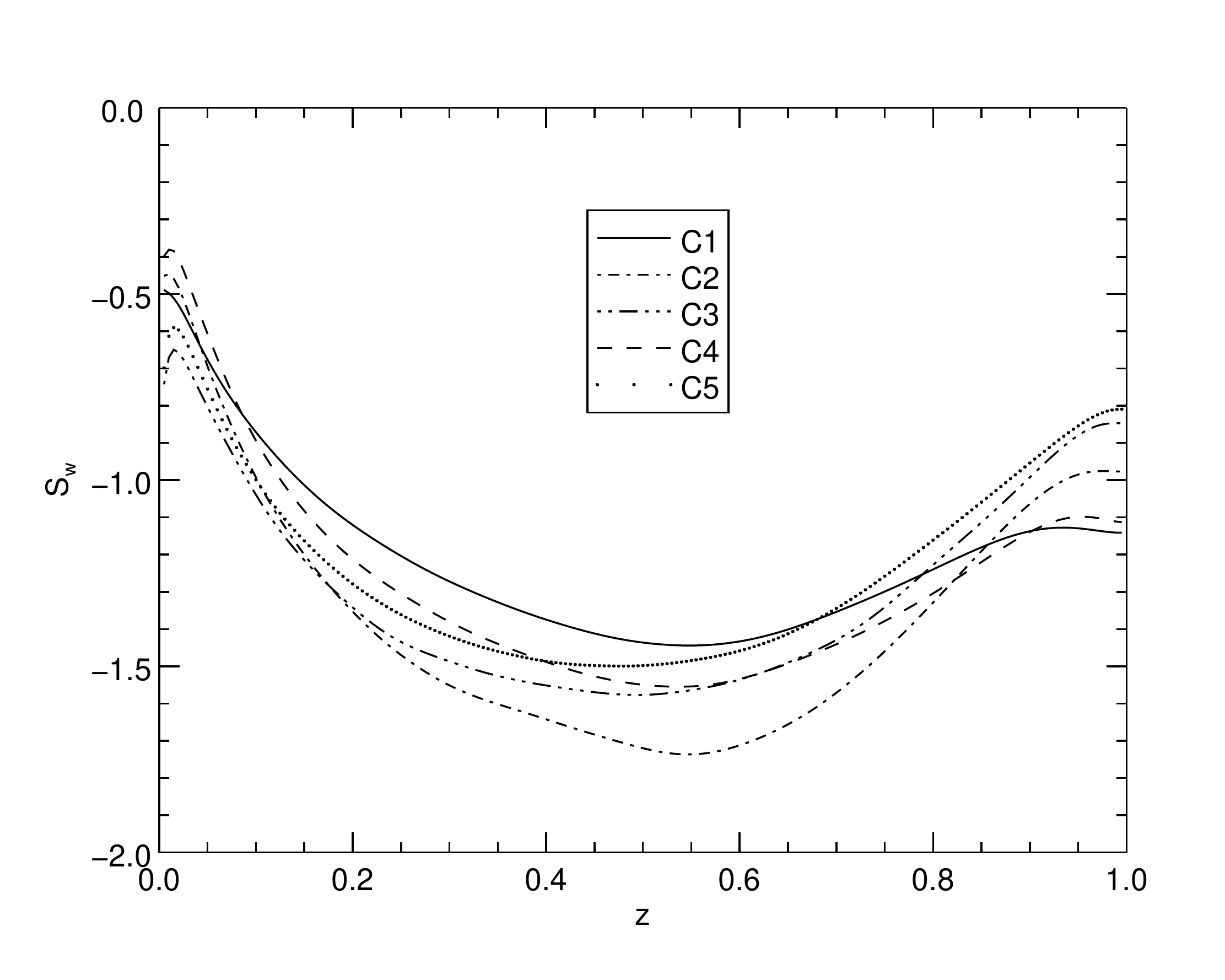}{0.5\textwidth}{(c)}
          \fig{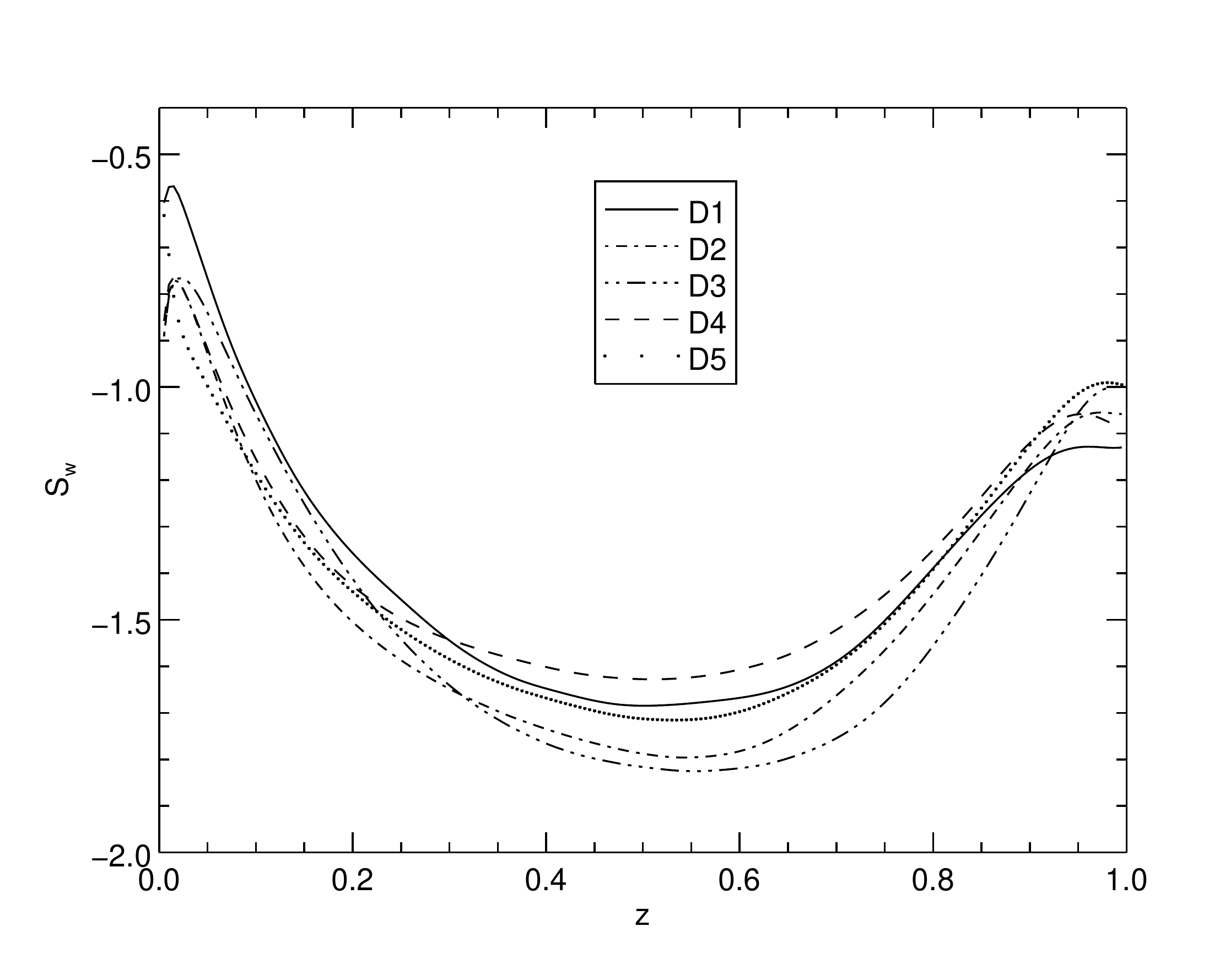}{0.5\textwidth}{(d)}
          }
\caption{The skewness $S_{w}$ as a function of depth. Panel (a)-(d) are four different groups with increasing depth of convection zone.\label{fig:Sw}}
\end{figure*}

\begin{figure*}
\gridline{\fig{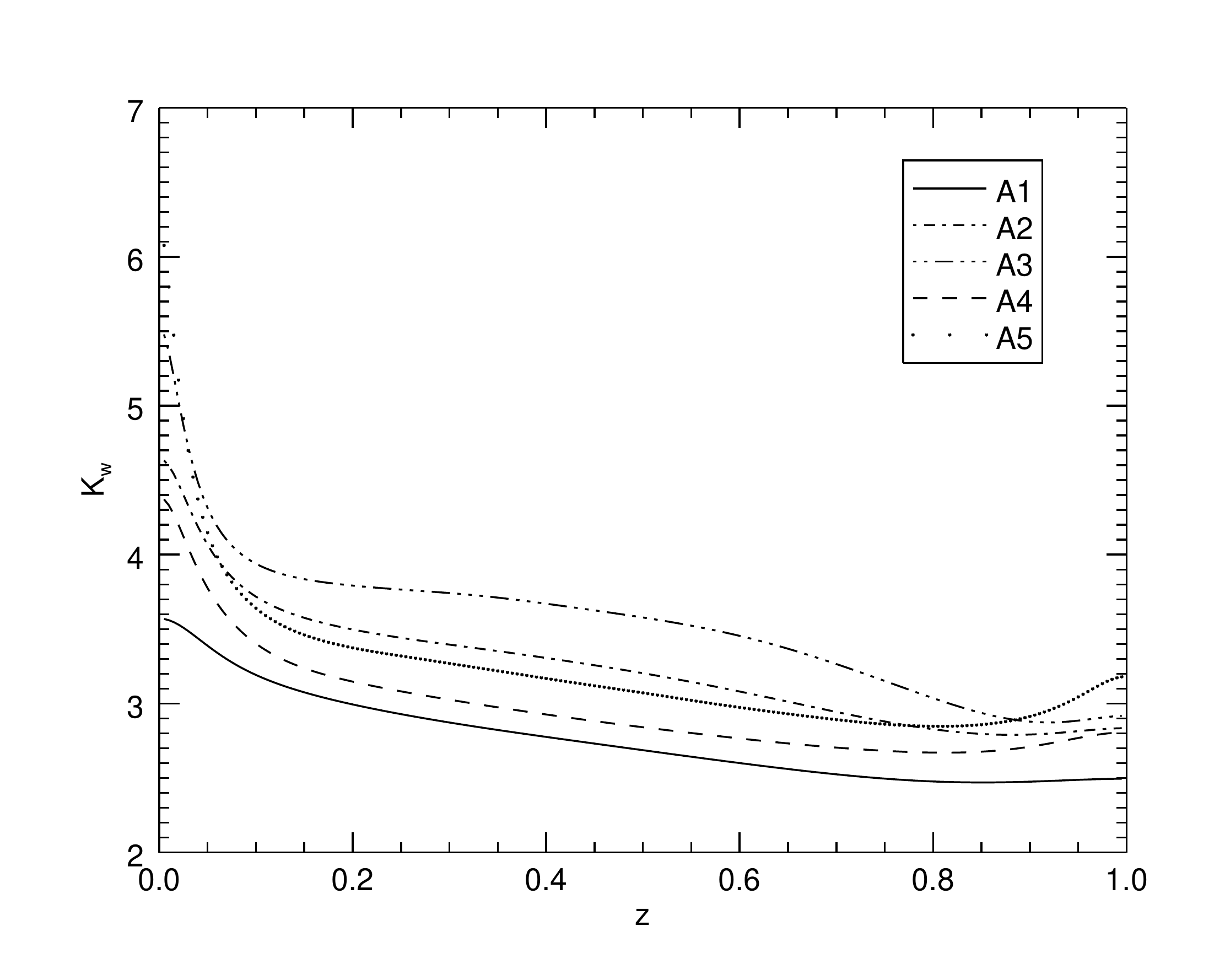}{0.5\textwidth}{(a)}
          \fig{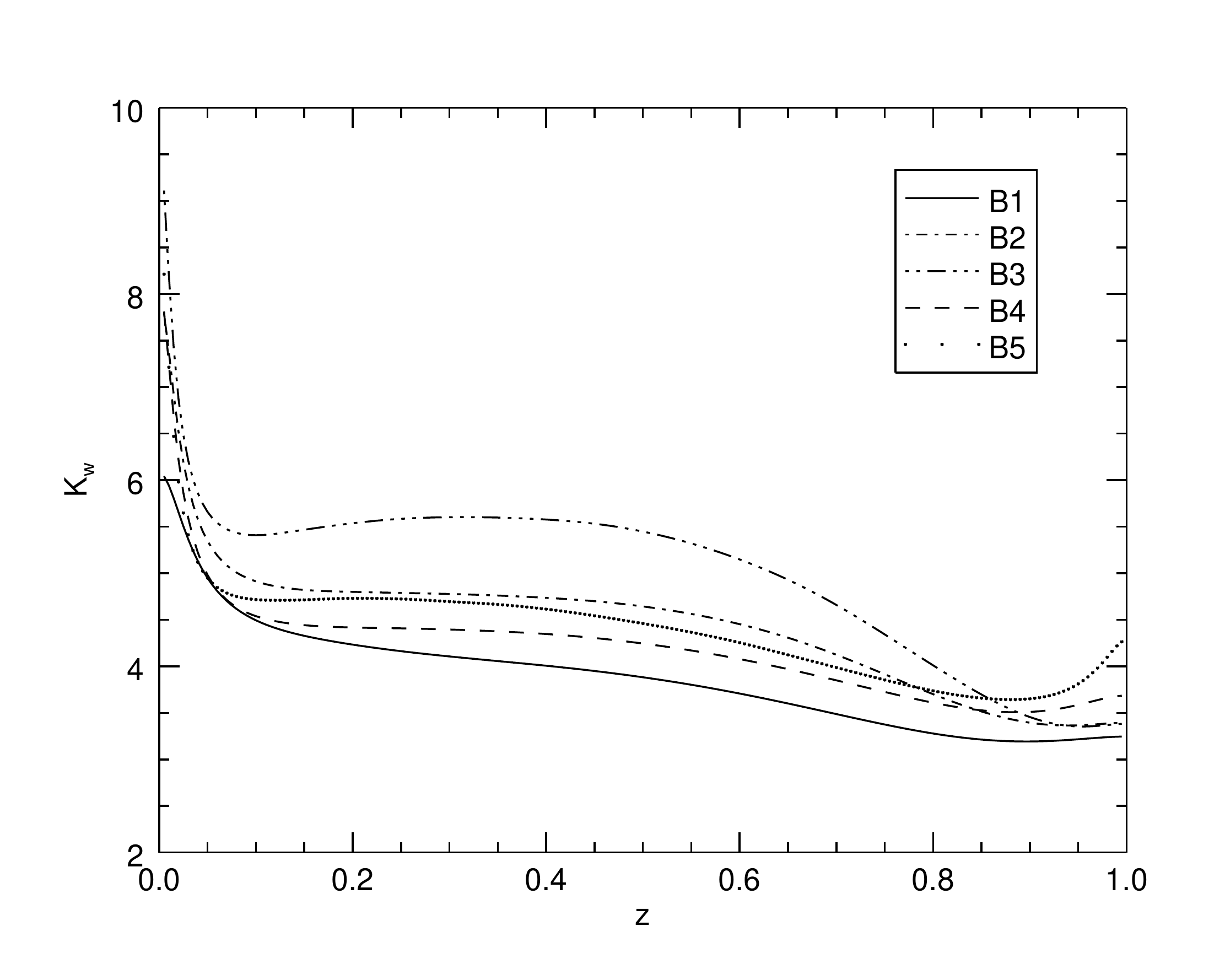}{0.5\textwidth}{(b)}
          }
\gridline{\fig{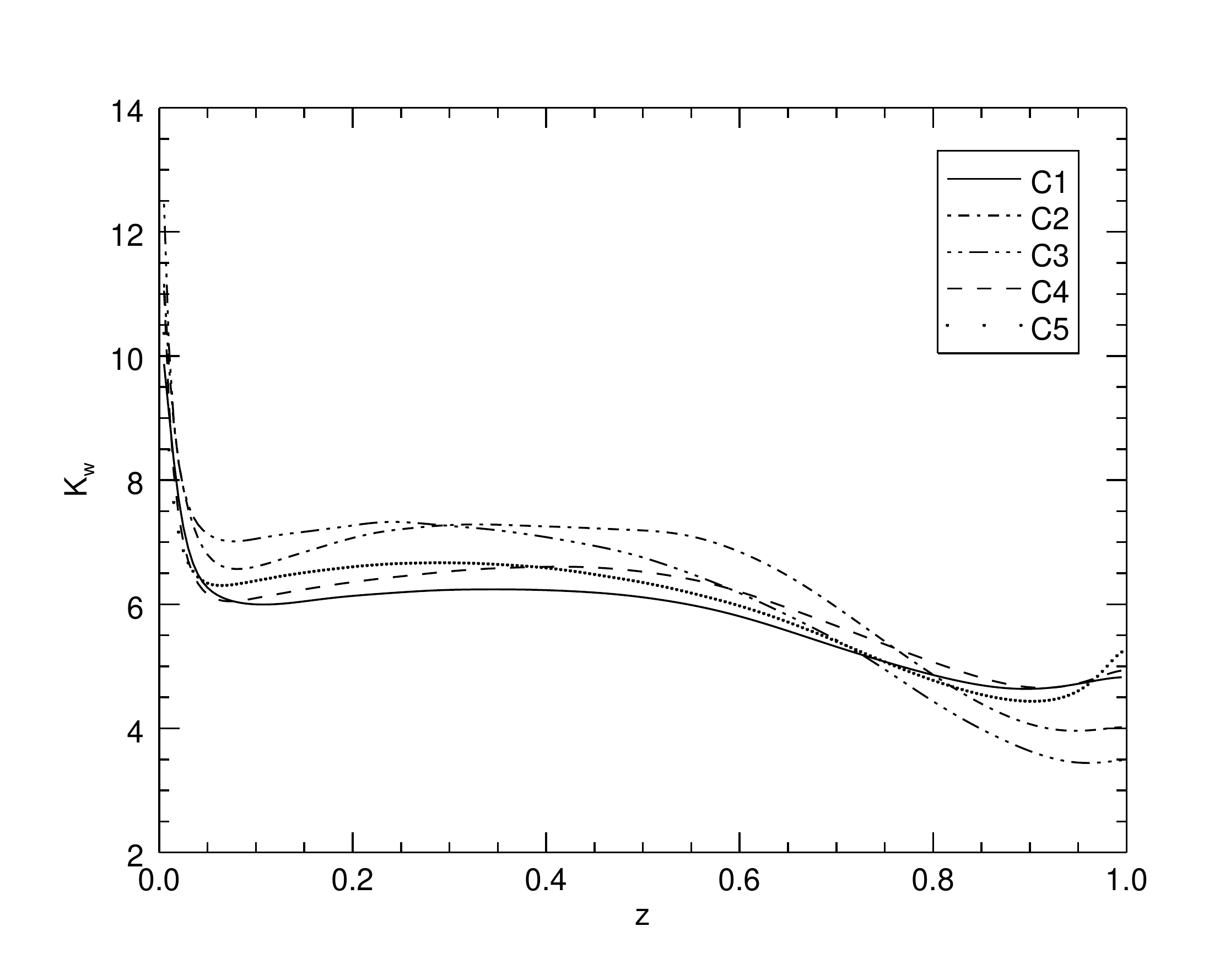}{0.5\textwidth}{(c)}
          \fig{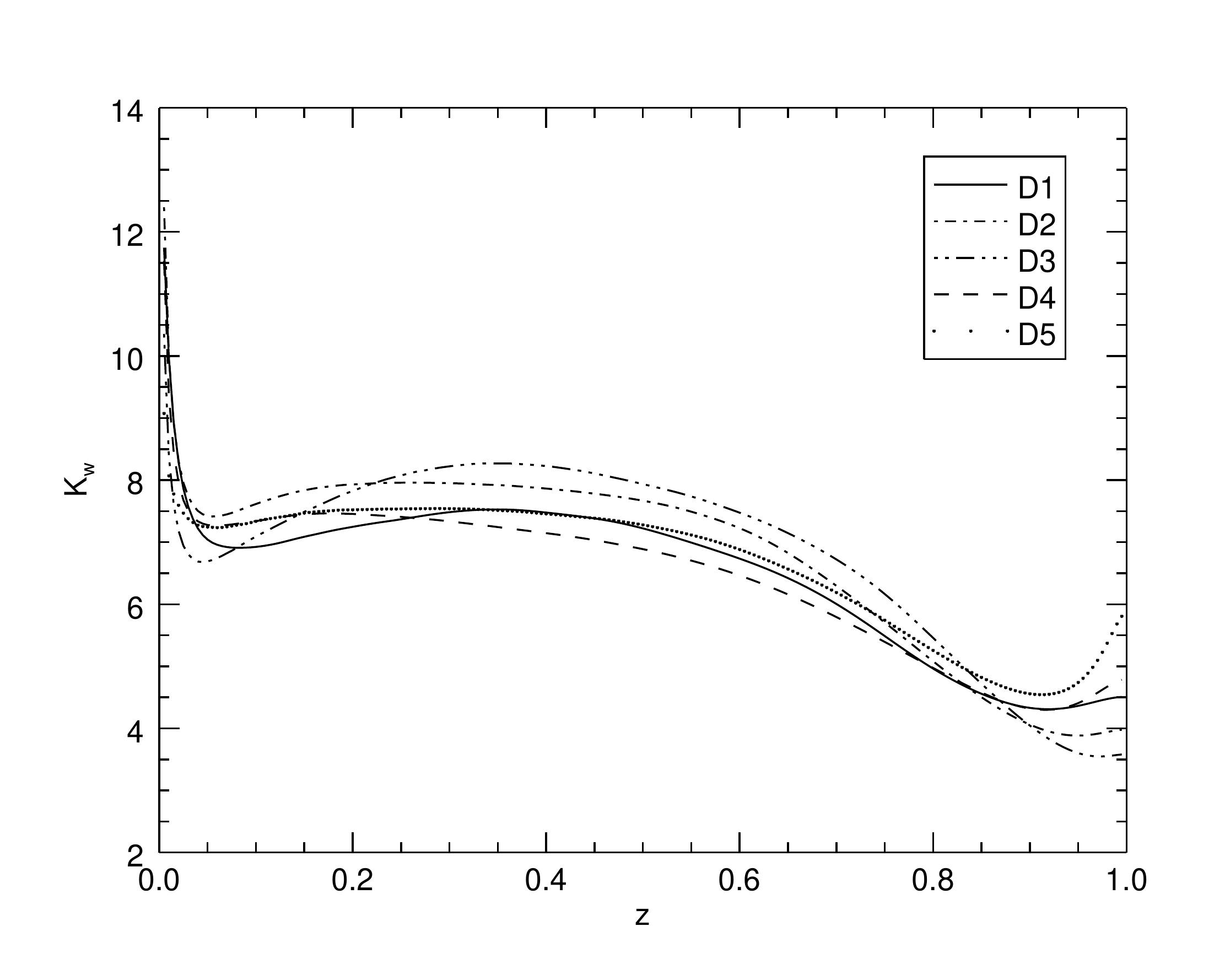}{0.5\textwidth}{(d)}
          }
\caption{The kurtosis $K_{w}$ as a function of depth. Panel (a)-(d) are four different groups with increasing depth of convection zone.\label{fig:Kw}}
\end{figure*}

\begin{figure*}
\gridline{\fig{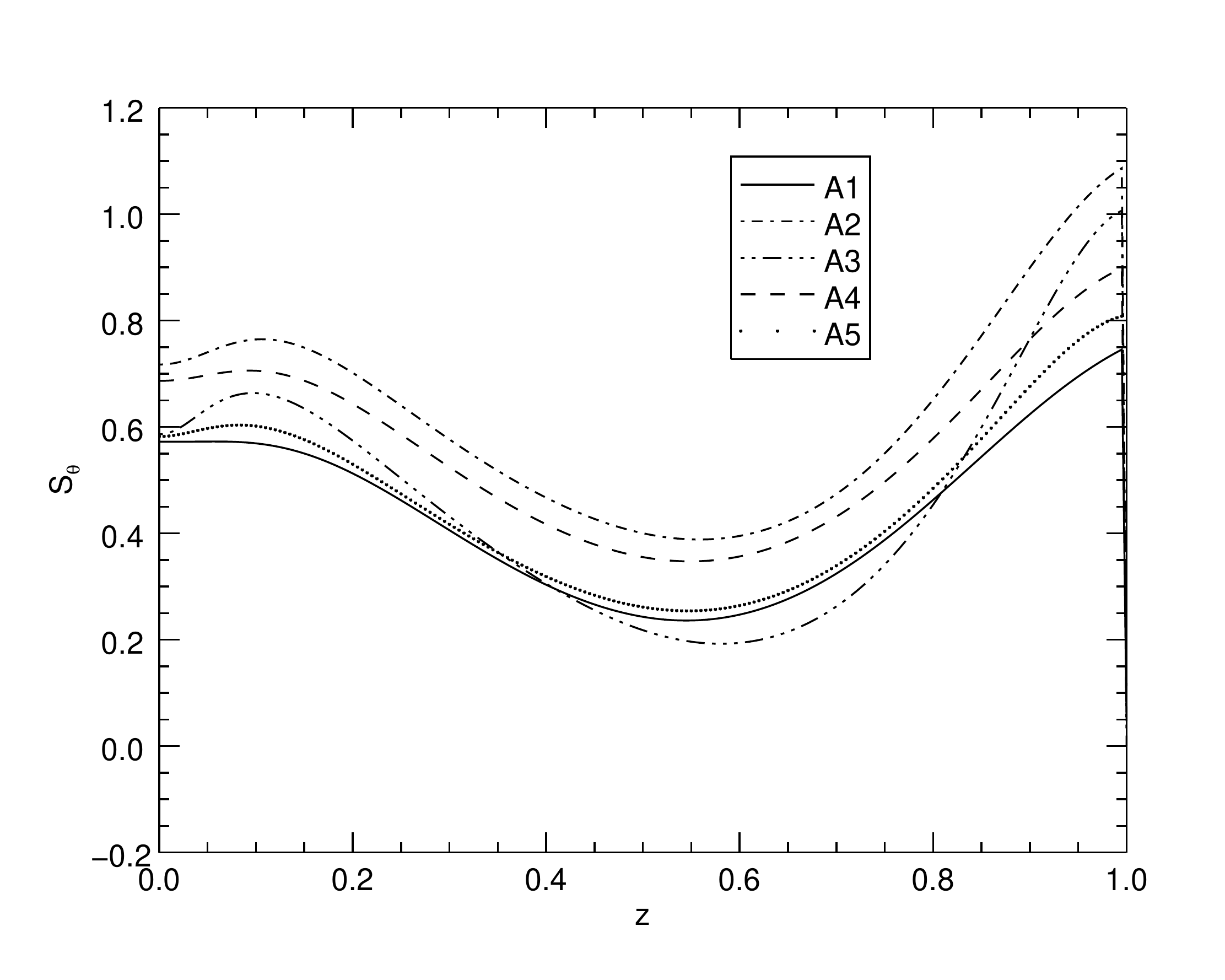}{0.5\textwidth}{(a)}
          \fig{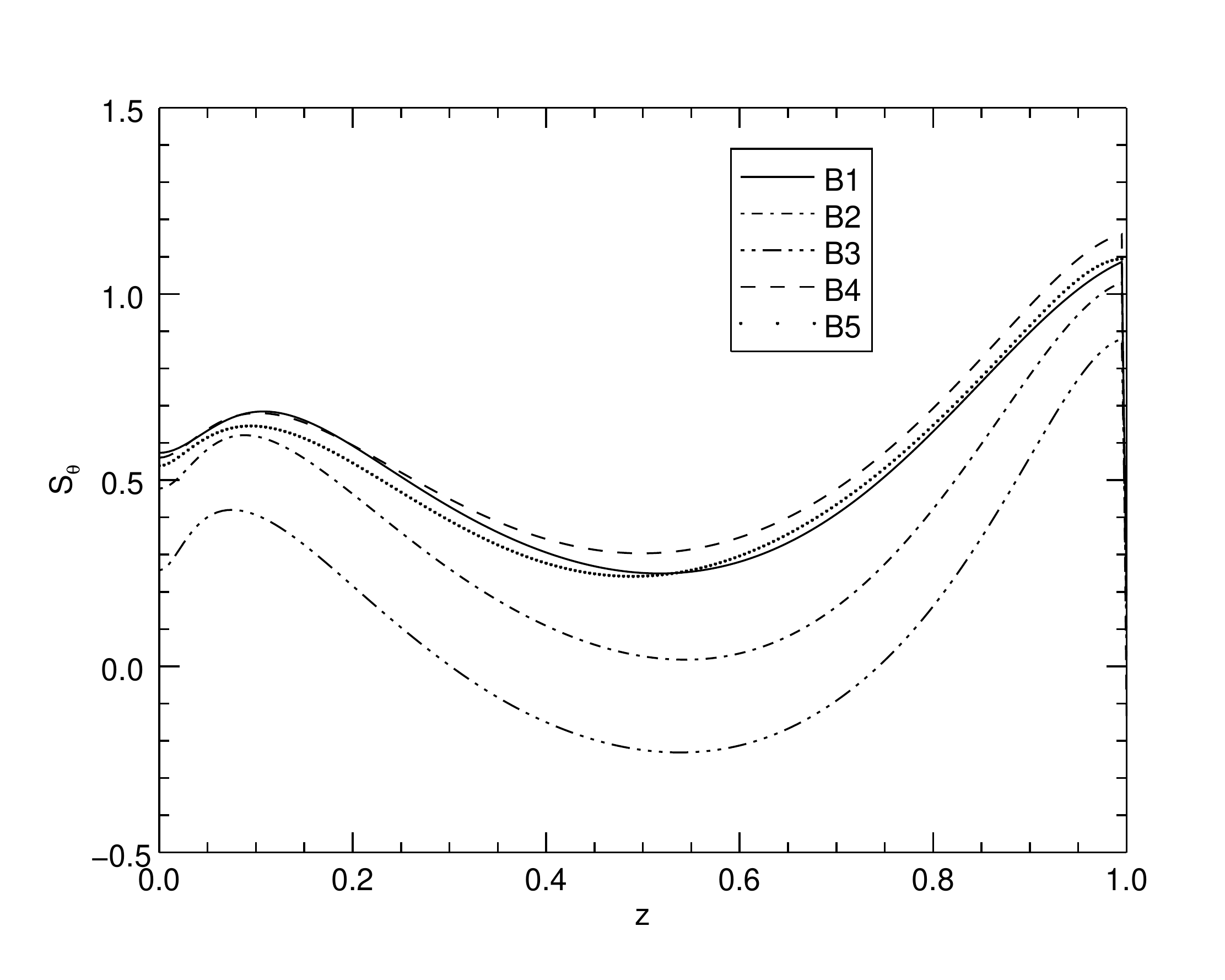}{0.5\textwidth}{(b)}
          }
\gridline{\fig{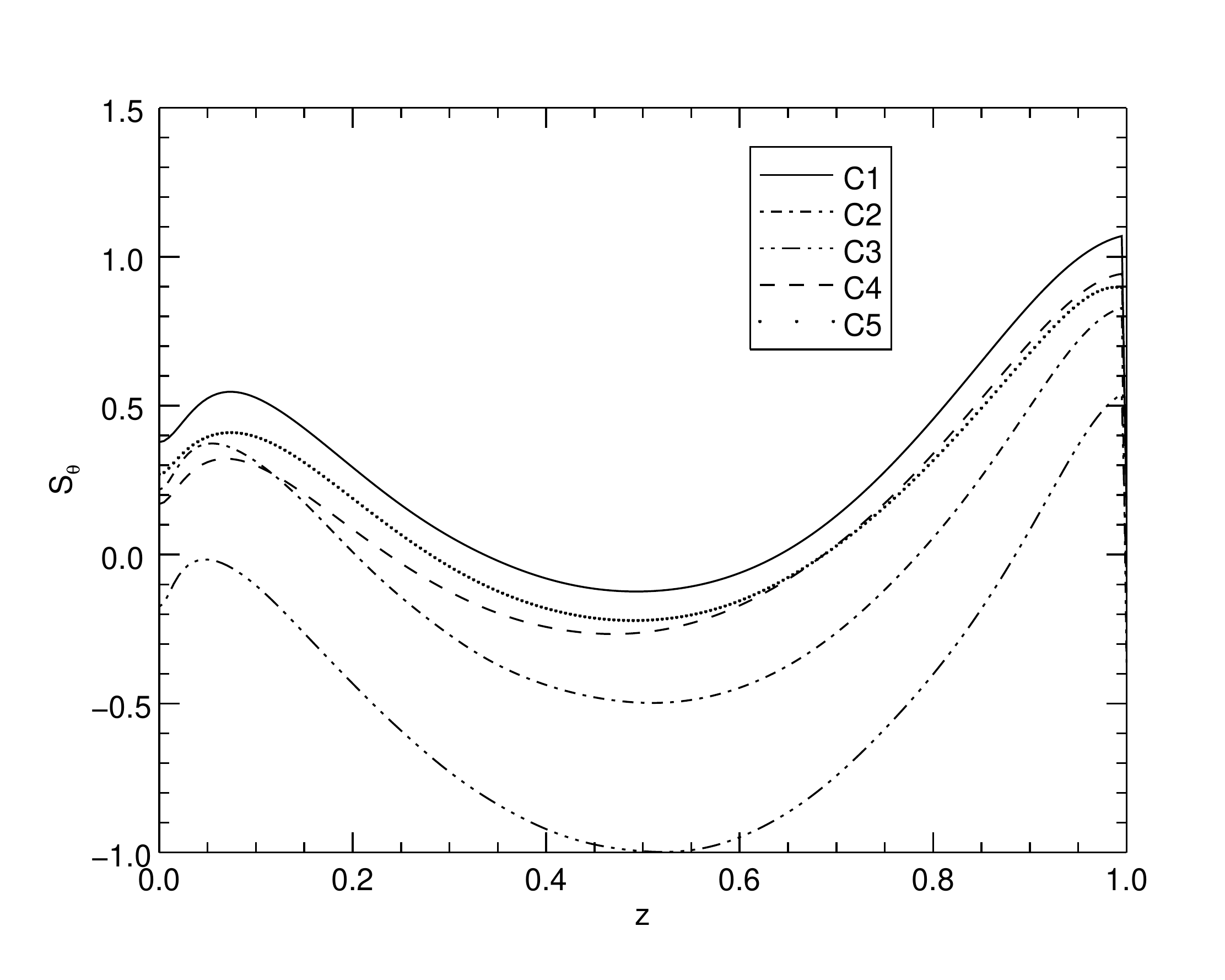}{0.5\textwidth}{(c)}
          \fig{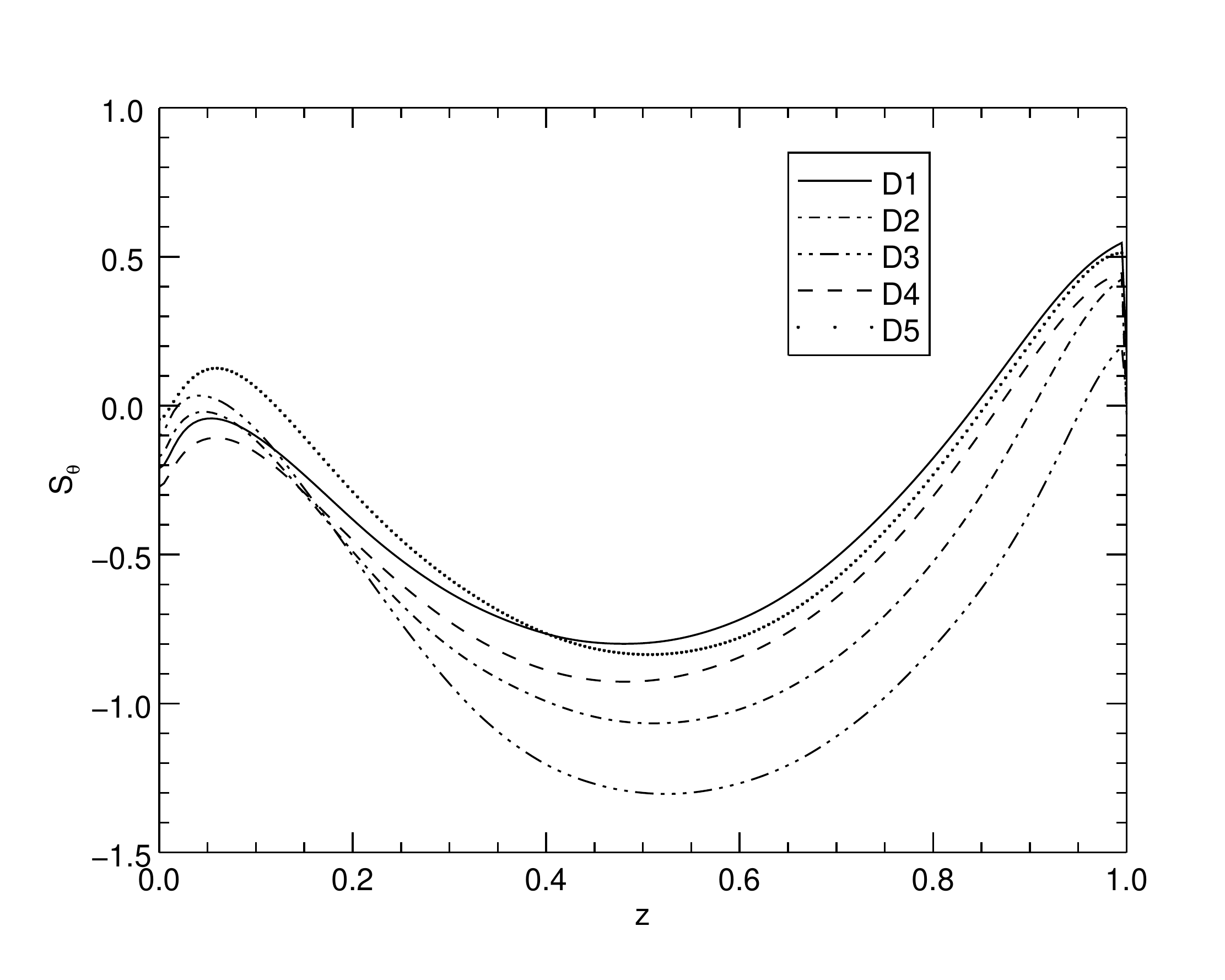}{0.5\textwidth}{(d)}
          }
\caption{The skewness $S_{\theta}$ as a function of depth. Panel (a)-(d) are four different groups with increasing depth of convection zone.\label{fig:St}}
\end{figure*}

\begin{figure*}
\gridline{\fig{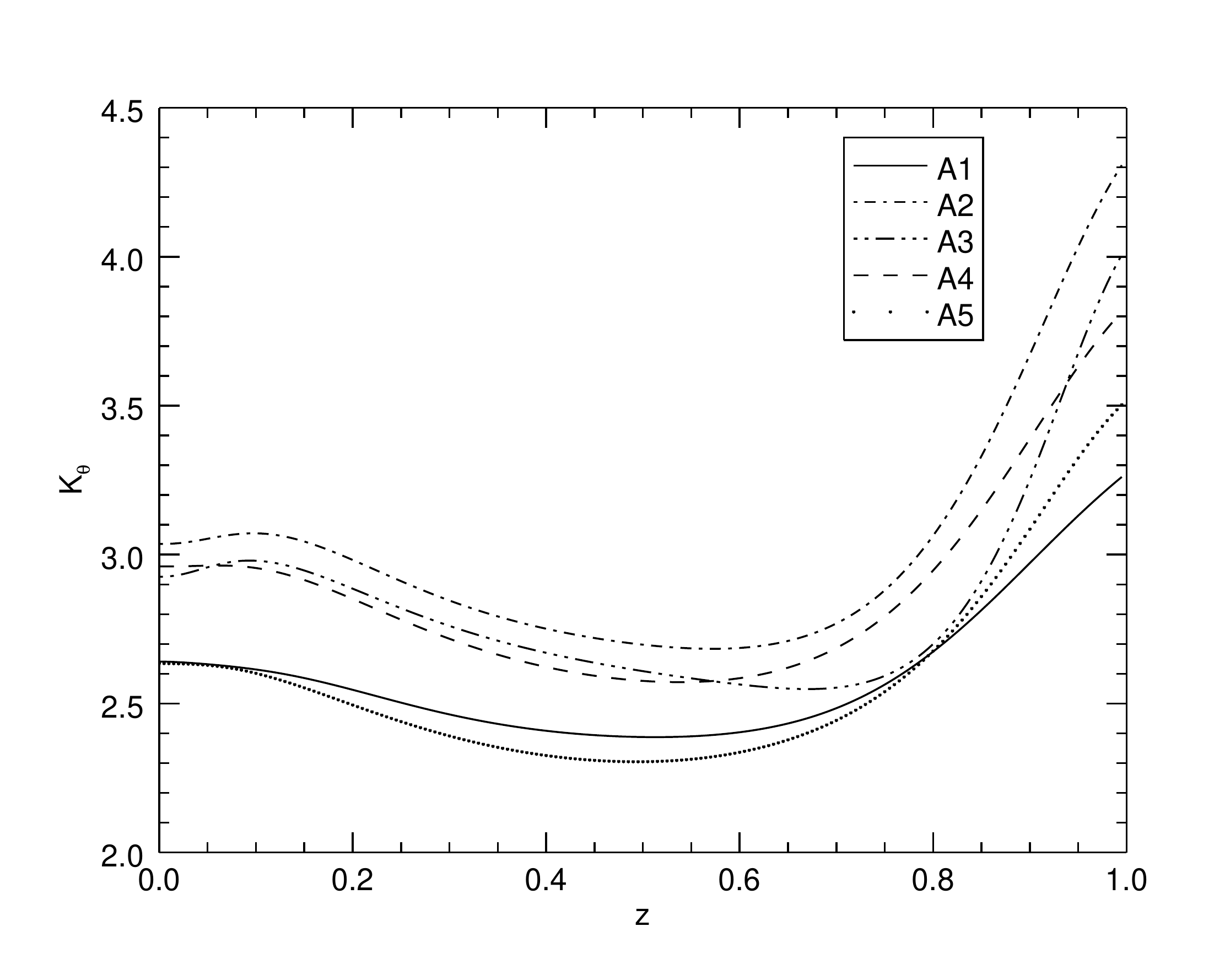}{0.5\textwidth}{(a)}
          \fig{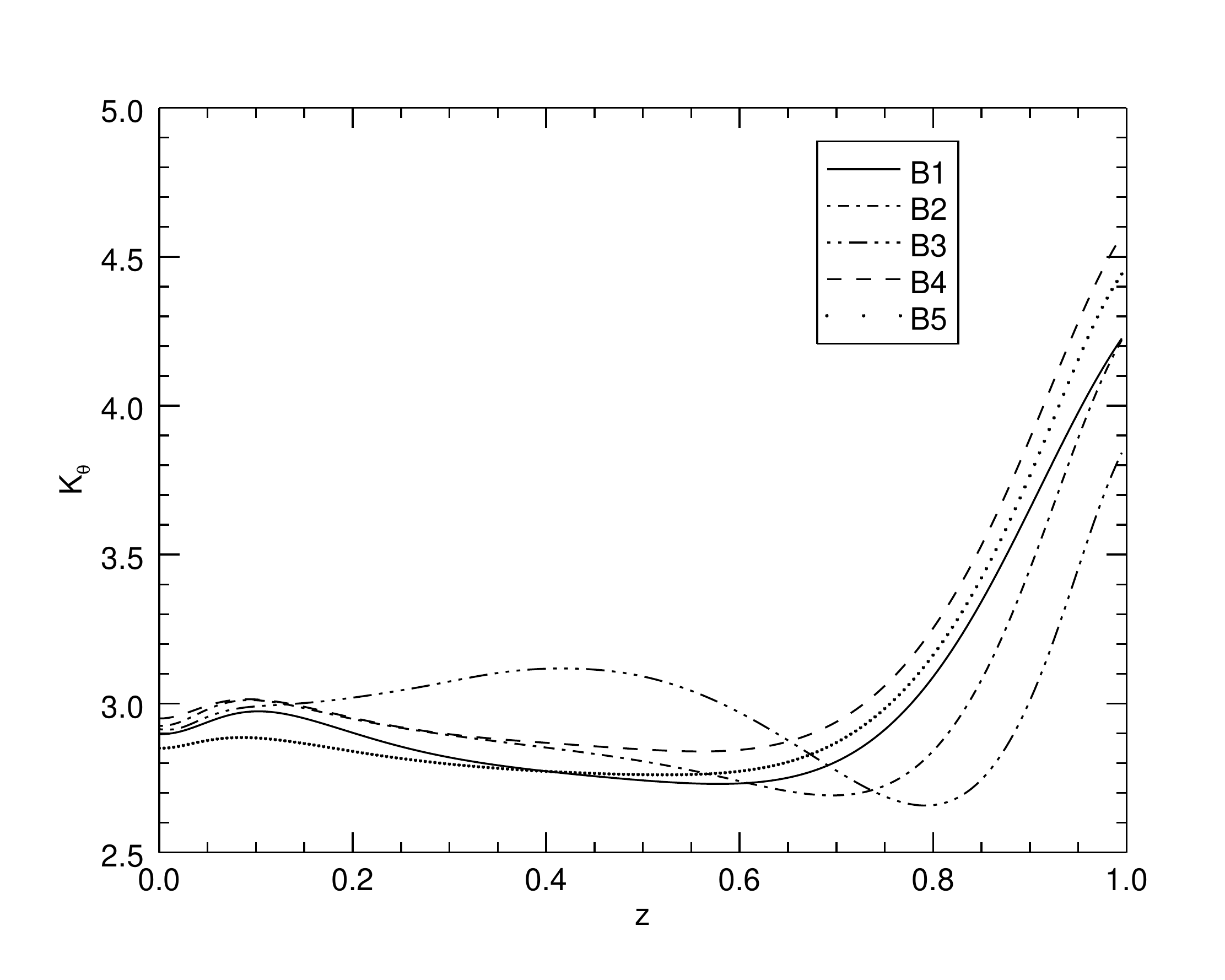}{0.5\textwidth}{(b)}
          }
\gridline{\fig{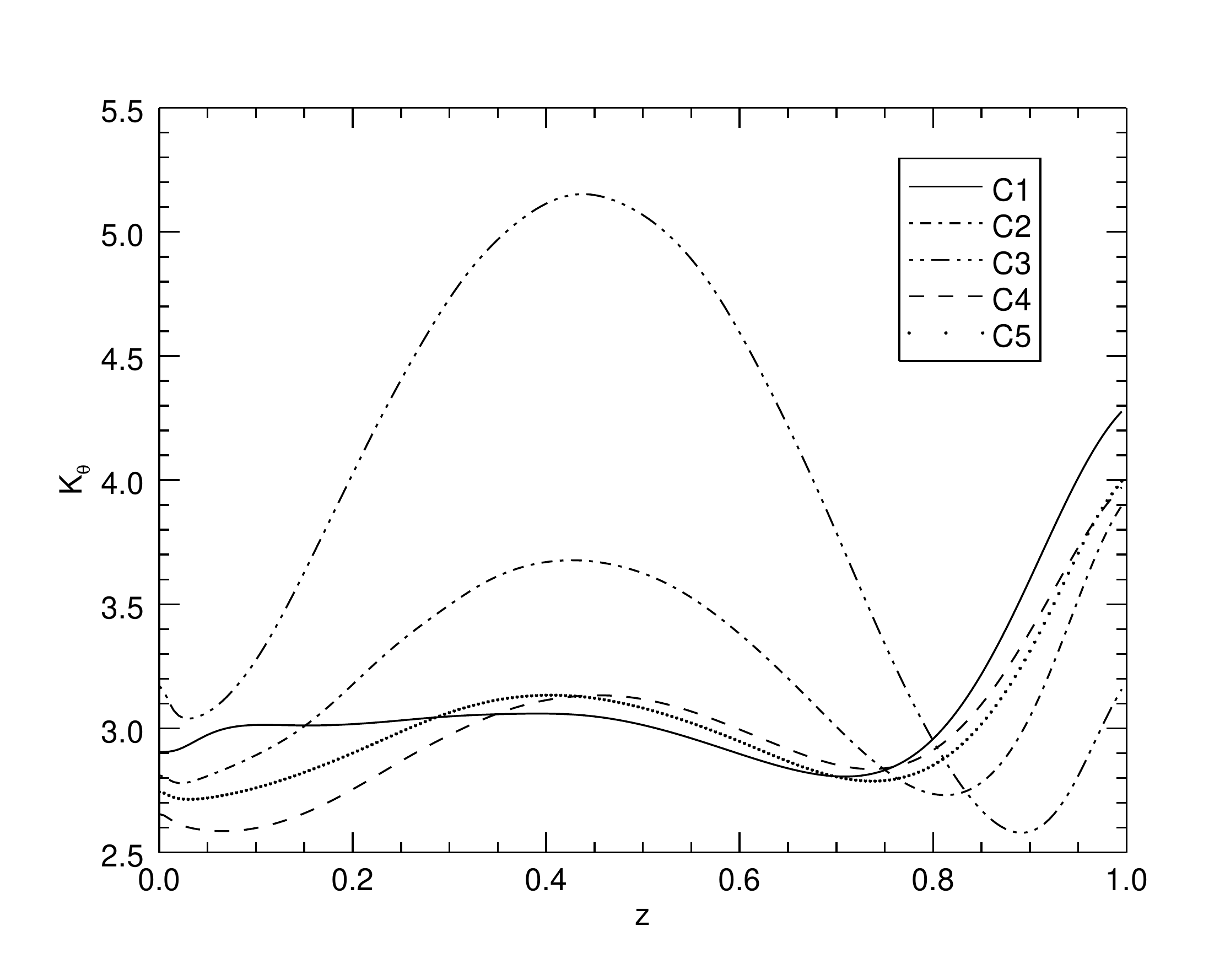}{0.5\textwidth}{(c)}
          \fig{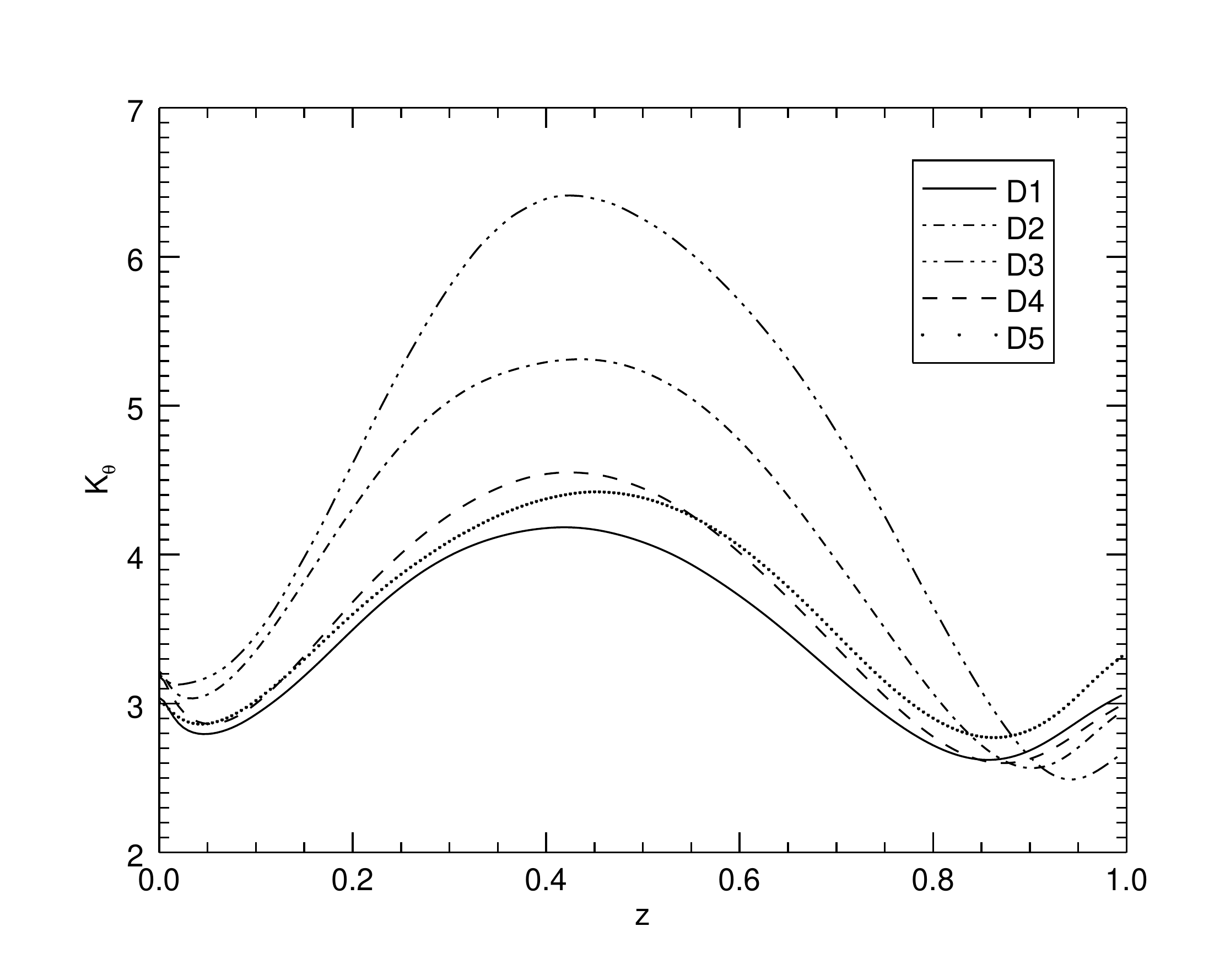}{0.5\textwidth}{(d)}
          }
\caption{The kurtosis $K_{\theta}$ as a function of depth. Panel (a)-(d) are four different groups with increasing depth of convection zone.\label{fig:Kt}}
\end{figure*}

We further consider the skewnesses and kurtoses of velocity and temperature perturbations in detail. Figure~\ref{fig:Sw} shows the skewness $S_{w}$ as a function of depth. In all the simulation cases, $S_{w}$ is always negative throughout the simulation domain. Negative $S_{w}$ means that the downward flows are much stronger than upward flows, and this difference is more significant in the deeper convection zone. Within each panel of the figure, we see that $S_{w}$ is largely affected by P\'{e}clet number. When the P\'{e}clet number increases, $S_{w}$ decreases and the vertical velocities are likely to be more asymmetric. Figure~\ref{fig:Kw} shows the kurtosis $K_{w}$ as a function of depth. Strongly affected by the boundary effects, $K_{w}$ increases rapidly at the boundary regions. Since the boundary region is less of interest, we mainly discuss the results of $K_{w}$ at the rest of simulation domain. In the upper part of the simulation domain ($z>0.5$), $K_{w}$ almost changes with $S_{w}^2$ at the same direction. In this region the flow is in a pattern of cellular network of strong downward lanes, hence the large values of $K_{w}$ are mainly attributed to the asymmetry of flow. On the other hand, in the lower part of the simulation domain ($z<0.5$), $K_{w}$ is no longer positively correlated with $S_{w}^2$. Thus the large values of $K_{w}$ in this region are most likely contributed from perturbations among the drafts. Figure~\ref{fig:St} plots the skewness $S_{\theta}$. Unlike $S_{w}$, the skewness $S_{\theta}$ is not always negative. In some cases ($Pe<50$), $S_{\theta}$ is positive throughout the whole simulation domain. In such cases, the temperature variations of hot fluids are much larger than cold fluids. In the cases of deep convection zones (cases D1-D5), $S_{\theta}$ is negative in most of the simulation domain just like $S_{w}$. It is anticipated since negative temperature fluctuations are highly correlated with strong downward flows, as we have seen in the flow patterns. The different signs of $S_{\theta}$ between shallow and deep convection zones can be explained through the effects of thermal boundary conditions. In our numerical setting, the convection zone is heated from the bottom and cooled at the top. $S_{\theta}$ is positive (negative) when the heating (cooling) effect dominates. In the deep convection zone, $S_{\theta}$ is negative because the cooling effect from the top dominates. In the shallow convection zone, however, the cooling and heating effects directly interact, and $S_{\theta}$ could be positive if the heating effect is more significant.   Figure~\ref{fig:Kt} plots the kurtosis $K_{\theta}$. Compared with $K_{w}$, the kurtosis of temperature perturbation $K_{\theta}$ is smaller. In the cases with $Pe<50$, $K_{\theta}$ is smaller or close to 3 in most of the simulation domain. It seems that quasi-normal approximation has a fair performance for the temperature perturbation in such cases. When $Pe>50$, $K_{\theta}$ exceeds 3 and worse performance of the quasi-normal approximation is anticipated.

\begin{figure*}
\gridline{\fig{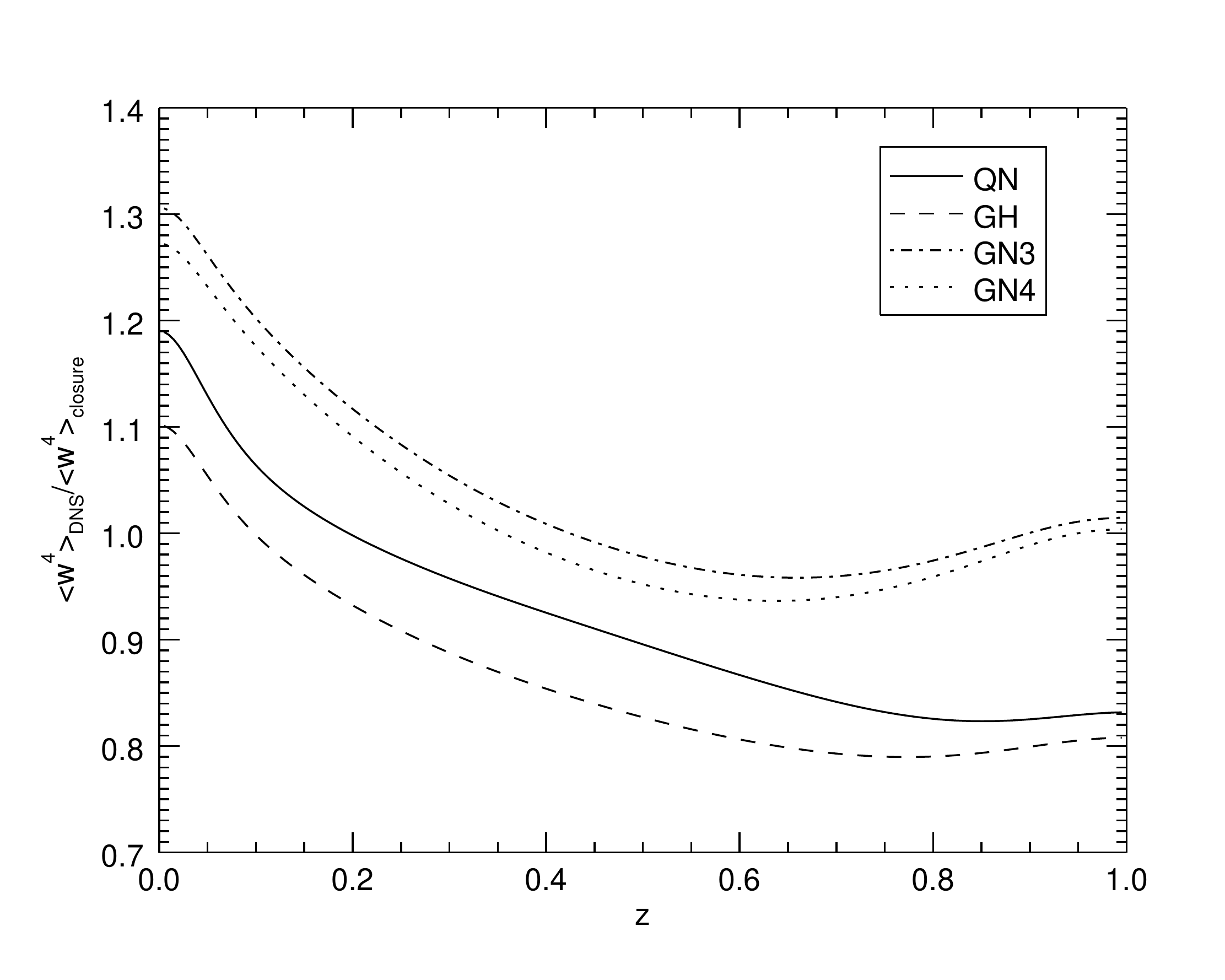}{0.5\textwidth}{(a)}
          \fig{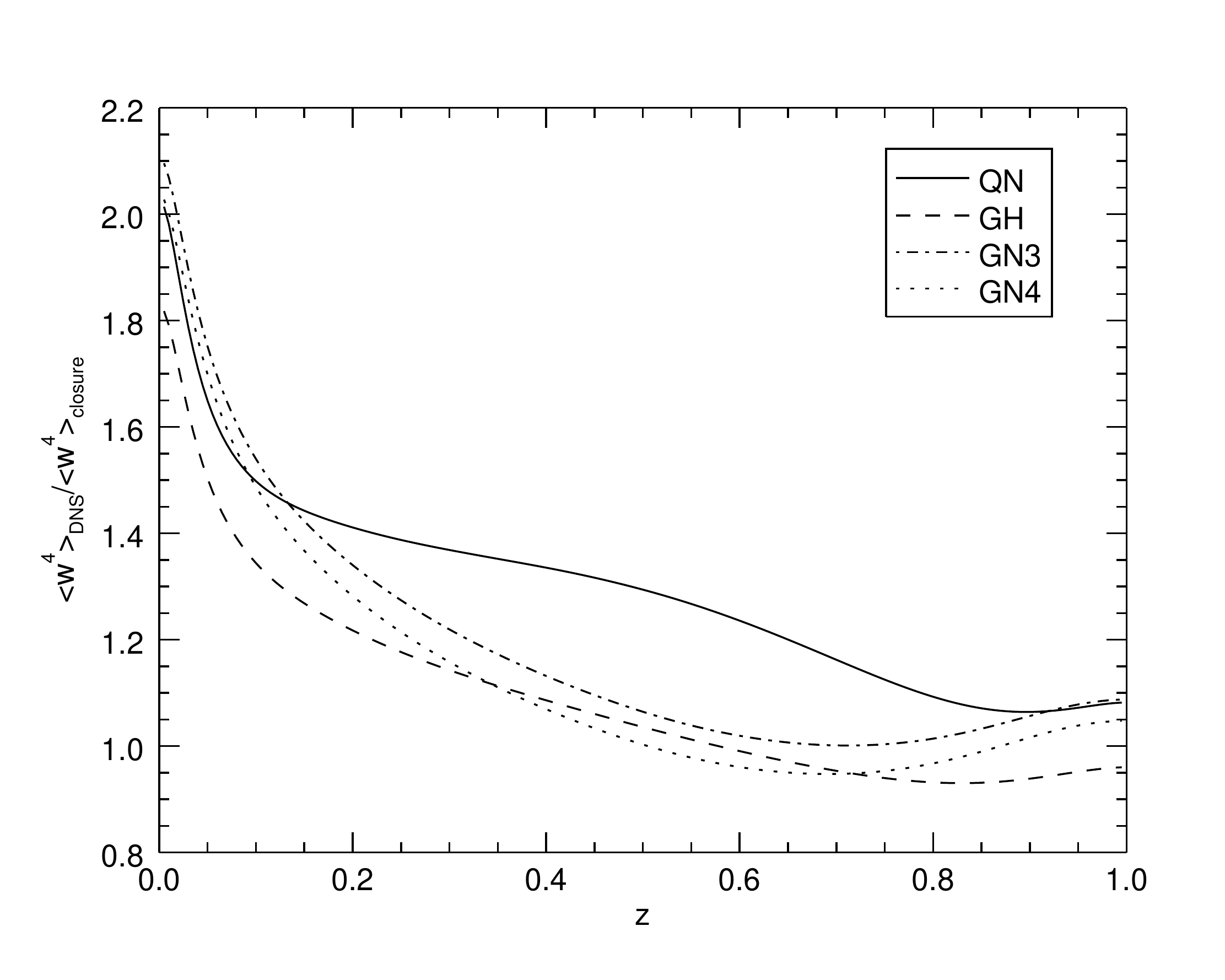}{0.5\textwidth}{(b)}
          }
\gridline{\fig{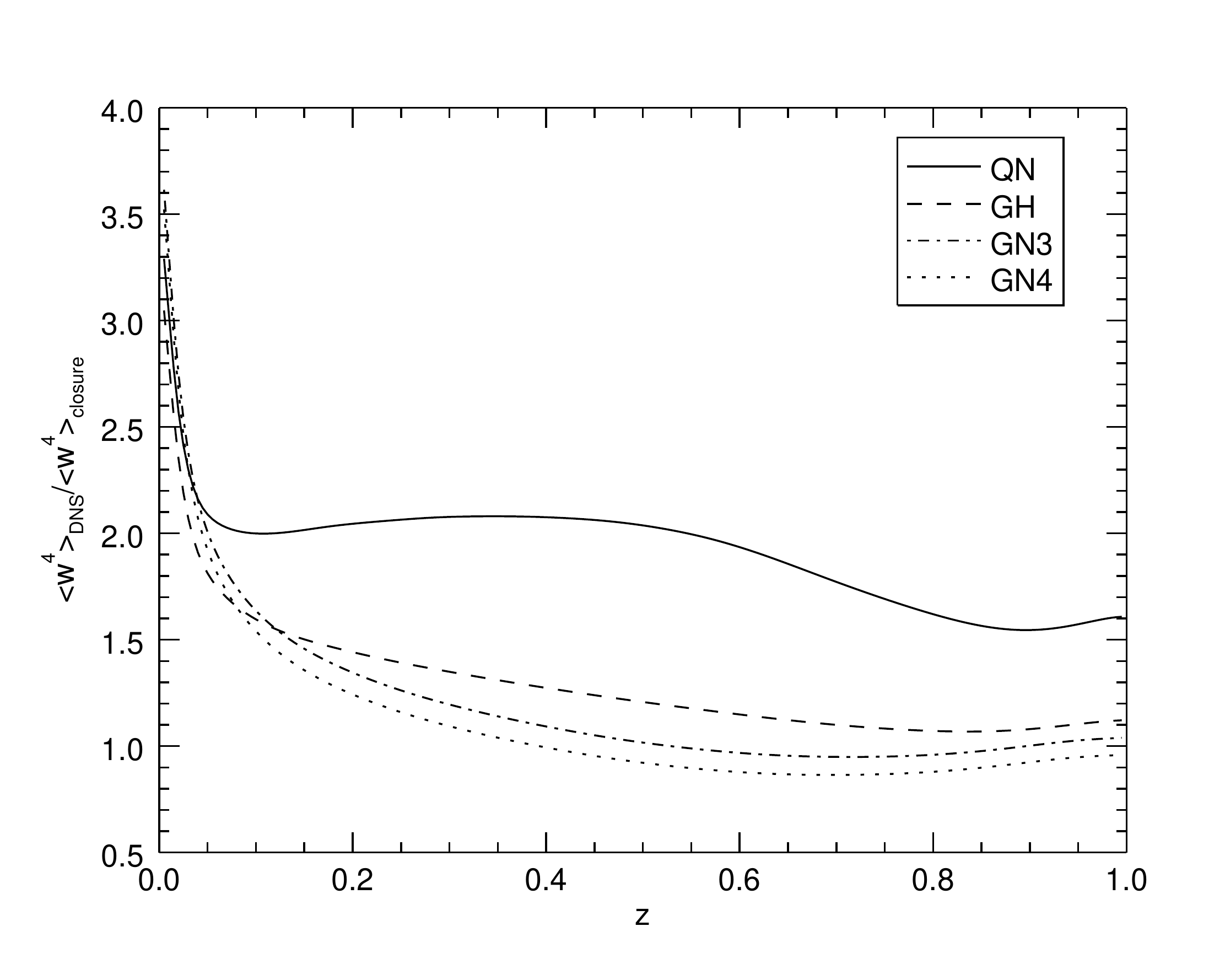}{0.5\textwidth}{(c)}
          \fig{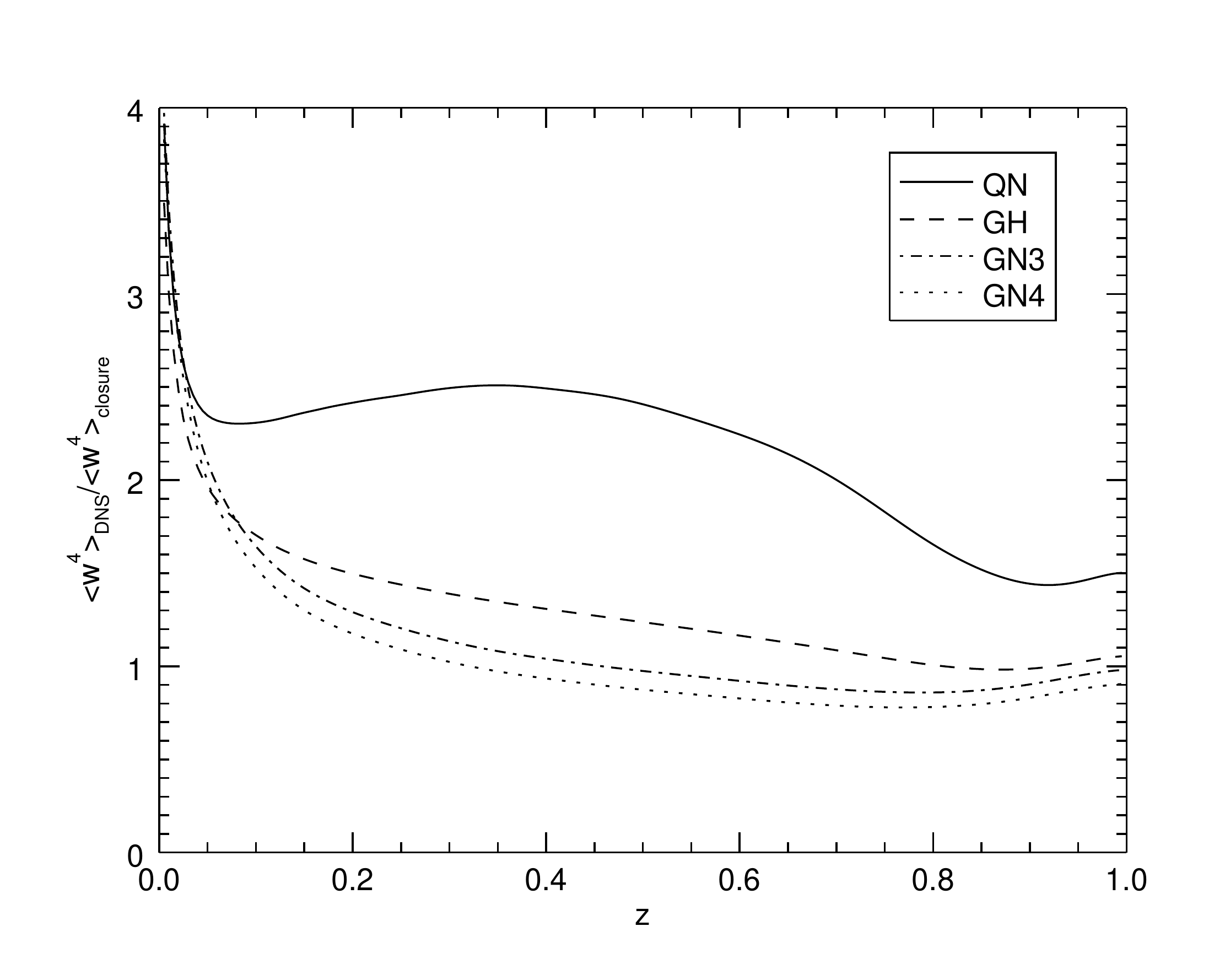}{0.5\textwidth}{(d)}
          }
\caption{The ratios of $\overline{w^4}$ from DNS to that of closure models. Panels (a)-(d) are the results of cases A1, B1, C1, and D1, respectively.\label{fig:w4}}
\end{figure*}

\begin{figure*}
\gridline{\fig{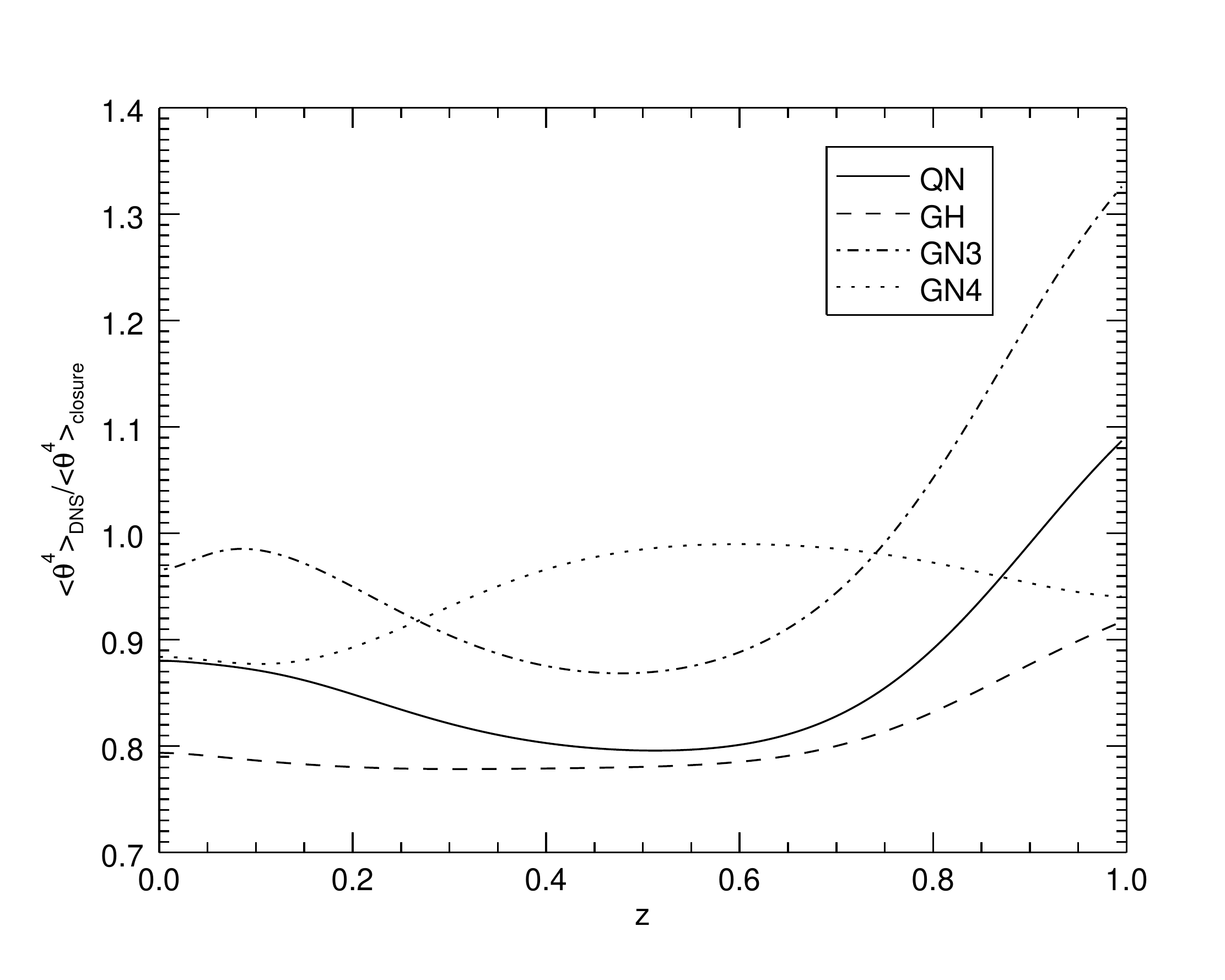}{0.5\textwidth}{(a)}
          \fig{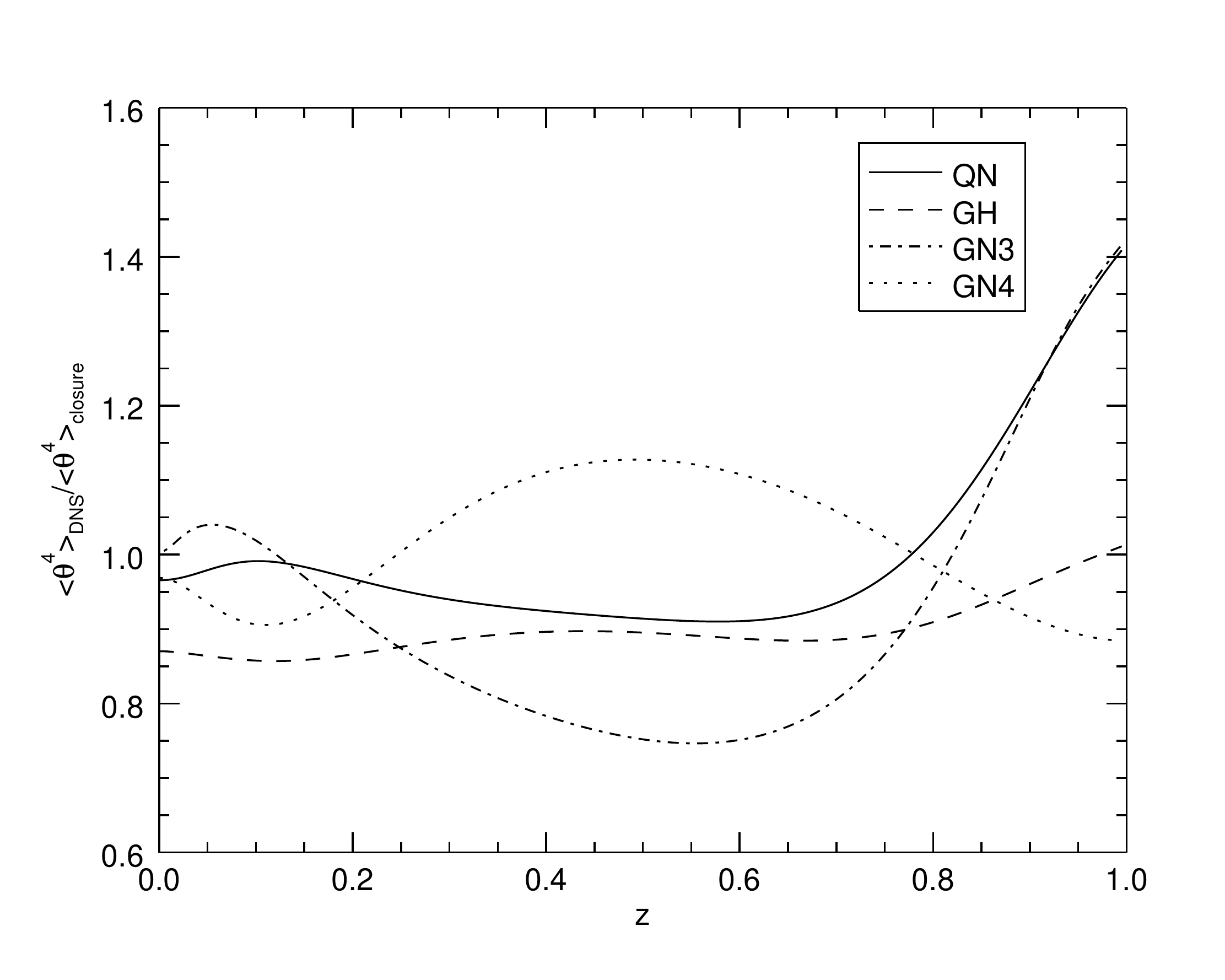}{0.5\textwidth}{(b)}
          }
\gridline{\fig{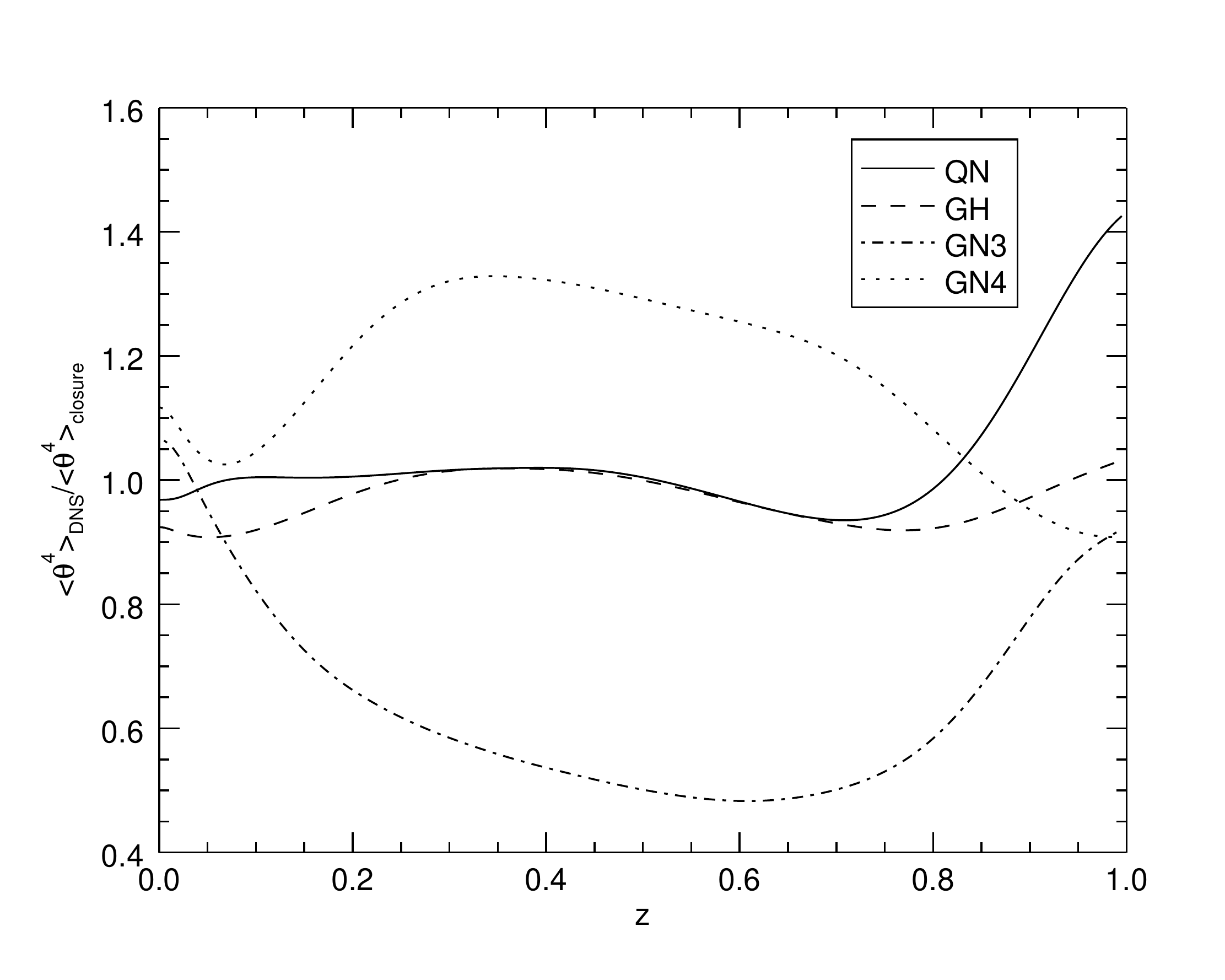}{0.5\textwidth}{(c)}
          \fig{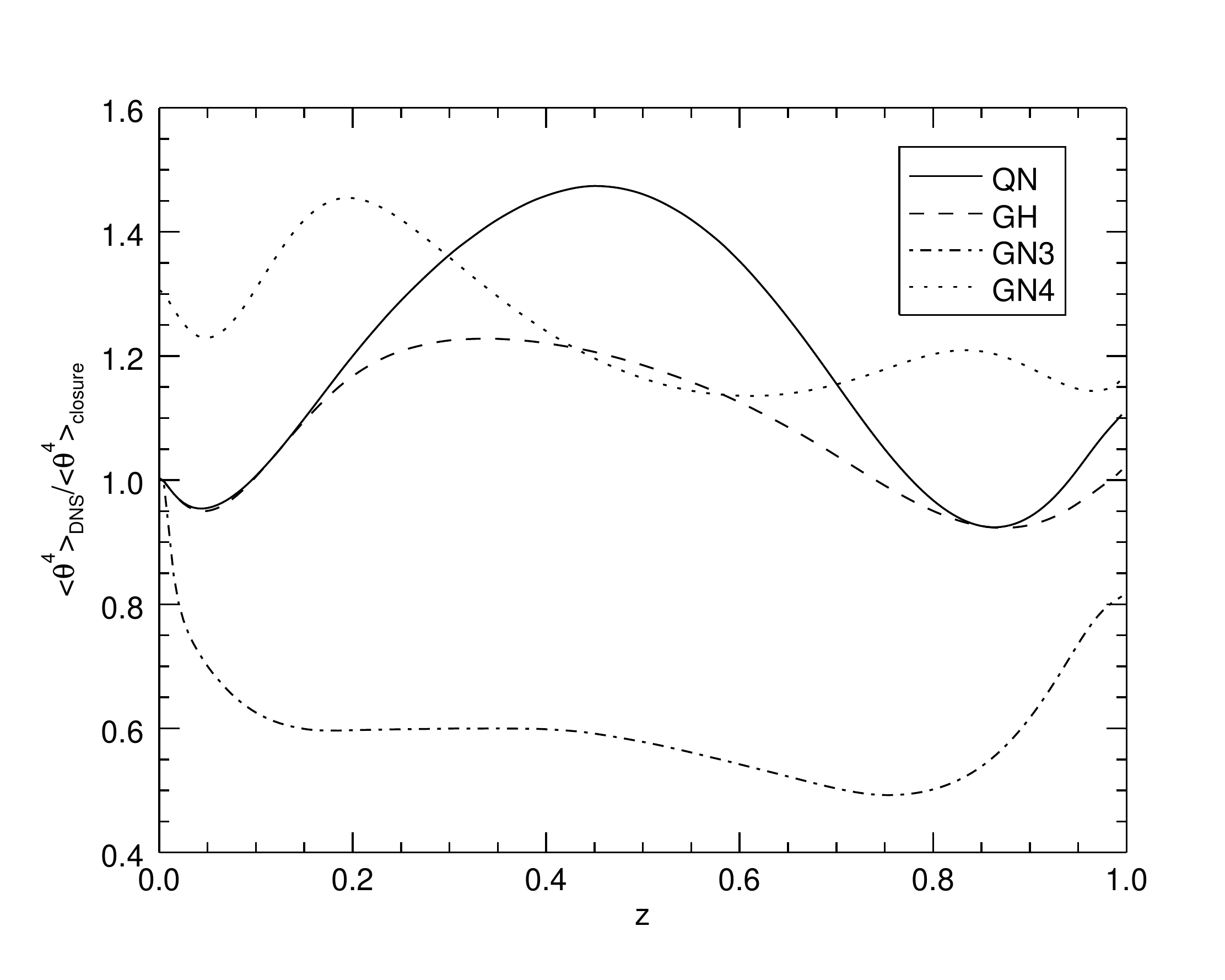}{0.5\textwidth}{(d)}
          }
\caption{The ratios of $\overline{\theta^4}$ from DNS to that of closure models. Panels (a)-(d) are the results of cases A1, B1, C1, and D1, respectively.\label{fig:t4}}
\end{figure*}

\begin{figure*}
\gridline{\fig{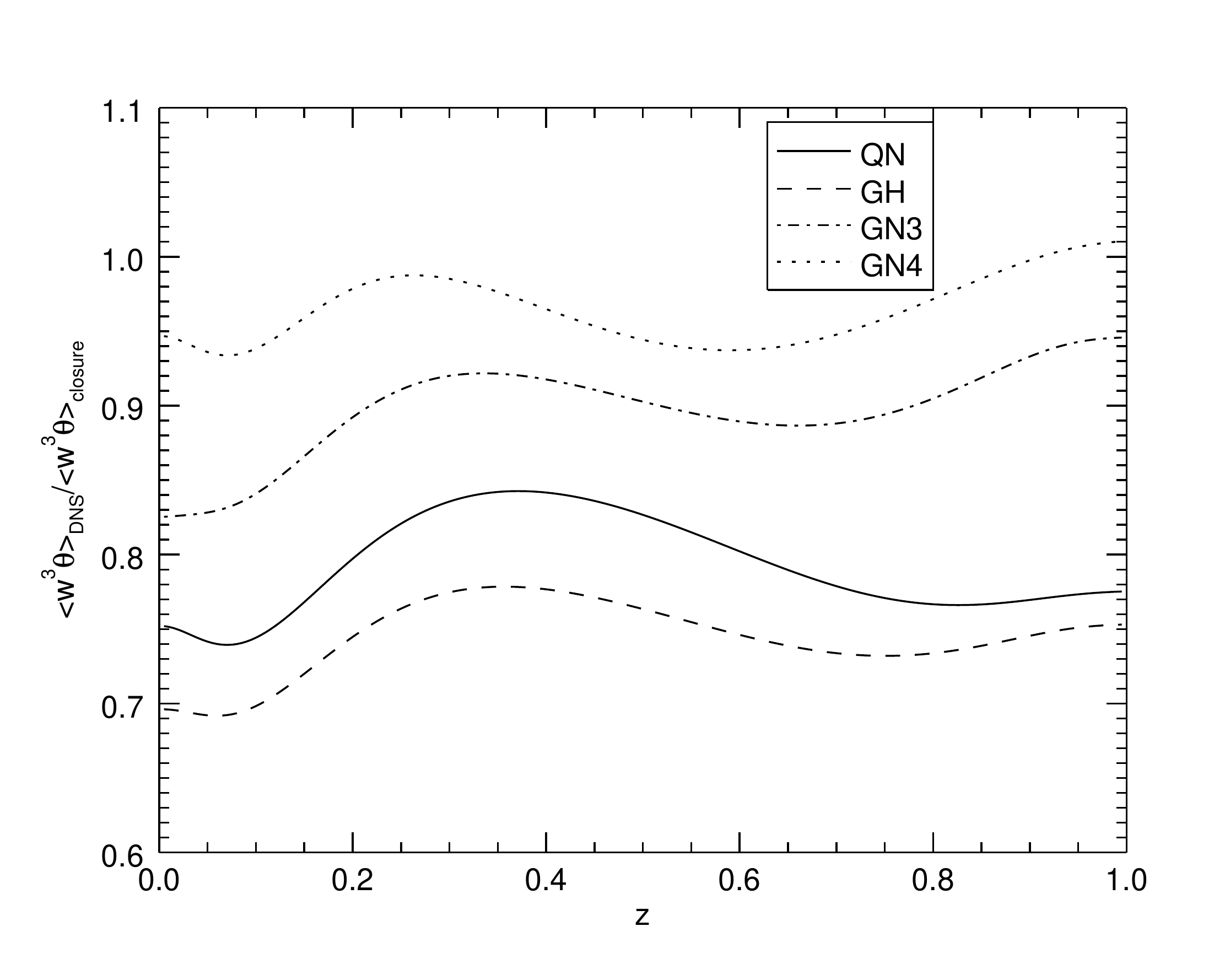}{0.5\textwidth}{(a)}
          \fig{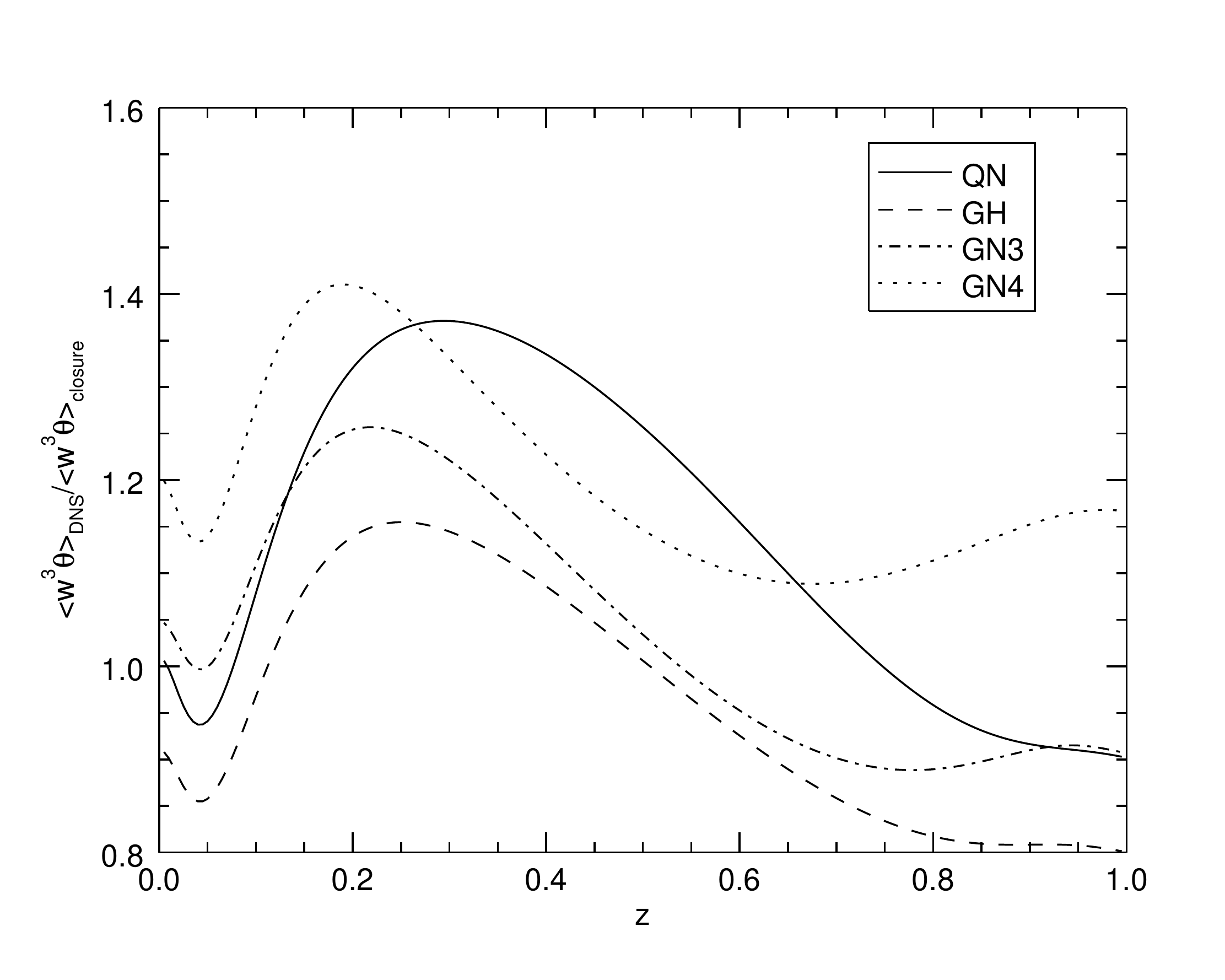}{0.5\textwidth}{(b)}
          }
\gridline{\fig{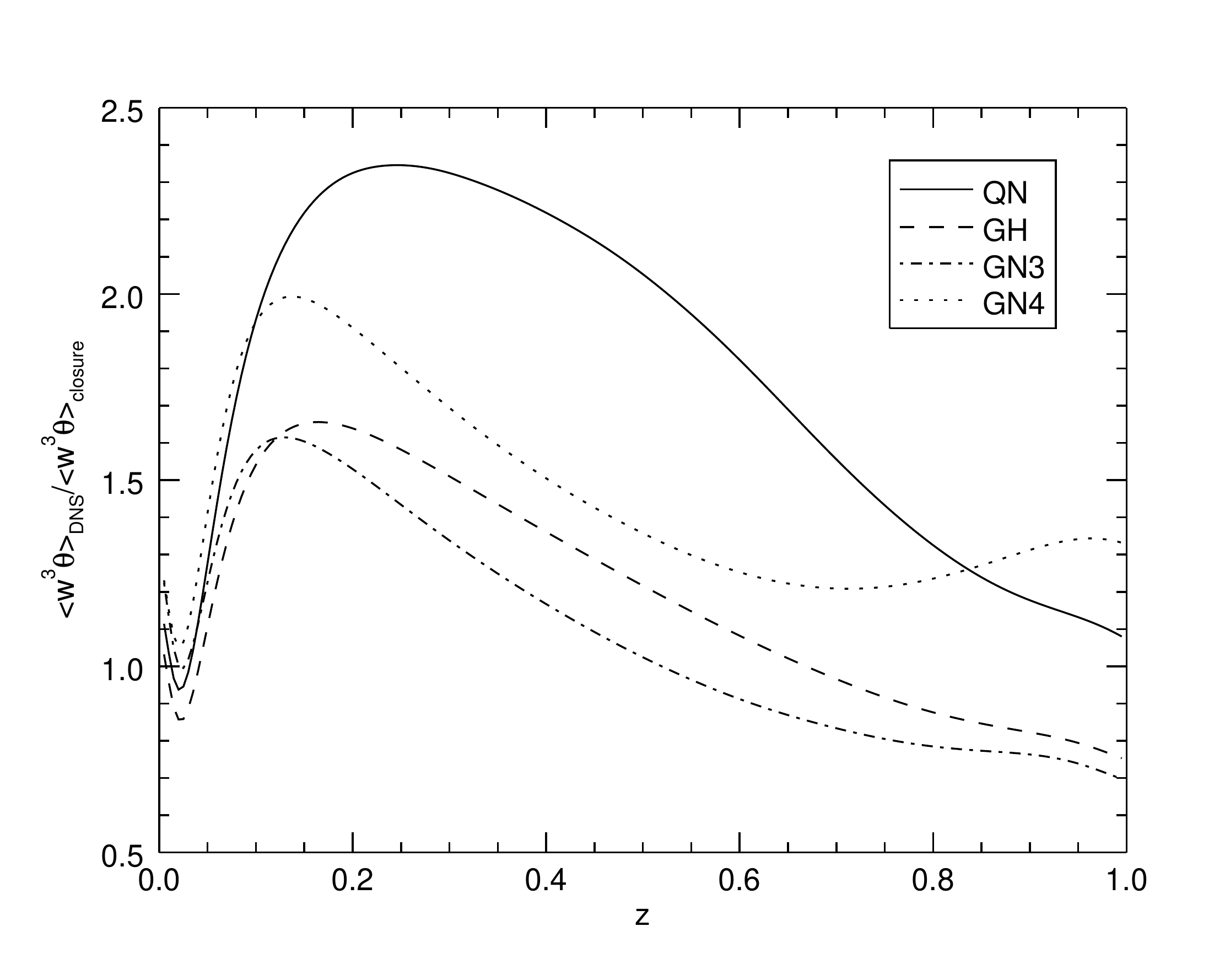}{0.5\textwidth}{(c)}
          \fig{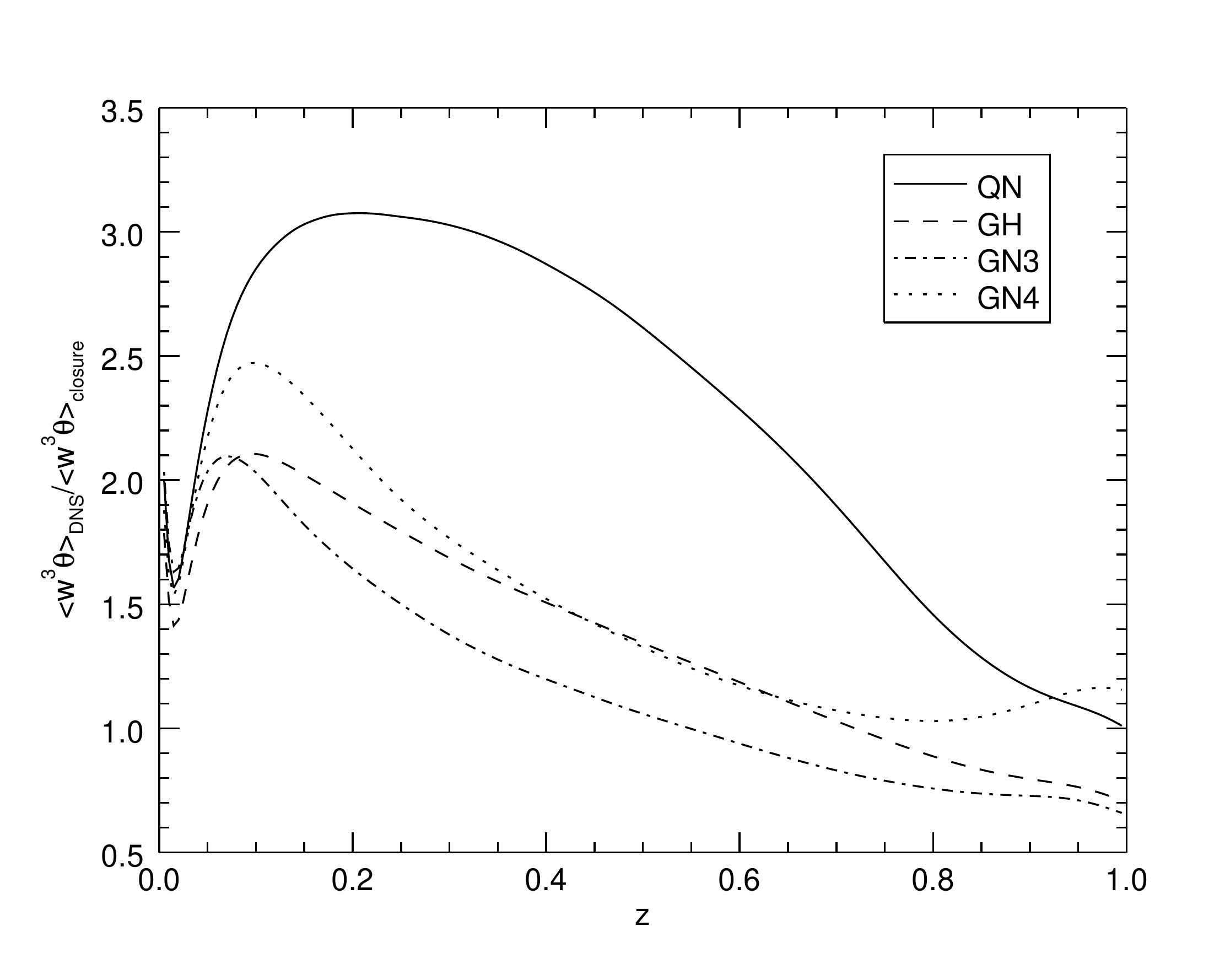}{0.5\textwidth}{(d)}
          }
\caption{The ratios of $\overline{w^3 \theta}$ from DNS to that of closure models. Panels (a)-(d) are the results of cases A1, B1, C1, and D1, respectively.\label{fig:w3t1}}
\end{figure*}

\begin{figure*}
\gridline{\fig{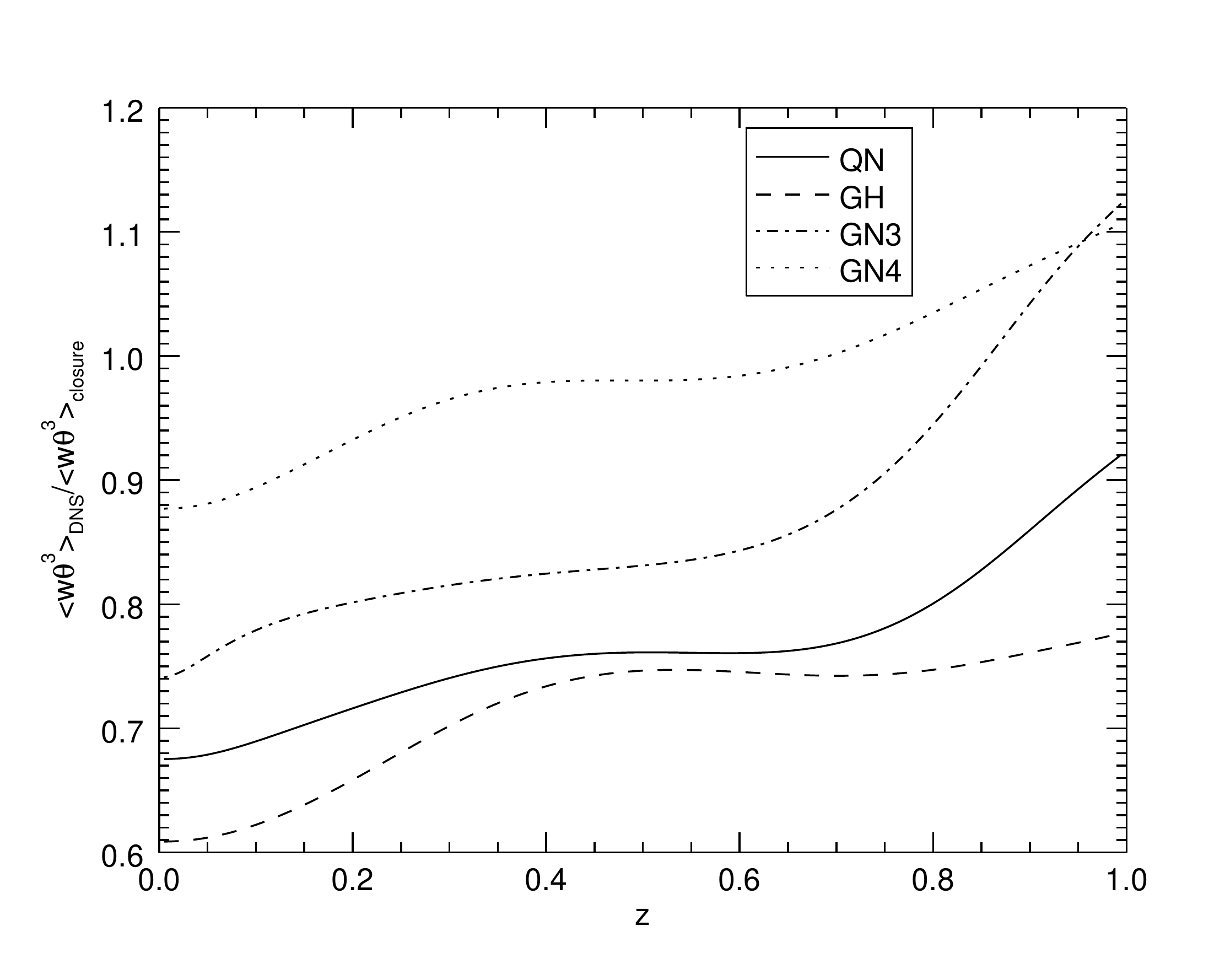}{0.5\textwidth}{(a)}
          \fig{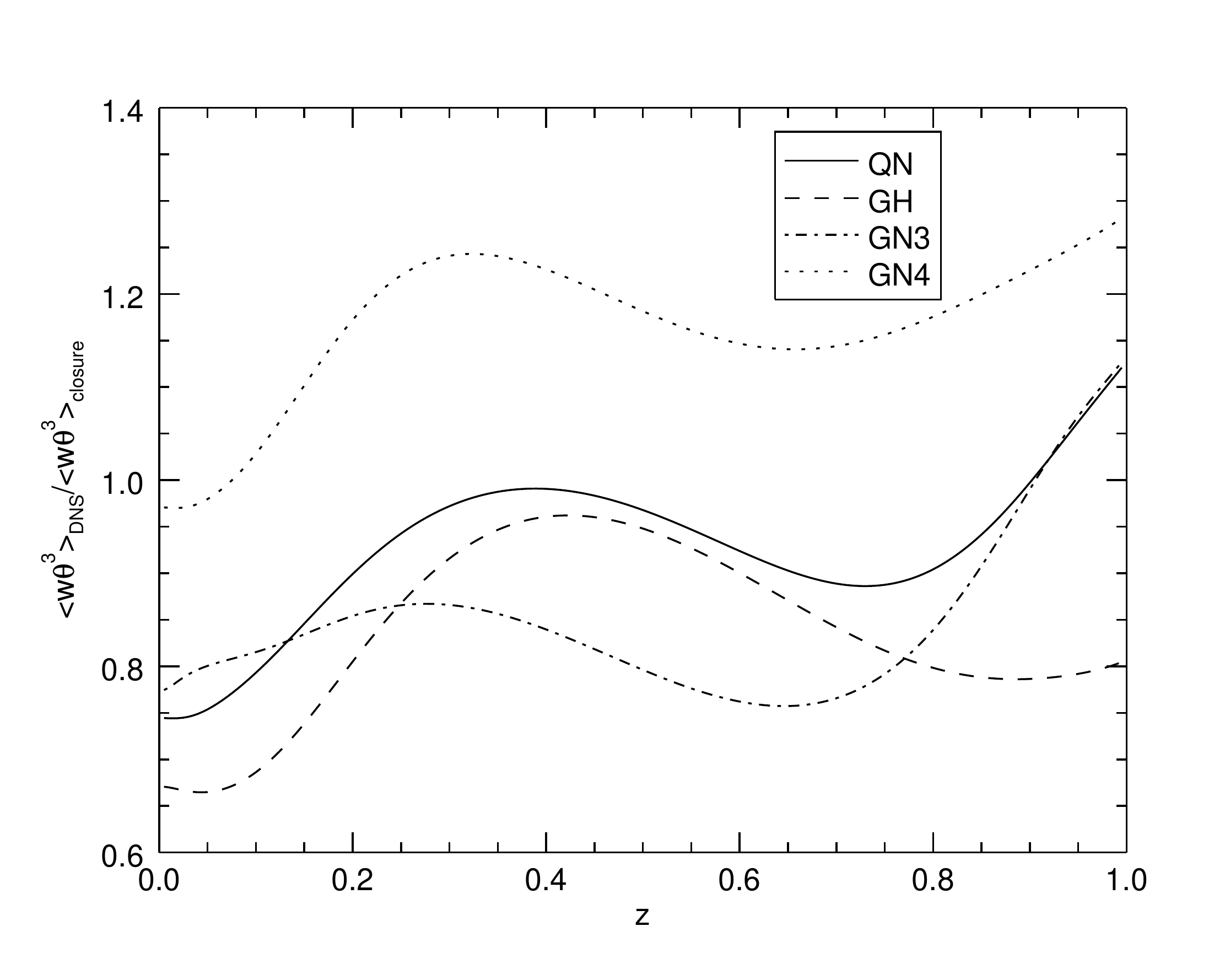}{0.5\textwidth}{(b)}
          }
\gridline{\fig{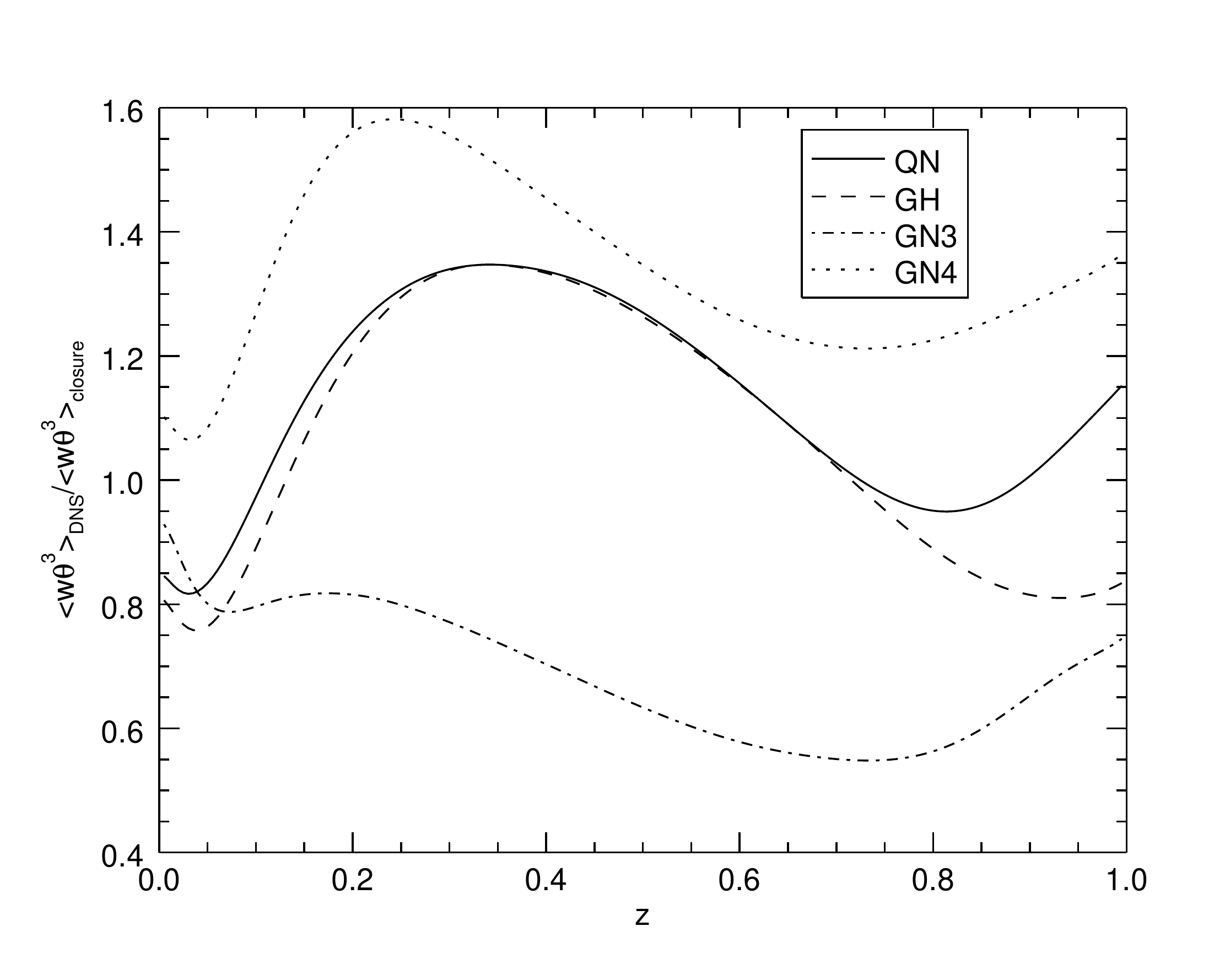}{0.5\textwidth}{(c)}
          \fig{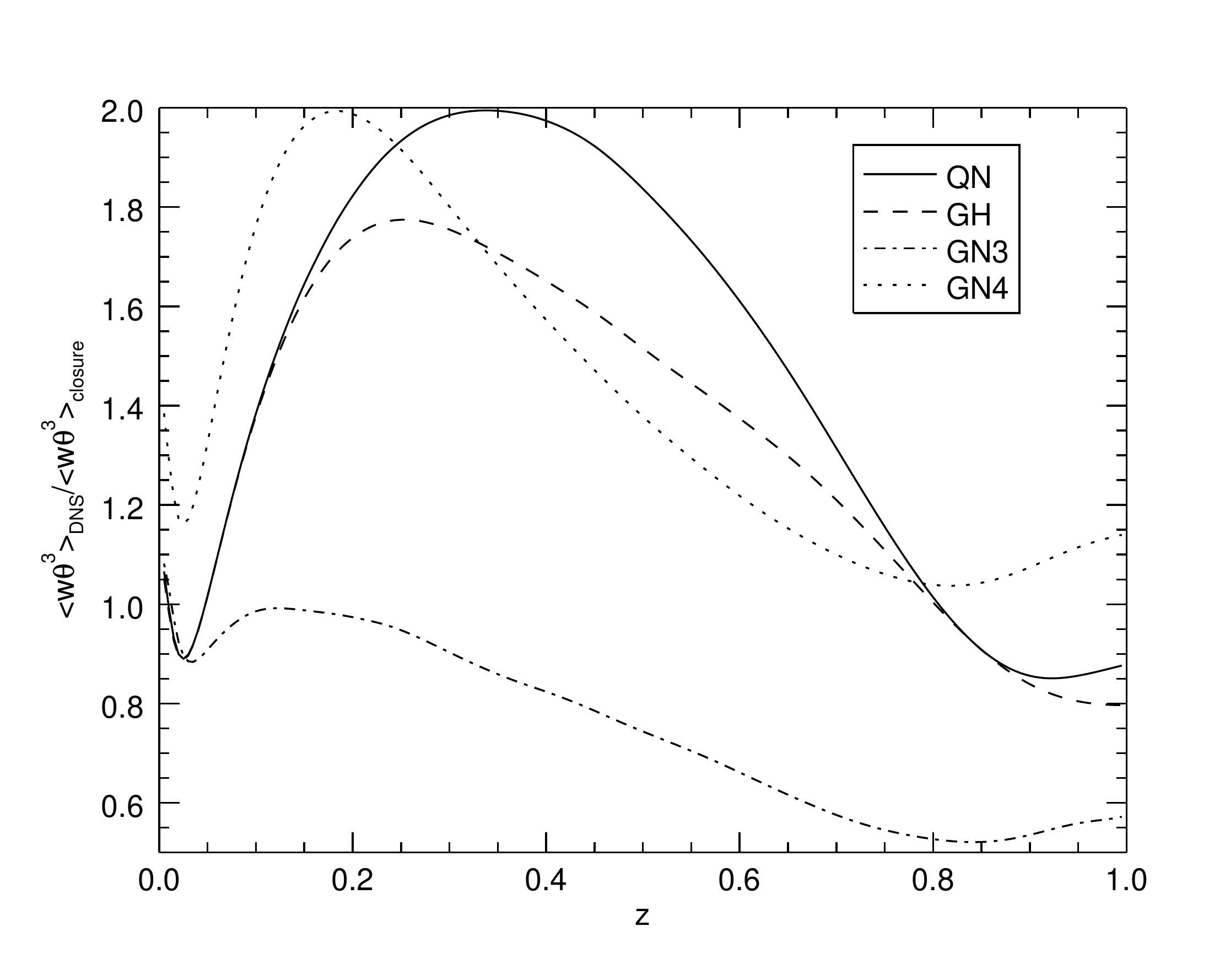}{0.5\textwidth}{(d)}
          }
\caption{The ratios of $\overline{w \theta^3}$ from DNS to that of closure models. Panels (a)-(d) are the results of cases A1, B1, C1, and D1, respectively.\label{fig:w1t3}}
\end{figure*}

\begin{figure*}
\gridline{\fig{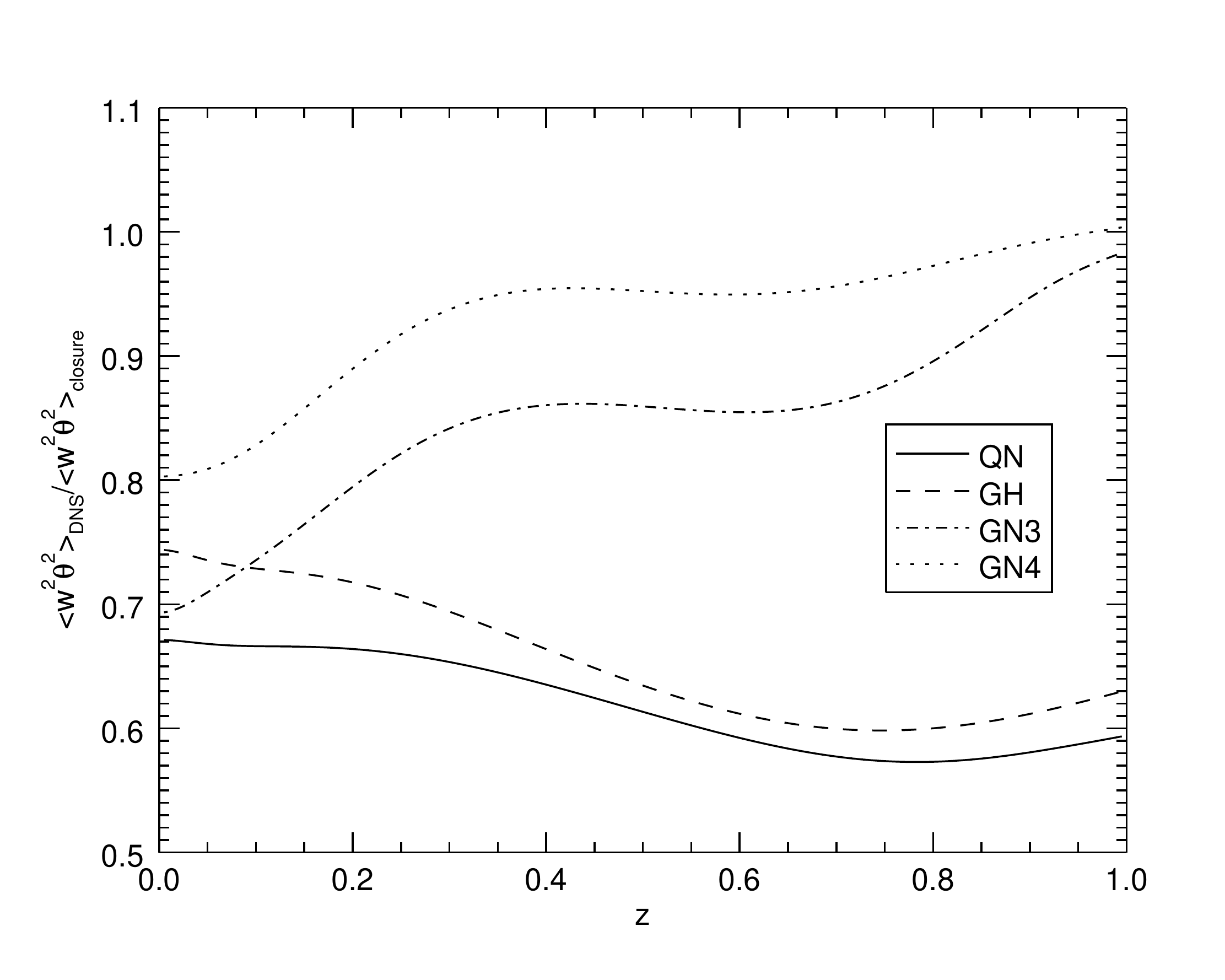}{0.5\textwidth}{(a)}
          \fig{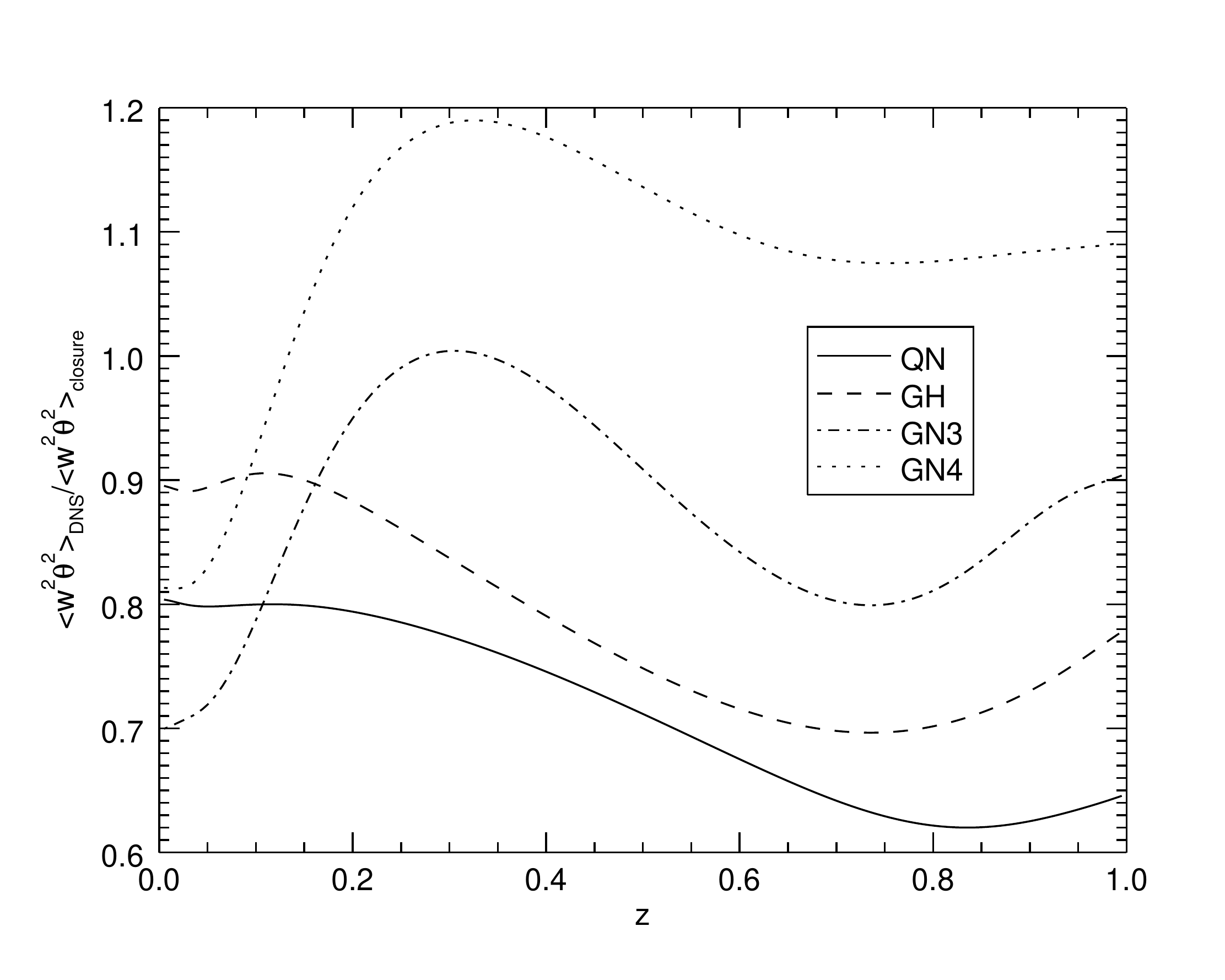}{0.5\textwidth}{(b)}
          }
\gridline{\fig{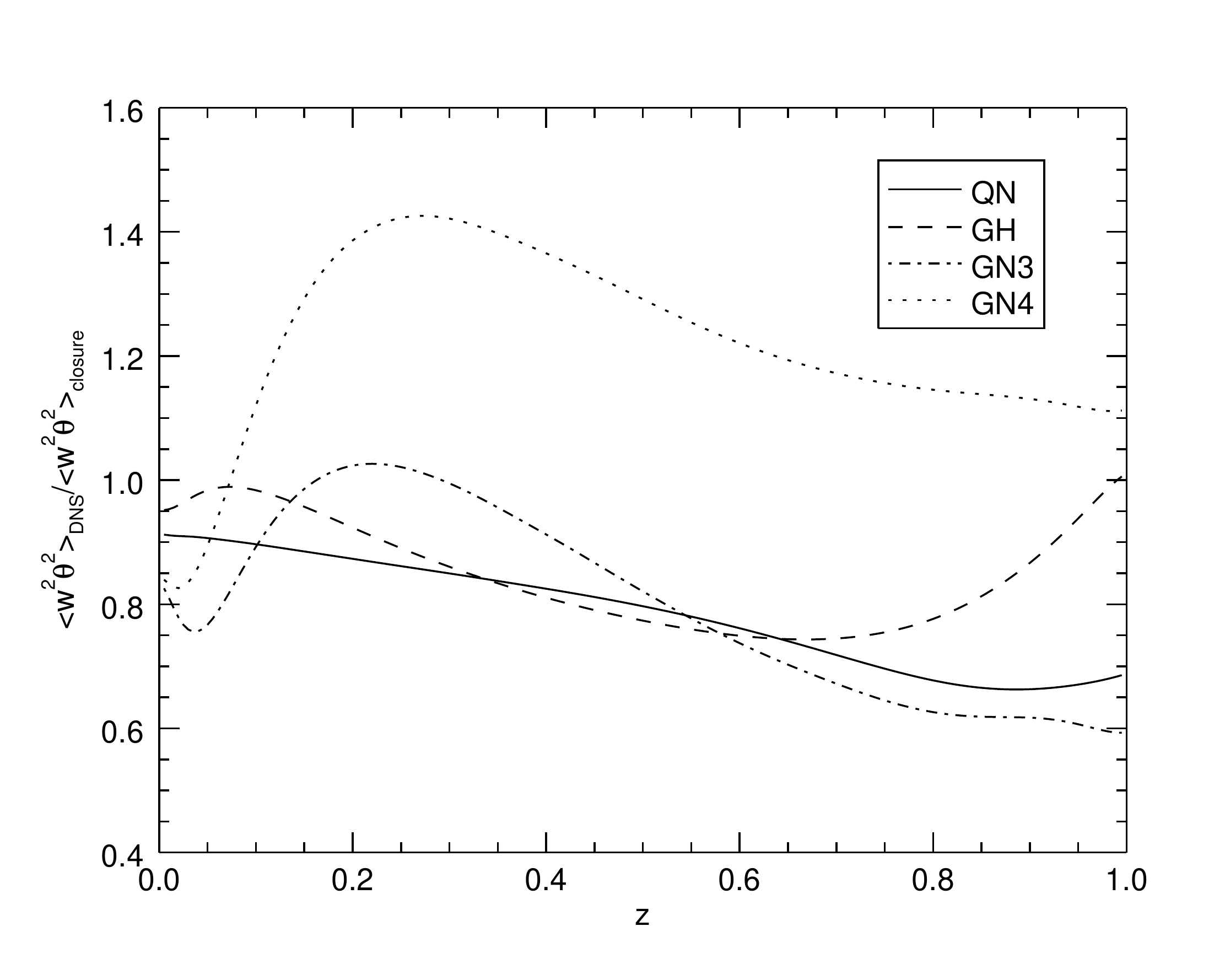}{0.5\textwidth}{(c)}
          \fig{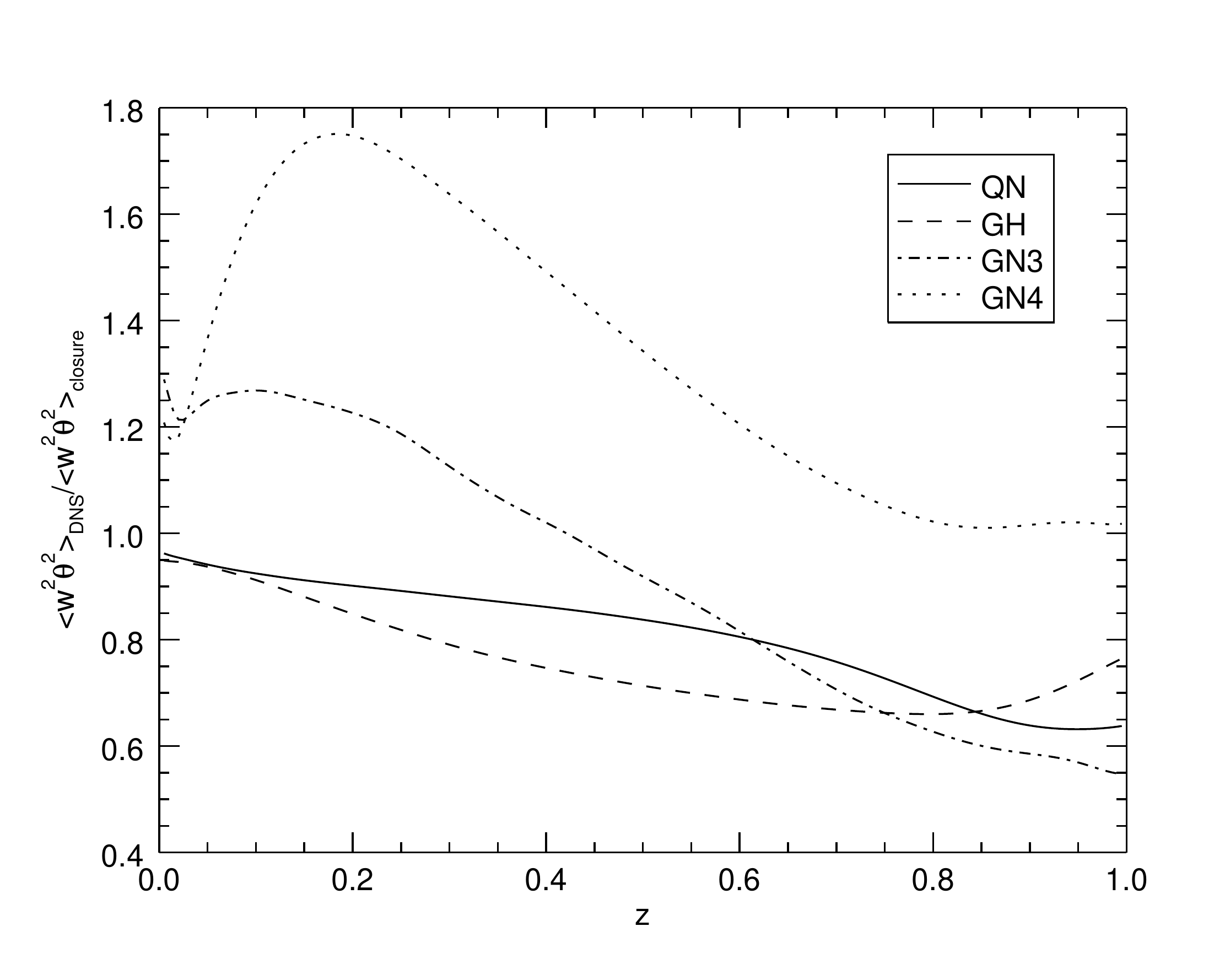}{0.5\textwidth}{(d)}
          }
\caption{The ratios of $\overline{w^2 \theta^2}$ from DNS to that of closure models. Panels (a)-(d) are the results of cases A1, B1, C1, and D1, respectively.\label{fig:w2t2}}
\end{figure*}

Now we compare the closure models with our simulation data. Figs.~\ref{fig:w4}-\ref{fig:w2t2} plot the ratios of the fourth-order moments computed from DNS to those predicted from different closure models for cases A1, B1, C1, and D1. The ratio is close to 1 if the closure model is perfect.
For the result of $\overline{w^4}$, quasi-normal approximation (QN) has good performance in the shallowest case A1, where the deviation of the ratio is less than 13 percent in most of the simulation domain. In the deep convection zone, however, QN has a very bad performance as the deviation reaches a factor up to 1.16 in case D1. GH has a better performance than QN except in the case A1, while the deviation still reaches 31 percent in the case D1. GN3 and GN4 performs good in all the cases of shallow and deep convection zones. The deviations are within 20 percents in most of the cases.
For the result of $\overline{\theta^4}$, GN3 has the worst performance in almost all the simulation cases. GN4 has better predictions in the cases of shallow convection zones, but the performance is worse than GH in the cases of deep convection zones.
For the result of $\overline{w^3\theta}$, GN3 has the best performance compared among all the closure models. The deviation of GN3 is below 10 percent in case A1 to 18 percent in case D1. QN overestimates the value of $\overline{w^3\theta}$ in case A1, and underestimates it in case D1. The deviation of QN reaches a factor of 1.27 in D1, thus quasi-normal approximation is bad for estimation of $\overline{w^3\theta}$. Compared to GN3, GN4 predicts a lower value of $\overline{w^3\theta}$. GH performs the second best in the deepest case D1, but the worst in the shallowest case A1.
For the result of $\overline{w\theta^3}$, GN4 has good performance in cases A1. QN, GH, and GN3 overestimate the values of $\overline{w\theta^3}$ in cases A1 and B1. For cases C1 and D1, none of these closure models makes satisfactory predictions. QN, GH, and GN4 underestimate, while GN3 overestimates $\overline{w\theta^3}$. The deviations are generally greater than 30 percent.
For the result of $\overline{w^2\theta^2}$, GN3 has the best performance among all the closure models. The deviation of GN3 is about 15 percent in case A1 and 18 percent in case D1. QN, GH, and GN3 overestimate the values of $\overline{w^2\theta^2}$, and GN4 underestimates them.

To better explain the difference between the closure models, we take a further step to compute the average values of the ratios of the fourth-order moments. To avoid unfavourable effect from the boundary condition, we calculate the statistical results in the simulation domain $z\in(0.4,0.7)$ (see appendix A for the discussion of boundary effects). Table~\ref{tab:table2} lists the average values for all the simulation cases and closure models. In the last four rows, we have also computed the averages and deviations among groups with shallow (groups A and B) and deep (groups C and D) convection zones. The deviation is computed by the root-mean-square values of the ratios away from 1. If the deviation is small, it means that the closure model is good and robust across different groups. As seen from the table, the QN model is the worst one among the four closure models, no matter which fourth-order moment to predict. It implies that the assumption on normal distributions of temperature perturbations or velocities are unsatisfactory. GN3 has good performances on prediction of $\overline{w^4}$ and $\overline{w^3\theta}$, but the performances on prediction of $\overline{\theta^4}$ and $\overline{w\theta^3}$ are bad. \citet{2007MNRAS..374...305} have drawn the same conclusion based on their numerical simulation data of the Sun and a K dwarf. The GN3 model is based on the assumptions underlying the simple mass-flux model, where deviations strictly correlate with up- and downflow. Both asymmetries on velocities and temperature perturbations are considered in the GH and GN4 models, while only asymmetry on velocities is considered in the GN3 model. The numerical results indicate that both asymmetries on velocities and temperature perturbations are important\citep{2007MNRAS..374...305}. As a result, GH and GN4 outperform GN3 as shown in the statistical results. The difference of the performance on prediction between GH and GN4 is very small. The coefficients of GN4 in Eq.(\ref{eq:GN4s}-\ref{eq:GN4e}) are calibrated from numerical simulations of turbulent motions with a shallow convection zone \citep{1993ApJS..89...361}. Our result has verified its predictive power in the cases with shallow convection zones (group A), since all the ratios of the fourth-order moments are close to 1. As the depth of the convection zone increases, GN4 underestimates $\overline{\theta^4}$, $\overline{w^3\theta}$, $\overline{w\theta^3}$, and $\overline{w^2\theta^2}$.

\begin{longrotatetable}
\begin{deluxetable}{c|cccc|cccc|cccc|cccc|cccc}
\tablewidth{700pt}
\tabletypesize{\scriptsize}
\tablecaption{The averaged values of the ratios of fourth-order moments computed from DNS to different closure formulas. \label{tab:table2}}
\tablehead{
    \multicolumn{1}{c}{}  & \multicolumn{4}{c}{$\overline{w^4}_{DNS}/\overline{w^4}_{closure}$} & \multicolumn{4}{c}{$\overline{\theta^4}_{DNS}/\overline{\theta^4}_{closure}$} & \multicolumn{4}{c}{$\overline{w^3\theta}_{DNS}/\overline{w^3\theta}_{closure}$} & \multicolumn{4}{c}{$\overline{w\theta^3}_{DNS}/\overline{w\theta^3}_{closure}$} & \multicolumn{4}{c}{$\overline{w^2\theta^2}_{DNS}/\overline{w^2\theta^2}_{closure}$} \\
 Case & QN & GH & GN3 & GN4 & QN & GH & GN3 & GN4 & QN & GH & GN3 & GN4 & QN & GH & GN3 & GN4 & QN & GH & GN3 & GN4
 }
\startdata
 A1 & 0.92 & 0.86 & 1.04 & 1.02 & 0.86 & 0.80 & 0.98 & 0.95 & 0.79 & 0.74 & 0.90 & 0.96 & 0.76 & 0.72 & 0.87 & 0.98 & 0.62 & 0.65 & 0.85 & 0.93 \\
 A2 & 1.08 & 0.99 & 1.17 & 1.13 & 1.00 & 0.88 & 1.09 & 0.95 & 0.94 & 0.86 & 1.01 & 1.08 & 0.88 & 0.77 & 0.95 & 1.12 & 0.68 & 0.74 & 0.92 & 1.02 \\
 A3 & 1.18 & 1.05 & 1.20 & 1.16 & 0.95 & 0.87 & 0.97 & 1.00 & 1.05 & 0.93 & 1.06 & 1.15 & 0.91 & 0.84 & 0.92 & 1.15 & 0.71 & 0.76 & 0.95 & 1.09 \\
 A4 & 0.99 & 0.94 & 1.14 & 1.12 & 0.96 & 0.86 & 1.11 & 0.96 & 0.85 & 0.81 & 0.98 & 1.03 & 0.81 & 0.73 & 0.94 & 1.04 & 0.66 & 0.70 & 0.92 & 0.99 \\
 A5 & 1.07 & 1.01 & 1.23 & 1.21 & 0.86 & 0.79 & 0.99 & 0.92 & 0.91 & 0.86 & 1.04 & 1.10 & 0.81 & 0.75 & 0.93 & 1.03 & 0.69 & 0.72 & 0.97 & 1.04 \\
 B1 & 1.29 & 1.10 & 1.19 & 1.14 & 1.00 & 0.90 & 0.92 & 1.01 & 1.14 & 0.96 & 1.04 & 1.20 & 0.93 & 0.84 & 0.85 & 1.17 & 0.71 & 0.78 & 0.88 & 1.08 \\
 B2 & 1.48 & 1.17 & 1.20 & 1.13 & 0.99 & 0.92 & 0.81 & 1.10 & 1.36 & 1.06 & 1.08 & 1.29 & 1.01 & 0.95 & 0.80 & 1.24 & 0.75 & 0.80 & 0.88 & 1.15 \\
 B3 & 1.67 & 1.21 & 1.18 & 1.10 & 1.00 & 0.97 & 0.72 & 1.20 & 1.62 & 1.17 & 1.12 & 1.37 & 1.11 & 1.08 & 0.77 & 1.30 & 0.77 & 0.79 & 0.89 & 1.21 \\
 B4 & 1.40 & 1.17 & 1.25 & 1.19 & 1.04 & 0.92 & 0.94 & 1.02 & 1.23 & 1.02 & 1.09 & 1.27 & 0.96 & 0.85 & 0.85 & 1.21 & 0.74 & 0.82 & 0.89 & 1.11 \\
 B5 & 1.46 & 1.19 & 1.26 & 1.20 & 1.01 & 0.91 & 0.88 & 1.03 & 1.32 & 1.07 & 1.12 & 1.34 & 0.95 & 0.86 & 0.82 & 1.20 & 0.75 & 0.82 & 0.89 & 1.14 \\
 C1 & 1.92 & 1.30 & 1.20 & 1.11 & 1.03 & 0.97 & 0.64 & 1.17 & 1.78 & 1.18 & 1.08 & 1.46 & 1.13 & 1.07 & 0.69 & 1.34 & 0.79 & 0.85 & 0.80 & 1.22 \\
 C2 & 2.11 & 1.30 & 1.15 & 1.06 & 1.08 & 1.03 & 0.59 & 1.24 & 2.10 & 1.26 & 1.10 & 1.45 & 1.30 & 1.24 & 0.68 & 1.36 & 0.80 & 0.79 & 0.82 & 1.24 \\
 C3 & 2.04 & 1.30 & 1.16 & 1.07 & 1.31 & 1.13 & 0.72 & 1.23 & 2.21 & 1.39 & 1.23 & 1.47 & 1.63 & 1.40 & 0.88 & 1.46 & 0.81 & 0.74 & 1.03 & 1.37 \\
 C4 & 2.00 & 1.30 & 1.19 & 1.10 & 1.00 & 0.96 & 0.59 & 1.18 & 1.94 & 1.25 & 1.12 & 1.47 & 1.16 & 1.12 & 0.67 & 1.31 & 0.80 & 0.83 & 0.82 & 1.23 \\
 C5 & 1.98 & 1.31 & 1.20 & 1.11 & 1.01 & 0.96 & 0.62 & 1.19 & 1.92 & 1.26 & 1.15 & 1.54 & 1.16 & 1.12 & 0.69 & 1.33 & 0.80 & 0.82 & 0.83 & 1.26 \\
 D1 & 2.16 & 1.31 & 1.15 & 1.06 & 1.13 & 1.03 & 0.59 & 1.18 & 2.27 & 1.36 & 1.18 & 1.50 & 1.47 & 1.33 & 0.75 & 1.41 & 0.81 & 0.76 & 0.92 & 1.33 \\
 D2 & 2.27 & 1.30 & 1.10 & 1.00 & 1.35 & 1.14 & 0.64 & 1.22 & 2.47 & 1.40 & 1.18 & 1.55 & 1.74 & 1.47 & 0.81 & 1.50 & 0.82 & 0.74 & 0.94 & 1.39 \\
 D3 & 2.28 & 1.30 & 1.10 & 1.00 & 1.54 & 1.20 & 0.71 & 1.20 & 2.52 & 1.42 & 1.20 & 1.54 & 1.93 & 1.50 & 0.88 & 1.53 & 0.83 & 0.72 & 1.00 & 1.42 \\
 D4 & 2.14 & 1.30 & 1.13 & 1.03 & 1.19 & 1.05 & 0.61 & 1.16 & 2.26 & 1.36 & 1.17 & 1.55 & 1.52 & 1.34 & 0.77 & 1.43 & 0.81 & 0.74 & 0.93 & 1.37 \\
 D5 & 2.22 & 1.31 & 1.13 & 1.03 & 1.19 & 1.08 & 0.60 & 1.24 & 2.33 & 1.37 & 1.17 & 1.59 & 1.51 & 1.37 & 0.74 & 1.42 & 0.82 & 0.76 & 0.89 & 1.35 \\
 \hline
avg AB & 1.25 & 1.07 & 1.19 & 1.14 & 0.97 & 0.88 & 0.94 & 1.01 & 1.12 & 0.95 & 1.04 & 1.18 & 0.91 & 0.84 & 0.87 & 1.14 & 0.71 & 0.76 & 0.90 & 1.08\\
avg CD & 2.11 & 1.30 & 1.05 & 0.96 & 1.18 & 1.06 & 0.63 & 1.20 & 2.18 & 1.33 & 1.16 & 1.51 & 1.46 & 1.30 & 0.76 & 1.41 & 0.81 & 0.78 & 0.90 & 1.32\\
dev AB & 0.34 & 0.13 & 0.20 & 0.15 & 0.07 & 0.13 & 0.13 & 0.08 & 0.28 & 0.14 & 0.08 & 0.22 & 0.13 & 0.19 & 0.14 & 0.17 & 0.30 & 0.25 & 0.10 & 0.11 \\
dev CD & 1.12 & 0.30 & 0.16 & 0.07 & 0.25 & 0.10 & 0.37 & 0.20 & 1.20 & 0.33 & 0.16 & 0.51 & 0.52 & 0.33 & 0.26 & 0.42 & 0.19 & 0.23 & 0.13 & 0.33\\
\enddata
\tablecomments{The averaged values are taken in the range $z\in(0.4,0.7)$. The last four rows show the averages and deviations of the ratios in the cases with shallow (groups A and B) and deep (groups C and D) convection zones. The deviations are evaluated by the root-mean-square values of the ratios away from 1.}
\end{deluxetable}
\end{longrotatetable}

\section{The 1D Nonlocal Model} \label{sec:model}
\subsection{Dynamic equations for second-order moments}
Based on the statistical turbulent theory, \citet{1978CAA..2...118,1989AA..209...126} has developed an isotropic nonlocal Reynolds stress model to study the convection in the interiors of stars. Several assumptions are made in this model. First, the fluid is assumed to be anelastic and the high frequency acoustic waves are filtered out. Second, fluid motions have no significant difference in vertical and horizontal directions. Third, downgradient approximations are used for the third-order moments in deducing the diffusive terms. Under these assumptions, dynamic equations for the second-order moments of hydrodynamic variables are derived from the compressible hydrodynamic equations. \citet{1992ApJ..392...218,1998ApJ..493...834} have argued that the turbulent model could be improved in several aspects. First, the turbulent fluid motion shall be considered as anisotropic instead of isotropic. Second, the dynamic equations could be extended to third-order moments so that fourth-order moments can be easily expressed as functions of lower-order moments through quasi-normal approximations. Third, it is better to adopt a nonlocal model of the dissipation rate $\epsilon$. Taking these into consideration, \citet{1997ApJS..108...529} has improved his model including the dynamic equations of third-order moments and anisotropic treatment of turbulent flows. The dissipation rate, however, is still treated locally for the sake of simplicity. For the convenience of comparing results with three-dimensional numerical simulations in rectangular regions, we rewrite the dynamic equations of Xiong's model in the Cartesian coordinates :
\begin{eqnarray}
\frac{\partial \overline{u^2}}{\partial t}+D_{f}(\overline{u^2})-\frac{2}{3}{\beta}g (\overline{w\frac{\theta}{T}}) +\frac{4\eta_{e}}{\sqrt{3}}\frac{\rho g}{c_{1,w^2}P}(\overline{u^{2}})^{3/2}&=&0~,\\
\frac{\partial (\overline{\frac{\theta^2}{T^2}})}{\partial t}+D_{f}(\overline{\frac{\theta^2}{T^2}}) +2 (\frac{\partial \ln T}{\partial z}-\nabla_{ad}\frac{ \partial \ln P}{\partial z})(\overline{w\frac{\theta}{T}})+2\sqrt{3}\eta_{e}\frac{\rho g}{c_{1,\theta^2}P}[(\overline{u^2})^{1/2}+u_{c}](\overline{\frac{\theta^2}{T^2}})&=&0~,\\
\frac{\partial (\overline{w\frac{\theta}{T}})}{\partial t}+D_{f}(\overline{w\frac{\theta}{T}})+(\frac{\partial \ln T}{\partial z}-\nabla_{ad}\frac{\partial \ln P}{\partial z})(\overline{w^2})-{\beta}g (\overline{\frac{\theta^2}{T^2}})+\sqrt{3}\eta_{e}\frac{\rho g}{c_{1,w\theta}P}[3(\overline{u^2})^{1/2}+u_{c}](\frac{\overline{u^2}}{\overline{w^2}})^{1/2}(\overline{w\frac{\theta}{T}})&=&0~,\\
\frac{\partial \overline{w^2}}{\partial t}+D_{f}(\overline{w^2})-2{\beta}g (\overline{w\frac{\theta}{T}}) +\frac{4\eta_{e}}{\sqrt{3}}\frac{\rho g}{c_{1,w^2}P}(\overline{u^2})^{1/2}[(1+c_{3})\overline{w^2}-c_{3}\overline{u^2}]&=&0~,
\end{eqnarray}
where the symbol overline represents the temporal and horizontal average of the quantity;
$H_{p}$ is the pressure scale height; $P$ is the pressure; $\rho$ is the density; $\beta$ is the expansion coefficient of gas; $\eta_{e}$ is the Heisenberg eddy coupling constant; $\nabla_{ad}$ is the adiabatic temperature gradient; $u_c=\frac{9}{4}\frac{\kappa_{T}}{c_{1,w\theta}\rho c_{p}H_{p}}$ is a variable associated with temperature, pressure, and conductivity in the fluid; $\kappa_{T}$ is the conductivity; $\overline{u^2}$ is the isotropic part of the Reynolds stress, and $(\overline{u^2})^{1/2}/u_{c}$ represents the effective P$\acute{e}$clet number which is approximately proportional to the ratio of energy carried by turbulent convection with respect to radiation; $c_{1,w^2},c_{1,\theta^2},c_{1,w\theta}$ are three coefficients which determine the decay length scale of the convective motions with respect to the pressure scale height; $c_{3}$ determines the isotropic level of the fluid motions; $D_{f}$ represents the diffusive terms in the dynamic equations.
\begin{eqnarray}
D_{f}(\overline{u^2})&=&\frac{1}{\rho}\frac{\partial}{\partial z}(\rho\overline{wu^2})~,\\
D_{f}(\overline{\frac{\theta^2}{T^2}})&=& \frac{1}{\rho}\frac{\partial}{\partial z}(\rho\overline{w\frac{{\theta^2}}{T^2}})~,\\
D_{f}(\overline{w\frac{\theta}{T}})&=& \frac{1}{\rho}\frac{\partial}{\partial z}(\rho\overline{w^2\frac{{\theta}}{T}})~,\\
D_{f}(\overline{w^2})&=&\frac{1}{\rho}\frac{\partial}{\partial z}(\rho\overline{w^3})~.
\end{eqnarray}
Thus, the dynamic equations consist of four equations of the second-order moments of hydrodynamic variables, and the nonlocality is represented by the diffusive terms.

\subsection{Local steady solution}
When the local time derivative terms and diffusive terms are removed from the l.h.s. of the above equations, the turbulent model degenerates to
the local mixing length model. Obviously, $\overline{u^{2}}=\overline{\frac{\theta^2}{T^2}}=\overline{w\frac{\theta}{T}}=\overline{w^2}=0$ is the trivial solution of the local limit steady equations. The non-trivial solution of the local mixing length model can be obtained through solving the following algebraic equations :
\begin{eqnarray}
\overline{u^{2}}&=&\frac{1}{6\eta_{e}^{2}}(\frac{3+c_{3}}{1+c_{3}})^{1/2} [c_{1,w\theta}\frac{(\frac{3+c_{3}}{1+c_{3}})(1+\frac{u_{c}}{(\overline{u^2})^{1/2}})c_{1,w^2}+2c_{1,\theta^2}}{(1+\frac{u_{c}}{(\overline{u^2})^{1/2}})+2}] (\frac{{\beta}{P}}{{\rho}})(1+\frac{u_{c}}{(\overline{u^2})^{1/2}})^{-1}(\nabla-\nabla_{ad})~,\label{eq:u2}\\
\overline{\frac{\theta^2}{T^2}}&=&\frac{1}{3\eta_{e}^2}(\frac{3+c_{3}}{1+c_{3}})^{1/2}\frac{c_{1,\theta^2}}{c_{1,w^2}}[c_{1,w\theta}\frac{(\frac{3+c_{3}}{1+c_{3}})(1+\frac{u_{c}}{(\overline{u^2})^{1/2}})c_{1,w^2}+2c_{1,\theta^2}}{(1+\frac{u_{c}}{(\overline{u^2})^{1/2}})+2}](1+\frac{u_{c}}{(\overline{u^2})^{1/2}})^{-2}(\nabla-\nabla_{ad})^{2}~,\\
\overline{w\frac{\theta}{T}}&=&\frac{\sqrt{2}}{6\eta_{e}^{2}}(\frac{3+c_{3}}{1+c_{3}})^{3/4}\frac{1}{c_{1,w^2}}[c_{1,w\theta}\frac{(\frac{3+c_{3}}{1+c_{3}})(1+\frac{u_{c}}{(\overline{u^2})^{1/2}})c_{1,w^2}+2c_{1,\theta^2}}{(1+\frac{u_{c}}{(\overline{u^2})^{1/2}})+2}]^{3/2}(\frac{{\beta}{P}}{{\rho}})^{1/2}(1+\frac{u_{c}}{(\overline{u^2})^{1/2}})^{-3/2}(\nabla-\nabla_{ad})^{3/2}~,\\
\overline{w^2}&=&\frac{3+c_{3}}{1+c_{3}}\overline{u^2}~,\label{eq:w2}
\end{eqnarray}
in the convectively unstable zone ($\nabla-\nabla_{ad}>0$), and $\overline{u^{2}}=\overline{\frac{\theta^2}{T^2}}=\overline{w\frac{\theta}{T}}=\overline{w^2}=0$ in the convectively stable zone ($\nabla-\nabla_{ad}\leq 0$), respectively. $\nabla$ is the temperature gradient and $\nabla_{ad}$ is the adiabatic temperature gradient. After some manipulation, equation~(\ref{eq:u2}) can be rewritten as the following cubic equation if $(\overline{u^{2}})^{1/2}$ is taken as the unknown variable :
\begin{equation}
[(\overline{u^{2}})^{1/2}]^3+\frac{4}{3}u_{c}[(\overline{u^{2}})^{1/2}]^2+\frac{1}{3}(u_{c}^2-\frac{3+c_{3}}{1+c_{3}}\frac{c_{1,w^2}}{c_{1,\theta^2}}D)[(\overline{u^{2}})^{1/2}]-\frac{1}{3}(\frac{3+c_{3}}{1+c_{3}}\frac{c_{1,w^2}}{c_{1,\theta^2}}u_{c}+2)D=0~.
\end{equation}
where $D=\frac{1}{6\eta_{e}^{2}}(\frac{3+c_{3}}{1+c_{3}})^{1/2} c_{1,w\theta}c_{1,\theta^2}(\frac{{\beta}{P}}{{\rho}})(\nabla-\nabla_{ad})>0$ in the convectively unstable zone. The roots of the cubic equation, can be either three real, or one real and a complex conjugate pair. Let us assume the three roots are $u_{1},u_{2},u_{3}$ respectively, then the following constraints must be satisfied :
\begin{eqnarray}
&&u_{1}+u_{2}+u_{3}=-\frac{4}{3}u_{c}<0~,\\
&&u_{1}u_{2}u_{3}=\frac{1}{3}(\frac{3+c_{3}}{1+c_{3}}\frac{c_{1,w^2}}{c_{1,\theta^2}}u_{c}+2)D>0~.
\end{eqnarray}
On one hand, if the three roots are all real, then one of them must be positive and the other two are negative. On the other hand, if the roots are one real and a complex conjugate pair, then the real root must be positive as the multiplication of a complex conjugate pair is always positive. As a result, only one positive real root exists for equation~(\ref{eq:u2}), and it is the physical solution of the root mean square of convective velocity we are looking for.

\subsection{The diffusive terms}
The diffusive terms include the third-order moments of velocity and temperature perturbation. Multiplying the dynamic equations of second-order moments by $w$ and $\theta$ and taking average, we can obtain the dynamic equations for third-order moments :
\begin{eqnarray}
\frac{\partial \overline{wu^2}}{\partial t}+\frac{1}{\rho}\frac{\partial}{\partial z}(\rho\overline{w^2 u^2})-\beta g(\overline{u^2 \frac{\theta}{T}} +\frac{2}{3}\overline{w^2 \frac{\theta}{T}})-\overline{u^2}\frac{1}{\rho}\frac{\partial}{\partial z}(\rho\overline{w^2})-\frac{2}{3}\overline{w^2}\frac{1}{\rho}\frac{\partial}{\partial z}(\rho\overline{w^2})+2\sqrt{3}\eta_{e}\frac{\rho g}{c_{1,w^2}P}(\overline{u^{2}})^{1/2} \overline{wu^2}&=&0~,\\  % equation5
\frac{\partial (\overline{w\frac{\theta^2}{T^2}})}{\partial t} +\frac{1}{\rho}\frac{\partial }{\partial z}(\rho\overline{w^2\frac{\theta^2}{T^2}})  - \beta g \overline{\frac{\theta^3}{T^3}}+2(\frac{\partial \ln T}{\partial z}-\nabla_{ad}\frac{\partial \ln P}{\partial z})\overline{w^2 \frac{\theta}{T}} -\overline{\frac{\theta^2}{T^2}}\frac{1}{\rho}\frac{\partial}{\partial z}(\rho\overline{w^2})-2\overline{w\frac{\theta}{T}}\frac{1}{\rho}\frac{\partial}{\partial z}(\rho\overline{w\frac{\theta}{T}})&&\nonumber\\
+{\sqrt{3}\eta_{e}}\frac{\rho g}{c_{1,\theta^2}P}[3(\overline{u^{2}})^{1/2}+2u_{c}](\overline{w\frac{\theta^2}{T^2}})&=&0~,\\  % equation6
\frac{\partial (\overline{w^2\frac{\theta}{T}})}{\partial t}+\frac{1}{\rho}\frac{\partial}{\partial z}(\rho\overline{ w^3\frac{\theta}{T}})-2\beta g\overline{w \frac{\theta^2}{T^2}}
 +(\frac{\partial \ln T}{\partial z}-\nabla_{ad}\frac{\partial \ln P}{\partial z})(\overline{w^3})
  - \overline{w^2}\frac{1}{\rho}\frac{\partial}{\partial z} (\rho\overline{w \frac{\theta}{T}})-2\overline{w\frac{\theta}{T}}\frac{1}{\rho}\frac{\partial}{\partial z}(\rho w^2) &&\nonumber\\
 +\frac{2\eta_{e}}{\sqrt{3}}\frac{\rho g}{c_{1,w\theta}P}[3(\overline{u^{2}})^{1/2}+u_{c}][(1+c_{3})\overline{w^2 \frac{\theta}{T}}-c_{3}\overline{u^2 \frac{\theta}{T}}]&=&0~,\\ %equation 7
\frac{\partial \overline{w^3}}{\partial t}+\frac{1}{\rho}\frac{\partial}{\partial z}(\rho\overline{w^4})-3\beta g\overline{w^2 \frac{\theta}{T}}-3\overline{w^2}\frac{1}{\rho}\frac{\partial}{\partial z}(\rho\overline{w^2})+2\sqrt{3}\eta_{e}\frac{\rho g}{c_{1,w^2}P}(\overline{u^{2}})^{1/2}[(1+c_{3})\overline{w^3}-c_{3}\overline{wu^2}]&=&0~,\\ %equation 8
\frac{\partial (\overline{u^2\frac{\theta}{T}})}{\partial t}+\frac{1}{\rho}\frac{\partial}{\partial z}(\rho\overline{w u^2 \frac{\theta}{T}})-\frac{2}{3}\beta g\overline{w \frac{\theta^2}{T^2}}
 +(\frac{\partial \ln T}{\partial z}-\nabla_{ad}\frac{\partial \ln P}{\partial z})(\overline{w u^2})
 - \overline{u^2}\frac{1}{\rho}\frac{\partial}{\partial z} (\rho\overline{w \frac{\theta}{T}})-\frac{2}{3}\overline{w\frac{\theta}{T}}\frac{1}{\rho}\frac{\partial}{\partial z}(\rho w^2) &&\nonumber\\
 +\frac{2\eta_{e}}{\sqrt{3}}\frac{\rho g}{c_{1,w\theta}P}[3(\overline{u^{2}})^{1/2}+u_{c}](\overline{u^2\frac{\theta}{T}})&=&0~,\\ %equation 9
\frac{\partial (\overline{\frac{\theta^3}{T^3}})}{\partial t} +\frac{1}{\rho}\frac{\partial }{\partial z}(\rho\overline{w\frac{\theta^3}{T^3}}) +3(\frac{\partial \ln T}{\partial z}-\nabla_{ad}\frac{\partial \ln P}{\partial z})\overline{w \frac{\theta^2}{T^2}}
-3\overline{\frac{\theta^2}{T^2}}\frac{1}{\rho}\frac{\partial}{\partial z} (\rho\overline{w \frac{\theta}{T}})+{3\sqrt{3}\eta_{e}}\frac{\rho g}{c_{1,\theta^2}P}[(\overline{u^{2}})^{1/2}+u_{c}](\overline{\frac{\theta^3}{T^3}})&=&0~. % equation 10
\end{eqnarray}
Unfortunately, the third-order dynamic equations involve the fourth-order moments, therefore closure relations of the fourth-order moments are required to close the system. In the previous section, we have examined several closure relations of the fourth-order moments based on the results of numerical simulations. The feedback of the closure relations on the thermal structure, however, is not examined so far. On developing 1D model, the goodness of the closure relations should be further investigated by coupling the dynamic equations of moments with the equations of thermal structure.

\subsubsection{The down-gradient approximations}
One simple and popular method to close the system is choosing the down-gradient approximation. The down-gradient approximations assume that the third-order moments are expressed in a down-gradient form of second-order moments.
\begin{eqnarray}
D_{f}(\overline{u^2})&=&\frac{1}{\rho}\frac{\partial}{\partial z}(\rho\overline{wu^2})\approx -\frac{1}{\rho}\frac{\partial}{\partial z}[\frac{\sqrt{3}}{4}c_{2,w^2}\frac{P}{g} (\overline{w^2})^{1/2} \frac{\partial}{\partial z}\overline{u^2}]~,\\
D_{f}(\overline{\frac{\theta^2}{T^2}})&=& \frac{1}{\rho}\frac{\partial}{\partial z}(\rho\overline{w\frac{{\theta^2}}{T^2}})\approx -\frac{1}{\rho}\frac{\partial}{\partial z}[\frac{\sqrt{3}}{4}c_{2,\theta^2}\frac{P}{g} (\overline{w^2})^{1/2} \frac{\partial}{\partial z}\overline{\frac{\theta^2}{T^2}}]~,\\
D_{f}(\overline{w\frac{\theta}{T}})&=& \frac{1}{\rho}\frac{\partial}{\partial z}(\rho\overline{w^2\frac{{\theta}}{T}})\approx -\frac{1}{\rho}\frac{\partial}{\partial z}[\frac{\sqrt{3}}{4}c_{2,w\theta}\frac{P}{g} (\overline{w^2})^{1/2} \frac{\partial}{\partial z}\overline{w\frac{\theta}{T}}]~,\\
D_{f}(\overline{w^2})&=&\frac{1}{\rho}\frac{\partial}{\partial z}(\rho\overline{w^3})\approx -\frac{1}{\rho}\frac{\partial}{\partial z}[\frac{\sqrt{3}}{4}c_{2,w^2}\frac{P}{g} (\overline{w^2})^{1/2} \frac{\partial}{\partial z}\overline{w^2}]~,
\end{eqnarray}
where $c_{2,w^2}$, $c_{2,\theta^2}$, and $c_{2,w\theta}$ are three different diffusive coefficients. The down-gradient approximation only keeps part of terms in the dynamic equations for third-order moments. Under this approximation, three new coefficients $c_{2,w^2}$, $c_{2,\theta^2}$, $c_{2,w\theta}$ shall be estimated.

\subsection{Calibration of coefficients}
In the nonlocal model, the convective coefficients $c_{1,w^2}$, $c_{1,\theta^2}$, $c_{1,w\theta}$ and $c_{3}$ should be given. If down-gradient approximations are adopted, then three additional diffusive coefficients $c_{2,w^2}$,$c_{2,\theta^2}$ and $c_{2,w\theta}$ are required. We use the data from the 3D numerical simulations to calibrate these coefficients. In the following, we discuss the calibration of the convective and diffusive coefficients, respectively.

In the far field from the boundaries, the diffusive terms play less important roles in the second-order dynamic equations. Thus the solution in this region is likely to be close to the local steady solution. Then the coefficients $c_{1,w^2}$, $c_{1,\theta^2}$, $c_{1,w\theta}$, and $c_{3}$ can be calibrated from the equations~(\ref{eq:u2})-(\ref{eq:w2}) with the numerical data.

In the near field $z\rightarrow 1$ from the top boundary, we assume the variables have the following power law approximation in the region $z\in [z_{b},1]$
\begin{eqnarray}
&& \overline{u^2} = \frac{\overline{u_{b}^2}}{1-\xi_{b}^{\alpha_{1}}}(1-\xi^{\alpha_{1}})~,\\
&& \overline{\frac{\theta^2}{T^2}}= (\overline{\frac{\theta^2}{T^2}})_{b}(\xi/\xi_{b})^{\alpha_{2}}~,\\
&& \overline{w\frac{\theta}{T}}=(\overline{w\frac{\theta}{T}})_{b} (\xi/\xi_{b})^{\alpha_{3}}~,\\
&& \overline{w^2}=\overline{w_{b}^2} (\xi/\xi_{b})^{\alpha_{4}}~,
\end{eqnarray}
where $\xi=1-z/L_{z}$ is the distance to the upper bound, and the subscript $b$ represents the corresponding value at the boundary layer where the above power law approximation is valid. Substituting the above equations into the nonlocal model and keeping the leading terms, we have $\alpha_{1}=1,\alpha_{4}=2$, and
\begin{eqnarray}
&&c_{2,w^2}=\frac{16\eta_{e}}{3}\sqrt{\frac{1+c_{3}}{3+c_{3}}}\frac{1}{c_{1,w^2}}\xi_{b}~,\label{eq:c2ww}\\
&&c_{2,\theta^2}=\frac{8\eta_{e}}{\alpha_{2}^2}\sqrt{\frac{1+c_{3}}{3+c_{3}}}\frac{1}{c_{1,\theta^2}}\xi_{b}~,\label{eq:c2tt}\\
&&c_{2,w\theta}=\frac{12\eta_{e}}{\alpha_{3}^2}\frac{1+c_{3}}{3+c_{3}}\frac{1}{c_{1,w\theta}}\xi_{b}~.\label{eq:c2wt}
\end{eqnarray}
If $\xi_{b}$ is given, the diffusive coefficients can be directly computed from the calibrated convective coefficients. We estimate the location of the boundary layer by checking the first zero point of $D_{f}$ from the upper bound. After specifying the location of the boundary layer, the value of $\xi_{b}$ is calculated accordingly. It should be noted that the diffusive coefficients are estimated in an artificial zone originating from the top boundary. These coefficients may differ in free overshooting zones caused by thick radiative layer around the convection zone.

\begin{figure*}
\gridline{\fig{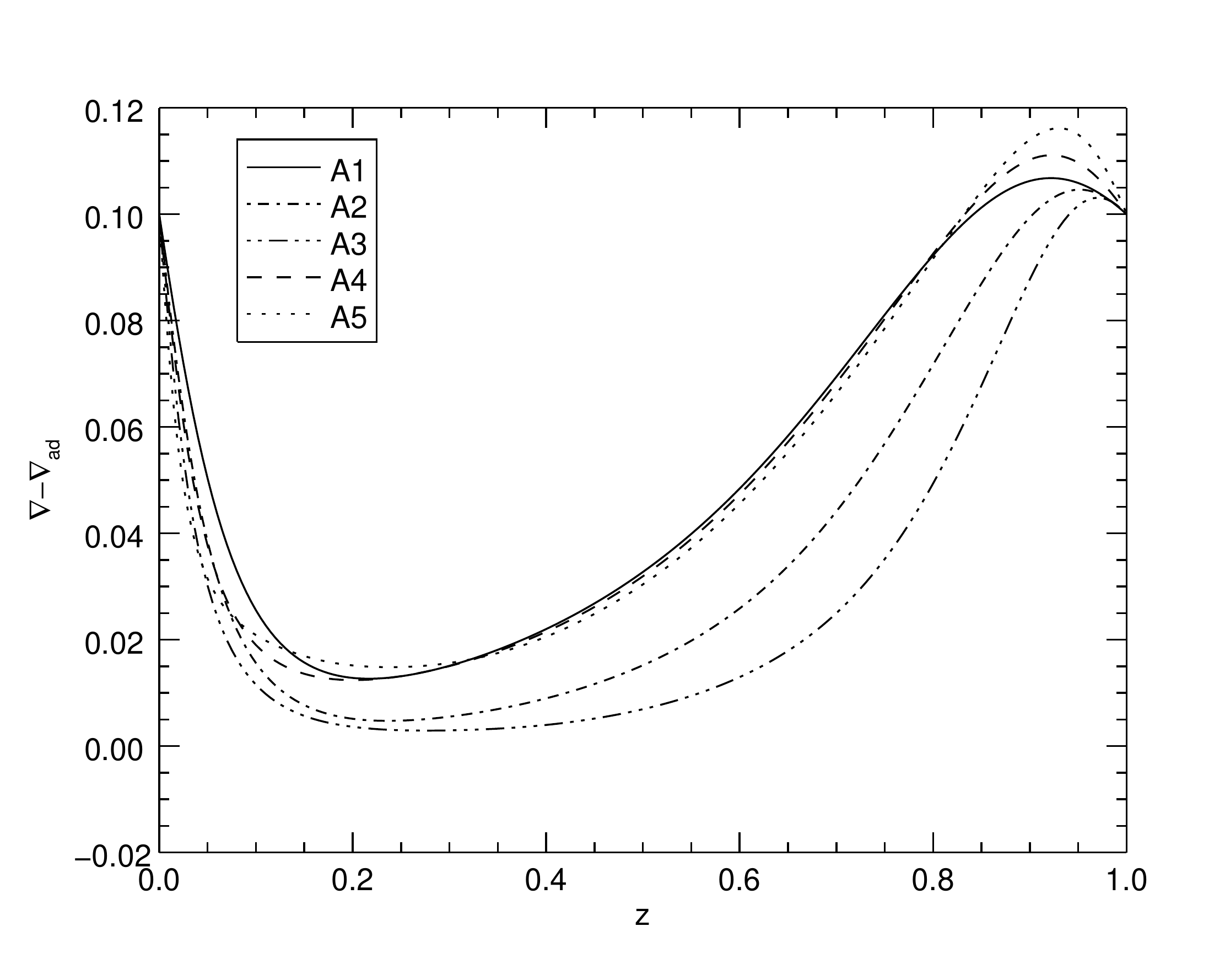}{0.5\textwidth}{(a)}
          \fig{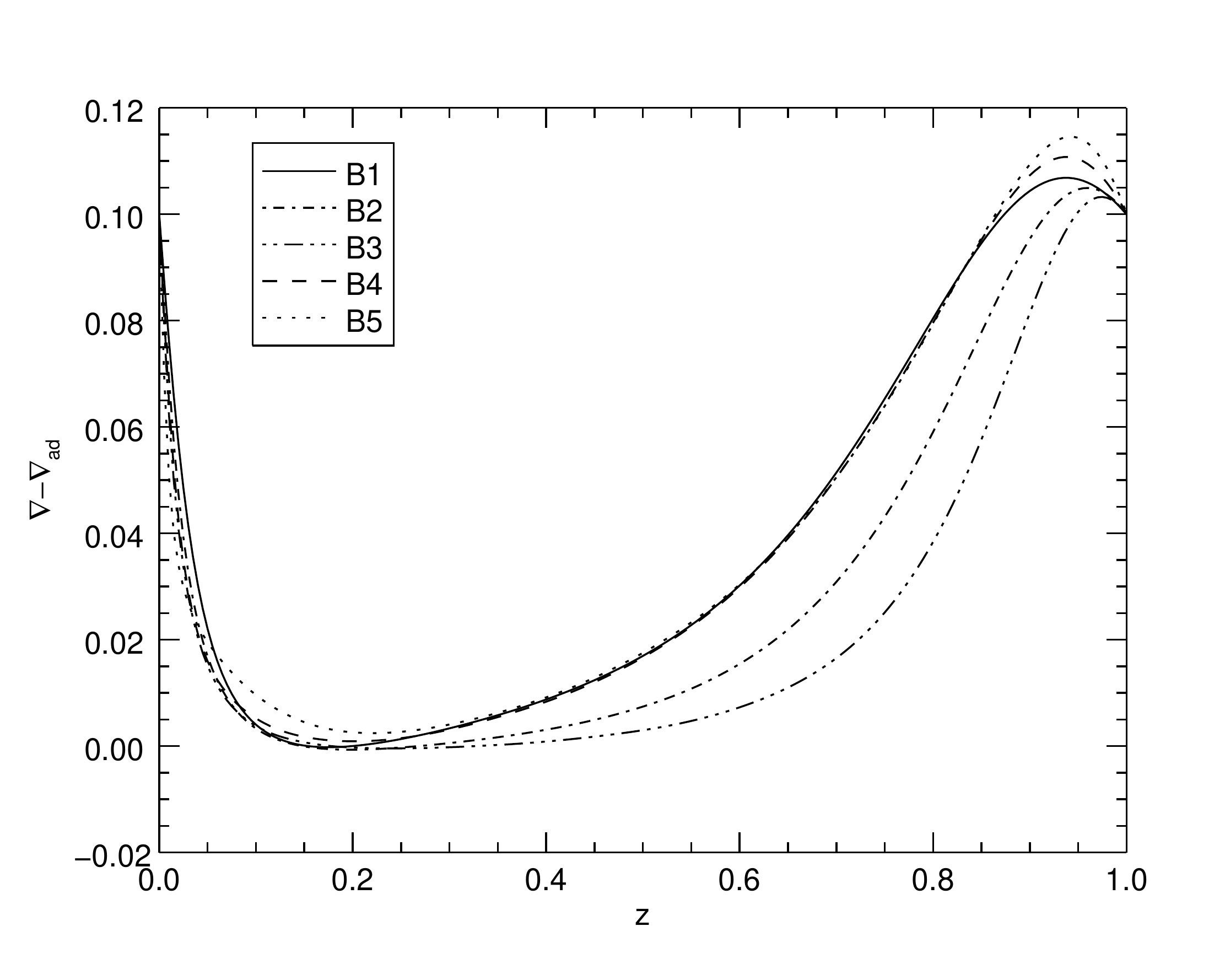}{0.5\textwidth}{(b)}
          }
\gridline{\fig{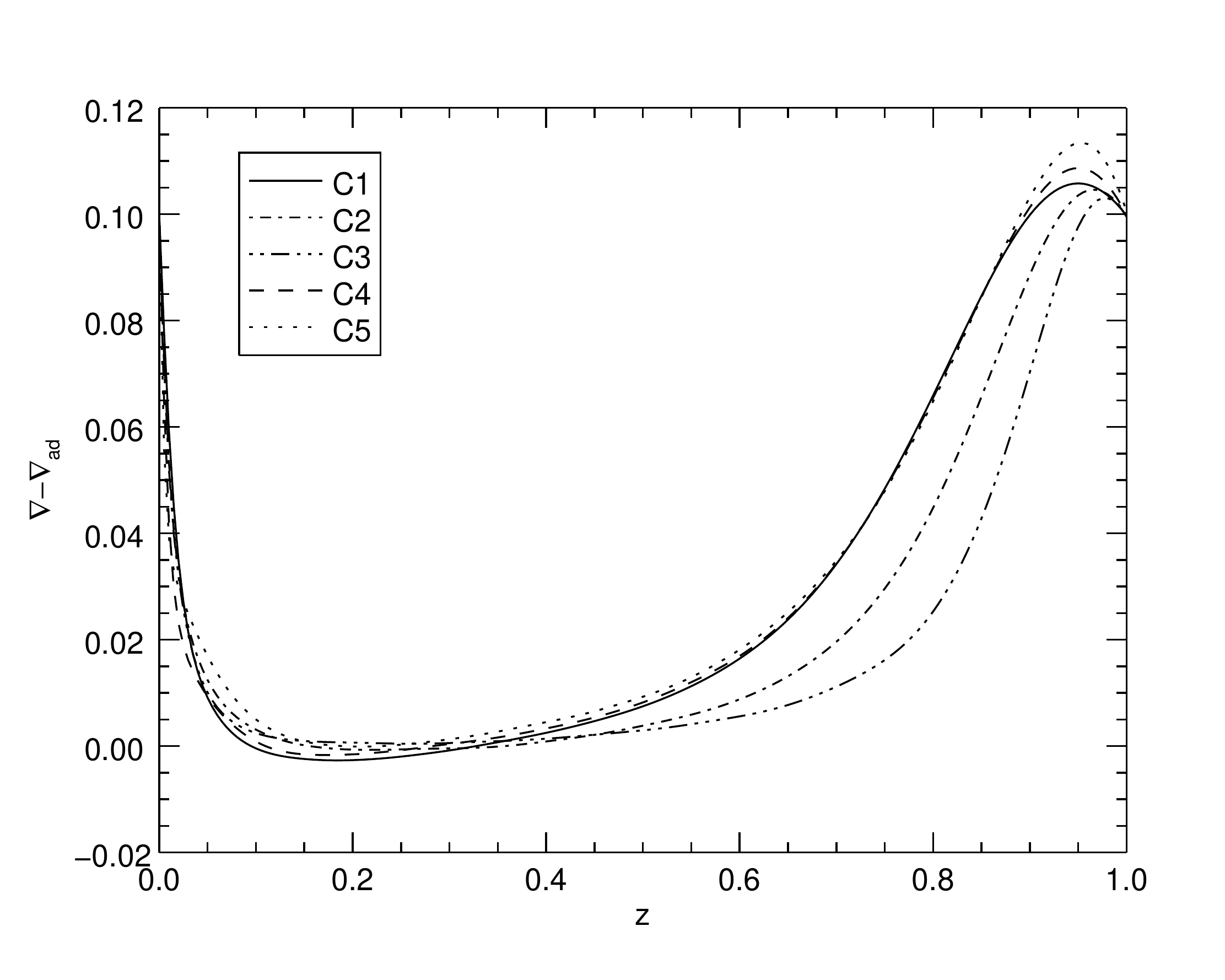}{0.5\textwidth}{(c)}
          \fig{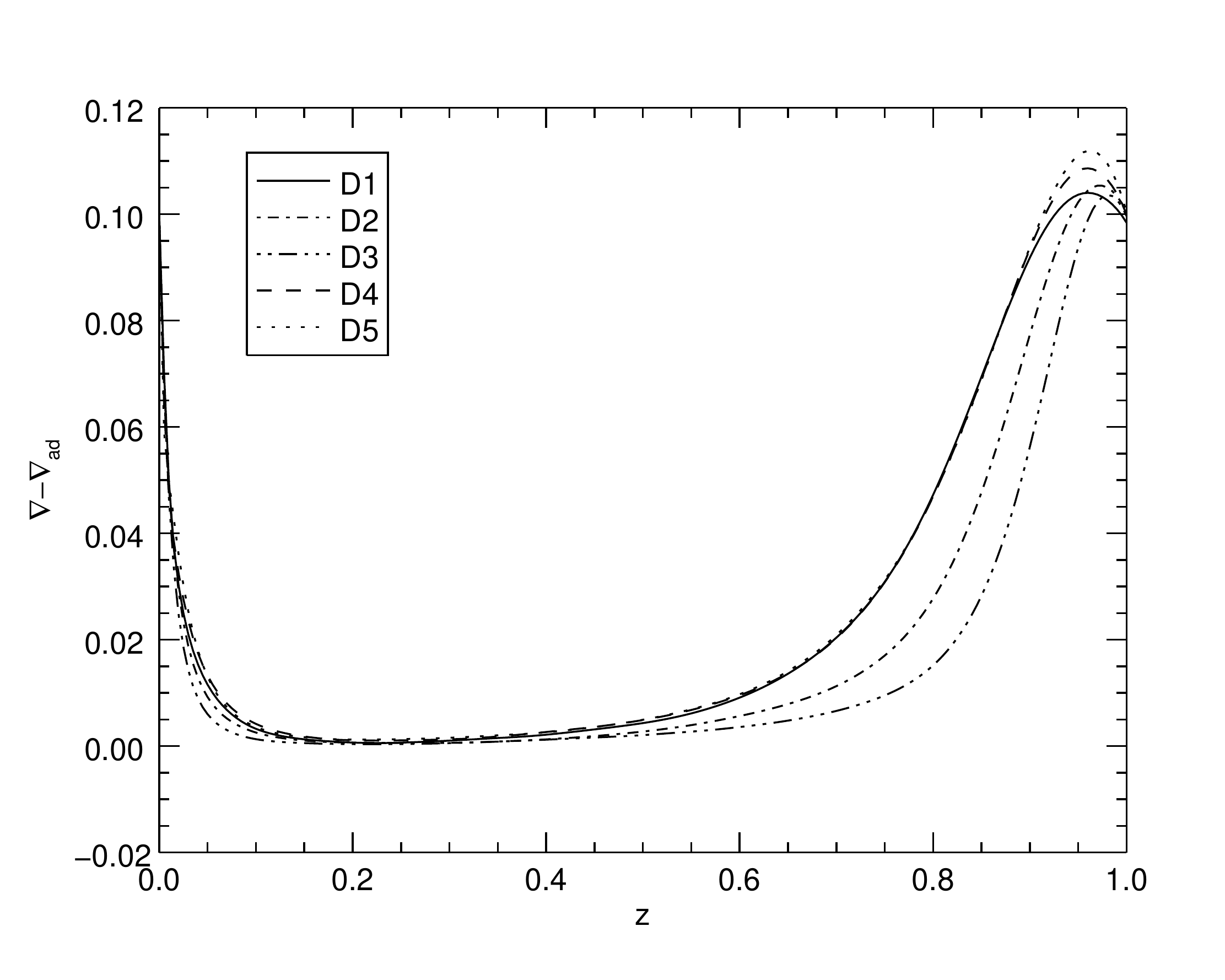}{0.5\textwidth}{(d)}
          }
\caption{The super-adiabatic temperature gradient obtained from the three-dimensional simulations. Panel (a)-(d) are four different groups with increasing depth of convection zone. \label{fig:superadiabatic}}
\end{figure*}

We use the previous 3D simulation data to calibrate the coefficients. To calibrate the turbulent coefficients $c_{1,.}$, the temperature gradient should be super-adiabatic. In some of our simulations, however, we find that the temperature gradient can be sub-adiabatic in the lower part of the simulation domain. Fig.~\ref{fig:superadiabatic} plots the averaged super-adiabatic temperature gradient $\nabla-\nabla_{ad}$ in the numerical simulations. In the panels (b) and (c), we see that $\nabla-\nabla_{ad}$ can be negative in the range $z\in (0.05,0.4)$. The sub-adiabatic temperature gradient in the convection zone is also mentioned in \citet{1992ApJL..389...L87,2000ApJs..127...159,2017ApJ..843...52}. \citet{1992ApJL..389...L87} gives an explanation for the existence of sub-adiabatic region. The low-entropy fluid parcels generated at the top of the convection zone are carried by the strong downward flows. The downward flows, if strong enough, are able to penetrate all the way into the bottom of the convection zone. Thus a sub-adiabatic region is created when the entropy of the parcels are lower than those of the surroundings at the bottom. In panel (a), both Pe and Re are small, and the low-entropy downward flows are not cool enough to create a sub-adiabatic region. One interesting finding is that the super-adiabatic temperature is positive for all the cases in panel (d), in which Pe and Re are larger than those of panels (b) and (c). It can be explained by two reasons. First, when the P$\acute{e}$clet number is higher, the energy transported by convection is more efficient. This leads to a flatter curve of $\nabla-\nabla_{ad}$ in the convection zone, as the temperature gradient is close to the adiabatic temperature in efficient convection. Second, the turbulent velocities and flux transportation at the lower region are influenced by the bottom boundary, which yields an adjustment of the thermal structure accordingly. The test on 3D simulations on solar convection \citep{2003MNRAS..340...923} has demonstrated that an undesirable effect of the impenetrable condition at the bottom produces an artificially high convective flux, and reduces the temperature gradient accordingly. To avoid the potential influence from the bottom boundary, the region $z\in(0,0.4)$ is simply excluded when calibrating parameters.
To calibrate $c_{1,.}$, we remove this sub-adiabatic region so that the local mixing length model could be applied. Apart from the sub-adiabatic region, we also exclude the region beside where $\nabla-\nabla_{ad}$ is too close to zero. In this region, the value of $\nabla-\nabla_{ad}$ can be easily contaminated by round-off error.  As a result, we choose the region $z\in(0.5,0.6)$ to calibrate these turbulent coefficients. We use $\overline{w^2}$ and $\overline{u^2}$ to calibrate $c_{3}$. As $c_{3}$ is non-negative, the ratio ${\overline{w^2}}/{\overline{u^2}}$ satisfies the constraint $1\leq {\overline{w^2}}/{\overline{u^2}} \leq 3$. In case D1, this constraint is satisfied in the region $z\in(0.23,0.48)$. However, it overlaps with the sub-adiabatic region, and we cannot calibrate both turbulent coefficients and $c_{3}$ within the same region. We assume that the velocities are not seriously affected by the sub-adiabatic temperature gradient, since its value is close to zero and the deceleration is very small. Therefore it is acceptable to calibrate $c_{3}$ using the data in the region where the constraint holds. To simplify it, we use the minimum value of $c_{3}$ in this region for the calibration.

Table~\ref{tab:table3} presents the calibrated coefficients from the simulation data. In addition, it also presents the correlations between the vertical velocity and temperature perturbation $cor[w,\theta]$, and the power law indices $\alpha_{2},\alpha_{3}$. From the table, we see that the turbulent coefficients $c_{1,w^2}$ and $c_{1,w\theta}$ do not vary too much across different cases. They have dispersions from one simulation to another of order 39 percent. The coefficient $c_{1,\theta^2}$, however, has a larger dispersion of order 289 percent. The large dispersion is mainly from the cases with shallower convection zones (groups A and B). If we only consider the cases with deep convection zones (groups C and D), the dispersion of $c_{2,\theta^2}$ decreases to 40 percent. From Eqs.(\ref{eq:c2ww}-\ref{eq:c2wt}), we know that the diffusive coefficients $c_{2,\cdot}$ are inversely proportional to the corresponding turbulent coefficients $c_{1,\cdot}$. Thus these parameters can be inferred from $c_{1,\cdot}$ after $\xi_{b}$ is estimated. The correlation coefficients between vertical velocity and temperature perturbation are at around 0.6. The power law indices $\alpha_{2}$ and $\alpha_{3}$ are almost close to 2. The last four rows of Table~\ref{tab:table3} give the averages and standard deviations of the estimated coefficients in the cases with shallow (groups A and B) and deep (groups C and D) convection zones. The coefficients of variation, defined as standard deviation divided by the average, are less than 15 percent in the deep groups. For the shallow groups, however, the coefficient of variation reaches 45 percent for $c_{1,\theta^2}$ and 31 percent for $c_{2,\theta^2}$, respectively. Thus, the estimated coefficients are more robust in the cases with deep convection zones. The robustness of the estimated coefficients in deep convection zone is important for the astrophysical implications, as real astrophysical convection zones are much deeper than the simulated ones in most situations.

\begin{deluxetable*}{ccccccccccc}[htb!]
\tablecaption{Estimated coefficients of Xiong's nonlocal model \label{tab:table3}}
\tablehead{
 Case & $c_{1,w^2}$ & $c_{1,\theta^2}$ & $c_{1,w\theta}$ & $c_{2,w^2}$ & $c_{2,\theta^2}$ & $c_{2,w\theta}$ & $c_{3}$  & $cor[w,\theta]$ & $\alpha_{2}$ & $\alpha_{3}$
}
\startdata
 A1 & 1.08 & 1.68 & 0.64 & 1.35 & 0.33 & 1.14 & 4.84  & 0.69 & 2.00 & 1.97 \\
 A2 & 1.15 & 2.59 & 0.63 & 1.42 & 0.24 & 1.36 & 6.67  & 0.62 & 1.98 & 1.96\\
 A3 & 1.17 & 3.62 & 0.61 & 1.52 & 0.20 & 1.60 & 7.54  & 0.59 & 1.94 & 1.92\\
 A4 & 1.27 & 1.85 & 0.66 & 1.20 & 0.31 & 1.18 & 5.52  & 0.65 & 2.00 & 1.97\\
 A5 & 1.39 & 1.97 & 0.70 & 1.18 & 0.31 & 1.26 & 7.95  & 0.65 & 2.00 & 1.96\\
 B1 & 1.41 & 3.18 & 0.75 & 1.15 & 0.19 & 1.14 & 7.30  & 0.62 & 2.00 & 1.96\\
 B2 & 1.37 & 4.59 & 0.73 & 1.29 & 0.15 & 1.32 & 7.80  & 0.59 & 1.97 & 1.94\\
 B3 & 1.32 & 6.76 & 0.73 & 1.42 & 0.11 & 1.48 & 8.72  & 0.59 & 1.92 & 1.89\\
 B4 & 1.43 & 3.50 & 0.74 & 1.17 & 0.18 & 1.19 & 6.85  & 0.60 & 2.00 & 1.96\\
 B5 & 1.47 & 3.51 & 0.74 & 1.18 & 0.19 & 1.26 & 8.04  & 0.59 & 2.00 & 1.95\\
 C1 & 1.49 & 5.83 & 0.74 & 1.17 & 0.11 & 1.28 & 8.81  & 0.57 & 1.98 & 1.94\\
 C2 & 1.32 & 6.54 & 0.73 & 1.40 & 0.11 & 1.41 & 7.60  & 0.60 & 1.95 & 1.90\\
 C3 & 1.41 & 4.67 & 0.85 & 1.38 & 0.18 & 1.41 & 9.11  & 0.65 & 1.87 & 1.83\\
 C4 & 1.41 & 5.21 & 0.76 & 1.24 & 0.13 & 1.25 & 7.36  & 0.59 & 1.99 & 1.93\\
 C5 & 1.46 & 4.92 & 0.76 & 1.24 & 0.14 & 1.34 & 7.86  & 0.59 & 1.98 & 1.91\\
 D1 & 1.37 & 5.73 & 0.79 & 1.36 & 0.13 & 1.38 & 9.45  & 0.62 & 1.97 & 1.88\\
 D2 & 1.41 & 5.66 & 0.81 & 1.37 & 0.14 & 1.47 & 8.93  & 0.62 & 1.90 & 1.83\\
 D3 & 1.45 & 5.14 & 0.83 & 1.39 & 0.18 & 1.69 & 10.7  & 0.63 & 1.81 & 1.73\\
 D4 & 1.50 & 4.86 & 0.88 & 1.26 & 0.15 & 1.28 & 9.46  & 0.64 & 1.96 & 1.86\\
 D5 & 1.49 & 5.14 & 0.85 & 1.28 & 0.15 & 1.35 & 8.92  & 0.63 & 1.96 & 1.85\\
 \hline
 avg AB & 1.31 & 3.33 & 0.69 & 1.29 & 0.22 & 1.29 & 7.12 & 0.62 & 1.98 & 1.95\\
 avg CD & 1.43 & 5.37 & 0.80 & 1.31 & 0.14 & 1.38 & 8.82 & 0.61 & 1.94 & 1.87\\
 dev AB & 0.13 & 1.52 & 0.05 & 0.13 & 0.07 & 0.15 & 1.19 & 0.03 & 0.03 & 0.03\\
 dev CD & 0.06 & 0.57 & 0.05 & 0.08 & 0.02 & 0.13 & 1.00 & 0.03 & 0.06 & 0.06\\
\enddata
\tablecomments{$c_{1,w^2},c_{1,\theta^2},c_{1,w\theta}$ are the convective coefficients. $c_{2,w^2},c_{2,\theta^2},c_{2,w\theta}$ are the diffusive coefficients. $c_{3}$ is the coefficient measuring the isotropic level of fluid motions.  $cor[w,\theta]=\overline{w\frac{\theta}{T}}/(\overline{w^2}\overline{\frac{\theta^2}{T^2}})^{1/2}$ represents the correlation coefficient between velocity and temperature perturbation. $\alpha_{2}$ and $\alpha_{3}$ are the power law indices of $\overline{\frac{\theta^2}{T^2}}$ and $\overline{w\frac{\theta}{T}}$, respectively. The last four rows give the averages and standard deviations of the estimated coefficients in the cases with shallow (groups A and B) and deep (groups C and D) convection zones.}
\end{deluxetable*}

\subsection{The equations of thermal structure}
Given the thermal structure variables, the second-order and third-order moments can be solved from the above dynamic equations. These moments, on the other hand, have feedback on the thermal variables. To obtain the thermal structure variables, we solve the following equations :
\begin{eqnarray}
\frac{\partial m}{\partial z}=\rho~,\\
\frac{\partial (P+\rho \overline{u^2})}{\partial z} = -\rho g~, \label{eq:balance}\\
F_{r}+F_{c}+F_{k}=F_{tot}~,
\end{eqnarray}
where $m$ is the mass, $g$ is the gravitational accelerate, $F_{r}$ is the conductive flux (or radiative flux)
\begin{equation}
F_{r}=- \kappa_{T} \frac{\partial T}{\partial z}~,
\end{equation}
$F_{c}$ is the convective flux
\begin{equation}
F_{c}= \rho c_{p}T \overline{w\frac{\theta}{T}}~,
\end{equation}
$F_{k}$ is the kinetic energy flux
\begin{equation}
F_{k}=\frac{3}{2}\rho \overline{wu^2}
\end{equation}
and $F_{tot}$ is the summation of conductive, convective, and kinetic energy fluxes.

In the numerical simulations, we impose boundary conditions that the fluid motions are impenetrable and stress free at the top and bottom, and the temperature is a constant at the top with a constant flux supplied at the bottom. To be consistent with the numerical simulations, we use the following boundary conditions for the Reynolds stress model: $F_{tot}$=constant, $m$=constant, $\partial_{z} \overline{u^2}=0$, $\partial_{z}{\overline{\theta^2}} =0$, $\overline{w\frac{\theta}{T}}=0$, $\overline{w^2}=0$ at the bottom; and $T$=constant, $m$=constant, $\partial_{z} \overline{u^2}=0$, $\overline{\theta^2 }=0$, $\overline{w\frac{\theta}{T}}=0$, $\overline{w^2}=0$ at the top.

\subsection{Numerical method and result}
We couple the dynamic equations for the second-order and third-order moments with the equations of the thermal structure. We discretize the spatial derivatives by a second-order staggered finite-difference scheme. On the temporary direction, we integrate the dynamic equations by a fully implicit method. In each time step, we solve the equations through Newton-Raphson method. The dynamic equations are evolved for a very long time till a steady state is achieved. The numerical method is robust when the third-order moments are approximated with the down-gradient form. However, for most of the cases, we could not find converged solutions when dynamic equations of the third-order moments and closure models for the fourth-order moments are included. The major reason is that the auto-correlation second-order moments ($\overline{u^2}$, $\overline{w^2}$, and $\overline{\theta^2}$) become negative during the time integration. These variables are also in the reaction term of the dynamic equations of second-order moments. Once the reaction terms are negative, the magnitudes of these variables would increase exponentially and blow up quickly as time increases. As a rescue strategy, the problem could be addressed by constraining the coefficients of the reaction and diffusion terms with positive values \citep{2001IJCFD..14...201}. This method has been successfully applied in the simulation of the turbulent k-epsilon model if the negative solutions are only a transient state. We have tried this method in our numerical models, and unfortunately still could not obtain a steady result. We conjecture that the solutions of these variables are negative if a steady state could be reached, and the instability is not from the numerical but the physical side. \citet{1996MNRAS..279...305} have found that the numerical scheme could be stabilized by introducing additional diffusion terms in the dynamic equations of third-order moments. This method, however, did not work in our simulations. The convection zones in our simulations are much deeper and the boundary conditions are different. It is not guaranteed that the auto-correlation second-order moments are always non-negative by introducing diffusion terms in the dynamic equations of the third order moments. \citet{2007IAUS..239...92} have also reported that the correlation moments appear off limits in the deep quasi-adiabatic zone, when the stationary limit of the dynamical equations of third order moments is closed by third order cross-correlations of the GH model. Since we could not obtain solutions when the dynamic equations of the third-order moments are included, we only report the numerical results of down-gradient approximation hereafter.

\begin{figure*}
\gridline{\fig{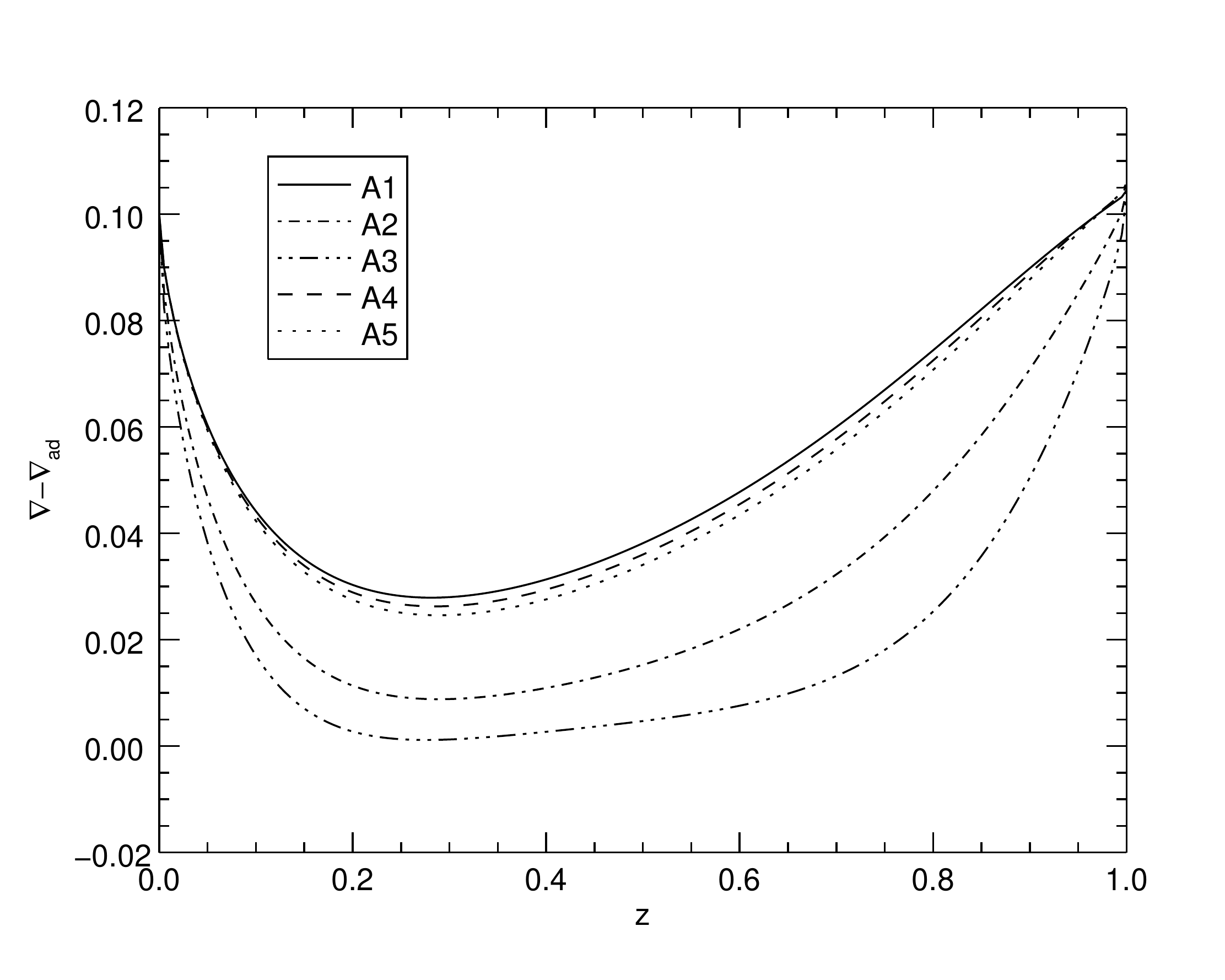}{0.5\textwidth}{(a)}
          \fig{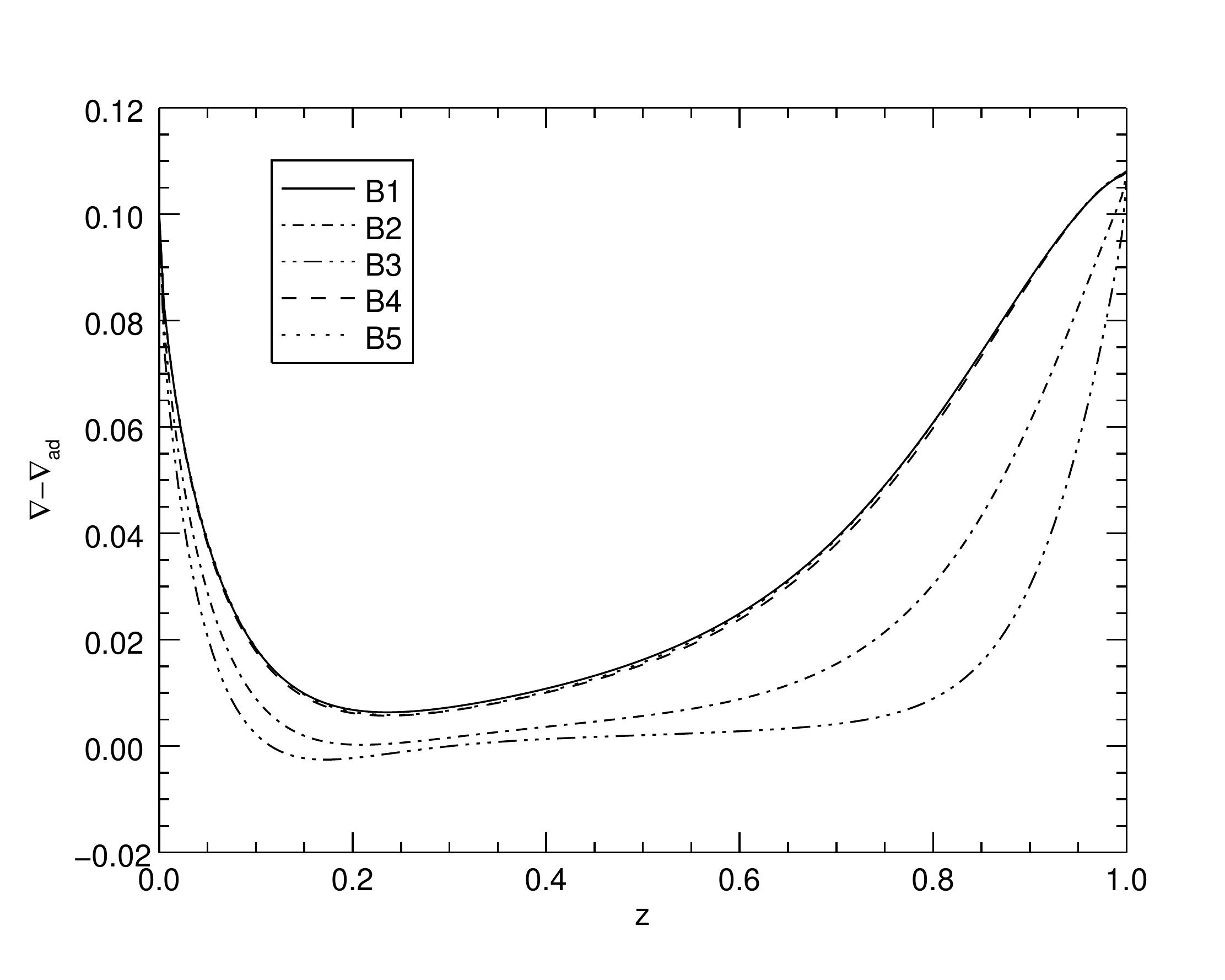}{0.5\textwidth}{(b)}
          }
\gridline{\fig{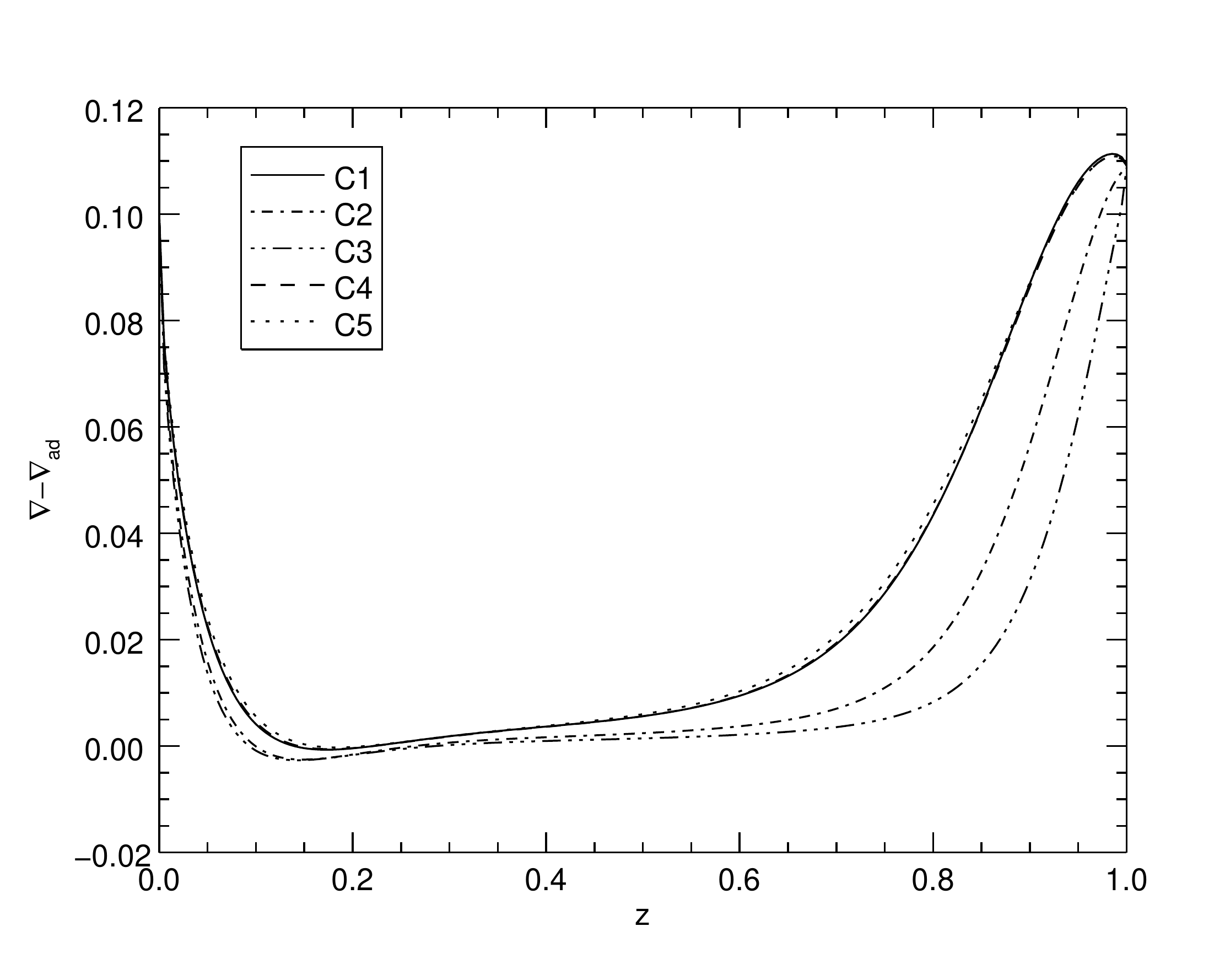}{0.5\textwidth}{(c)}
          \fig{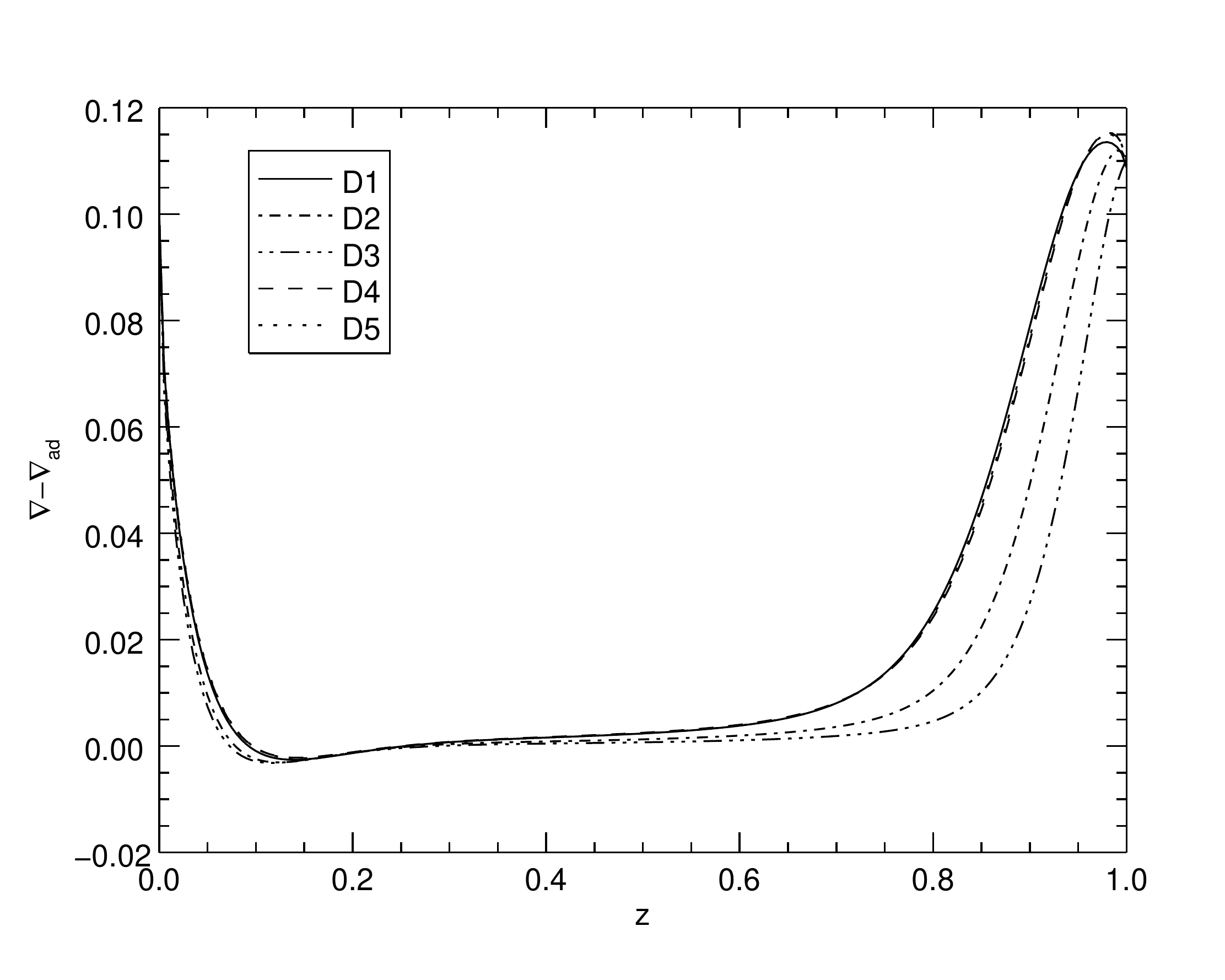}{0.5\textwidth}{(d)}
          }
\caption{The super-adiabatic temperature gradient obtained from the one-dimensional down-gradient approximation model. Panel (a)-(d) are four different groups with increasing depth of convection zone. \label{fig:superadiabatic1D}}
\end{figure*}

Figure~\ref{fig:superadiabatic1D} plots the numerical results for the super-adiabatic temperature gradient from the 1D down-gradient approximation model. The results show some agreements with those of 3D simulations. First, the super-adiabatic temperature gradient has a U-shape with a minimum value at the lower part of the convection zone. The P$\acute{e}$clet number has more significant effect on the structure of super-adiabatic temperature gradient than the Prandtl number. Second, in some cases with deeper convection zone, e.g. panels (c) and (d), there exists a bump at the top for $\nabla-\nabla_{ad}$. The turbulent pressure ($\rho \overline{u^2}$) cannot be ignored at the top where the Mach number can reach as high as 0.15. From equation~(\ref{eq:balance}), we can see that ${\partial P}/{\partial z}$ would decrease when ${\partial \rho \overline{u^2}}/{\partial z}$ increases. As a result, we can see the bump of the super-adiabatic temperature gradient at the top of convection. Third, in one-dimensional results, it is very interesting to find that the super-adiabatic temperature gradient also can be negative at the lower part of the convection zone. Apart from these agreements, there do exist some differences between 1D and 3D results. In the cases with shallower convection zone, 1D result produces a little bit higher value of $\nabla-\nabla_{ad}$. By varying the coefficients in a wide range, we have verified that this difference is not due to the error of calibrated coefficients. It is probably because the 1D model underestimates the kinetic energy flux \citep{2007IAUS..239...83,1996ApJ..466...372}, leading to a higher portion of energy transported by radiation (or conduction). Another difference is that the bump at the top is gone at the shallower cases in 1D results. In such case, the effect from the turbulent pressure is not strong enough for the appearance of the bump.

\begin{figure*}
\gridline{\fig{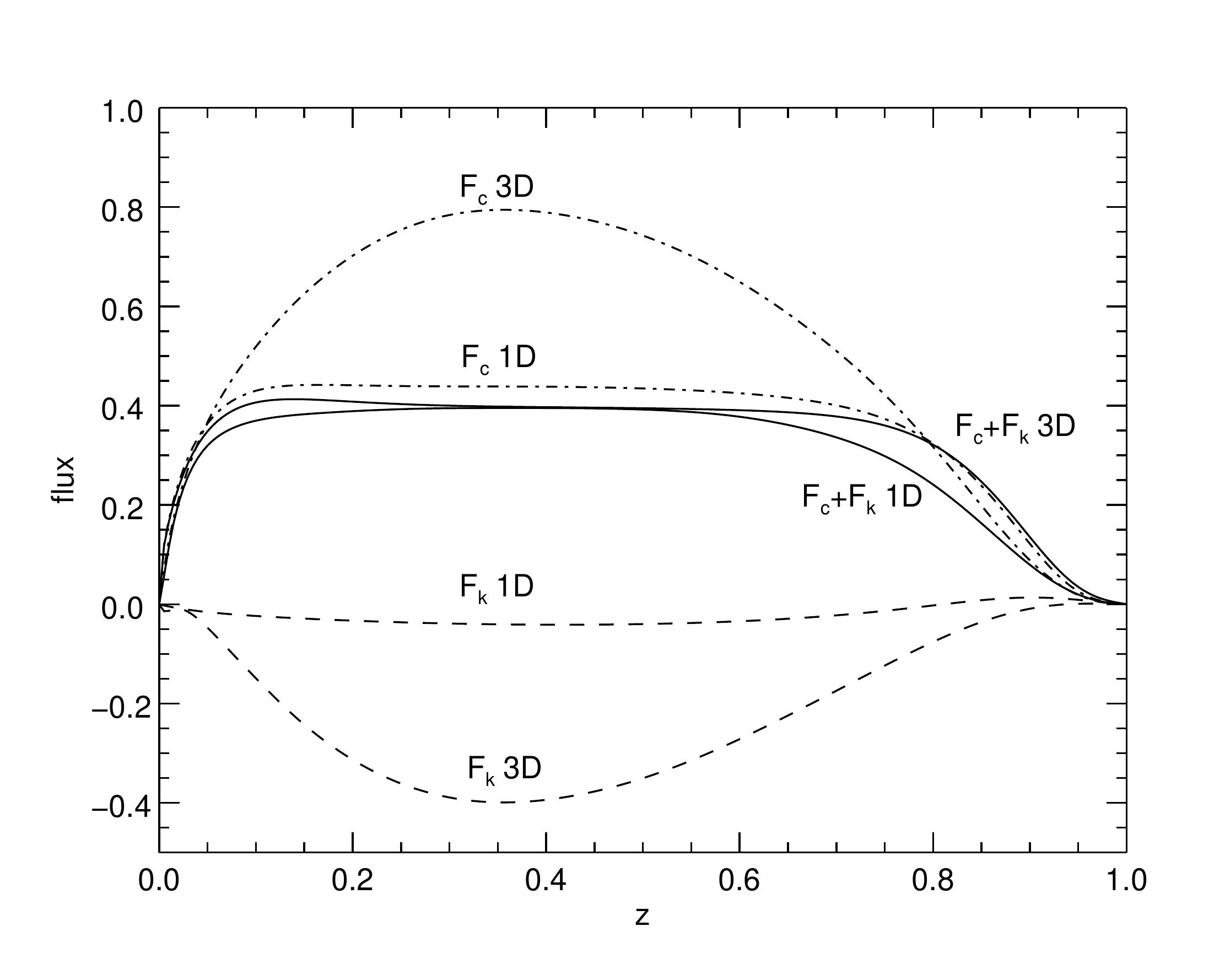}{0.5\textwidth}{(a)}
          \fig{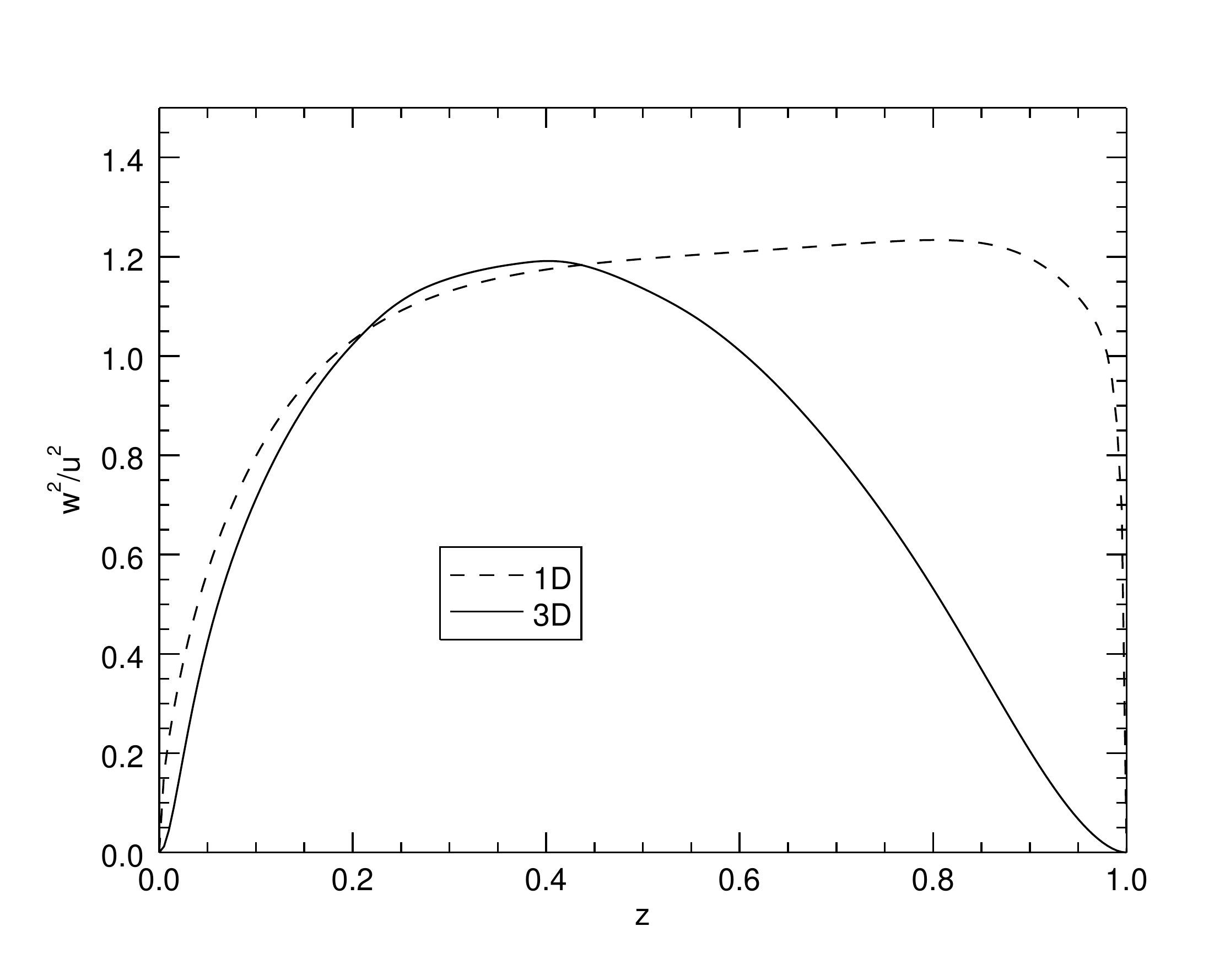}{0.5\textwidth}{(b)}
          }
\caption{Comparison of fluxes and anisotropic level between 1D and 3D results : (a)The convective flux $F_{c}$, kinetic energy flux $F_{k}$, and the summation of them. (b)The ratio of $\overline{{w^2}}$ to $\overline{{u^2}}$ as a function of depth. \label{fig:compare}}
\end{figure*}

Figure~\ref{fig:compare} compares the fluxes and anisotropic level between one- and three-dimensional results in case D1. The coefficients used in 1D model are calibrated from the result of 3D simulations. As we mentioned before, we calibrate the turbulent coefficients $c_{1,.}$ and the anisotropic coefficient $c_{3}$ in two different regions. From panel (b), we see that ${\overline{w^2}}/{\overline{u^2}}$ agrees well for 1D and 3D in the lower part of the convection zone. In the upper part, however, ${\overline{w^2}}/{\overline{u^2}}$ is much smaller in the 3D result. Thus the coefficient $c_{3}$ fitted from the 3D data is overestimated in the upper region. As a result, the turbulent coefficients $c_{1,.}$ will be overestimated when the fitted data is from the upper region. As $c_{1,w\theta}$ is overestimated, the convective flux $F_{c}$ obtained from 1D calculation is a little bit higher than that of 3D result (see panel (a)). The kinetic energy flux from 1D result is very small and almost negligible. This is inconsistent with the 3D result, where the kinetic energy flux is comparable to convective flux in magnitude. Similar results have been obtained in the simulations of efficient \citep{2007IAUS..239...83,1996ApJ..466...372} and inefficient \citep{2007IAUS..239...83} convection embedded in radiative layers. It seems that the down-gradient approximation induces a large amount of error when predicting the magnitudes of the third-order moments. The prediction on the second-order moments, however, is much better. In the 3D result, a moderate portion of convective flux is cancelled by the kinetic energy flux. The total flux carried by turbulent motions ($F_{c}+F_{k}$) of 1D result agrees well with that of 3D result in the lower region of the convection zone. The agreement in the upper region is not as good as the lower region, probably because the anisotropic levels between 1D and 3D results are quite different in the upper region.

\section{Conclusion}
In this paper, we have made a detailed comparison of nonlocal 1D turbulent models with the 3D direct numerical simulations. The 3D simulations are based on a semi-implicit spectral model in Cartesian geometry. We have performed a total of 20 simulation cases of efficient turbulent convection with varying parameters on the P$\acute{e}$clet number, the Prandtl number, and the depth of convection zone. We have analyzed the skewnesses and kurtoses of velocity and temperature perturbations with the simulation data. Both skewness and kurtosis are largely affected by the P$\acute{e}$clet number. $S_{w}$ and $S_{\theta}$ decrease to negative values in the middle of convection zone when the P$\acute{e}$clet number increases. It implies that the flow is more and more asymmetric and downward flow plays an important role when the P$\acute{e}$clet number increases. $K_{w}$ and $K_{\theta}$ could exceed 3 when the P$\acute{e}$clet number is greater than 50. In a normal distribution, the value of kurtosis should not be greater than 3. Thus the assumption of quasi-normal distributions of velocity and temperature perturbation is not good when the P$\acute{e}$clet number is large.

In one dimensional Reynolds stress convection models, closure assumptions are required for the completeness of the dynamic equations. One simple treatment is approximating the third-order moments as formulas of the second-order moments with down-gradient forms. Thus the dynamic equations are completed to the level of second-order moments. Or one can go further extending the dynamic equations to the level of third-order moments, and expressing the fourth-order moments with second-order moments. Based on the 3D simulation data, we have evaluated the performance of four different closure models of the fourth-order moments. The performance of the closure model varies among different cases and different fourth-order moments. In general, the QN model has the worst performance on prediction of the fourth-order moments compared with the rest of closure models. The GN3 model, due to its lack of a proper account of symmetries with respect to temperature perturbations, has bad performance in the prediction on $\overline{\theta^4}$, $w\overline{\theta^3}$. The difference of the predictive performance on the model GH and GN4 is small.

In the application of the Reynolds stress convection model, the dynamic equations of moments should be coupled with the equations of thermal structure. In this paper, we mainly focus on Xiong's nonlocal model. In particular, we consider two versions of Xiong's model. One adopts the down-gradient approximation and close the dynamic equations in the level of second-order moments. The other extends to the dynamic equations of third-order moments and close the fourth-order moments with the closure relations mentioned above. We rewrite Xiong's nonlocal model in Cartesian coordinates for the comparison with our 3D simulation results. In Xiong's model, turbulent coefficients $c_{1,.}$, diffusive coefficients $c_{2,.}$, and anisotropic coefficient $c_{3}$ are introduced. The turbulent coefficients $c_{1,.}$ and the anisotropic coefficient $c_{3}$ are calibrated with the algebraic formulas of local steady solutions in the far field. The diffusive coefficients $c_{2,.}$ are calibrated with the asymptotic power-law solutions in the near field of the top boundary. It is found that these coefficients vary with the depths of convection zones. Those coefficients calibrated from the cases of deeper convection zones are more robust.

With the calibrated coefficients, we solve the dynamic equations of moments with equations of the thermal structure together. For the dynamic equations of the third-order moments, we cannot obtain the steady solutions with any of the closure relations. The numerical instability may arise from the mathematical complexity, or even because that solution is unphysical \citep{2017LRCA..3...1}. We speculate that the numerical instability is not from the numerical side, but the physical side.  Through the process of time evolution, we find that the auto-correlation second-order moments tend to be negative, which are not physical solutions. As these variables turn to negative, the numerical calculations blow up quickly since the reaction terms are negative. For the model with down-gradient approximations of the third-order moments, on the other hand, the numerical method is quite stable in all the cases. Thus, we are able to gauge the 1D down-gradient approximation model with the 3D simulation result. The super-adiabatic temperature gradient shows some similarities among the 1D and 3D results. First, $\nabla-\nabla_{ad}$ has a U-shape with a minimum value in the lower part of the convection zone. Second, there exists a bump near the top of the convection zone when the P$\acute{e}$clet number is large, probably because the turbulent pressure is not ignorable there. Third, $\nabla-\nabla_{ad}$ can be negative in the convection zone, which can be attributed to the nonlocal transport of heat from the downward flows generated at the top. Apart from these similarities, the anisotropic level predicted by the 1D model is unsatisfactory. The value of $\overline{w^2}/\overline{u^2}$ predicted by the 1D model is around 1.0 near the top of the convection zone, which means the flow is almost isotropic there. Through the 3D simulation, however, we find that anisotropic level near the top is very high because of the boundary effect. Since most of real astrophysical flows are much deeper than our simulated cases, this disagreement has probably only very limited astrophysical implications. Another weakness of the 1D model is that its prediction on the magnitude of third-order moments is unsatisfactory. The predicted kinetic energy flux in the 1D result is too small compared with that of the 3D result. In general, the 1D down-gradient model have better performance on prediction of the second-order moments than the third-order moments. Improvements on prediction of the third-order moments are needed for the 1D model.

\acknowledgments

I sincerely thank an anonymous referee for the valuable comments and suggestions on the manuscript. Discussions on turbulent convection models with D.R. Xiong, Y. Li, K.L. Chan, and Q.S. Zhang are highly appreciated. I am indebted to X.P. Zhu and W.P. Lin for providing me with the computational resources of this work. This work has been supported by NSFC (No. 11503097, 11521101), Science and Technology Program of Guangzhou (No. 201707010006), FDCT of Macau(No. 119/2017/A3), and the Fundamental Research Funds for the Central University (No. 16lgpy44). The computations are carried out on the National Supercomputer Center in Guangzhou.

\appendix

\section{The boundary effects}
\begin{figure}[ht!]
\plotone{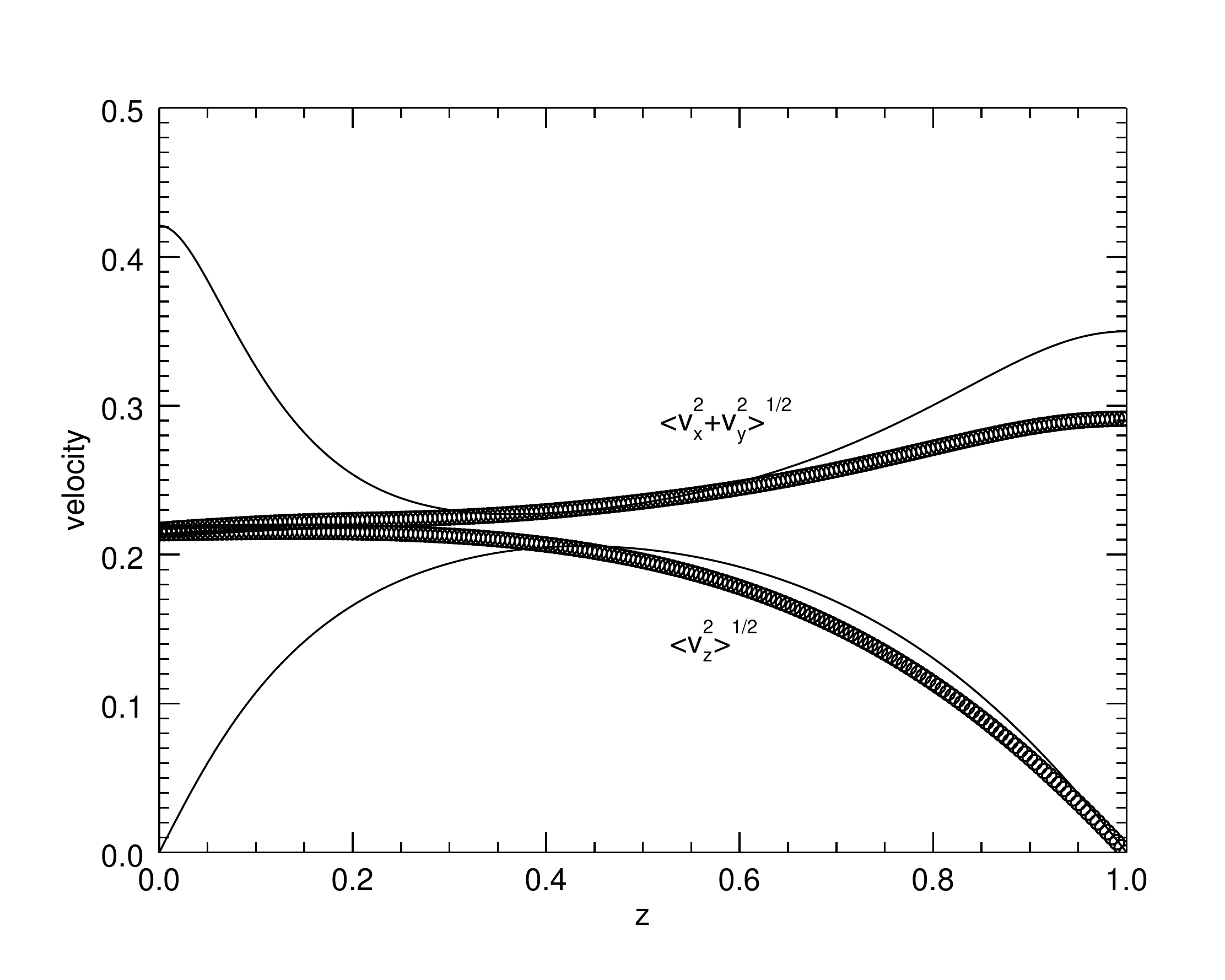}
\caption{The effect of the lower boundary on the velocity profiles. The solid lines show the horizontal and vertical velocities in case A1. The lines with cycle symbols plot the corresponding velocities when an additional convectively stable zone is added below the convection zone. \label{fig:boundaryEffect}}
\end{figure}

In our 3D simulations, impenetrable boundary conditions are applied at the top and bottom. \citet{2003MNRAS..340...923} have shown that the velocity and temperature correlations are strongly influenced by the lower impenetrable boundary condition at least 2 pressure scale heights away from the boundaries. \citet{2009MSAI..80...701} confirmed this conclusion by comparing different codes with various boundary conditions. The domain of our simulated cases ranges from 2.19 to 5.67 pressure scale heights. Thus, the lower half of the simulation domains can be strongly influenced by the lower boundary conditions. To alleviate the boundary effects on the statistical results, we exclude the data in the region $z\in(0,0.4)$ in our analysis. For the shallowest group A, the whole domain is slightly larger than 2 pressure scale heights. Thus, it is possible that the whole domain is affected by the lower boundary condition. To address this problem, we have performed an additional simulation by adding a convectively stable zone below the unstable zone in case A1. Fig.~\ref{fig:boundaryEffect} compares the horizontal and vertical velocities in the cases with and without the lower stable zone. From the figure, we can see that both the horizontal and vertical velocities at the bottom are strongly affected by the lower boundary condition. Thus, it is necessary to exclude the lower part of the domain in the analysis. The horizontal velocity at the top has also been slightly affected (about 15 percent). To obtain reliable statistical results, we mainly focus on the data in the middle of the computational domain $z\in(0.4,0.7)$, where turbulent motions are less influenced by the boundary effect.

\end{document}